\documentclass[%
 reprint,
 amsmath,amssymb,
 aps,
prd,
]{revtex4-2}

\usepackage{dcolumn}
\usepackage{bm}
\usepackage[T1]{fontenc}
\usepackage[utf8]{inputenc}
\usepackage{graphicx} 
\usepackage{amsmath,amsfonts,amssymb,amsthm,bm}
\usepackage{mathtools}
\usepackage{mathrsfs}
\usepackage{pzccal}
\usepackage{textcomp}
\usepackage[dvipsnames]{xcolor}
\usepackage{tensor}
\usepackage{bm,upgreek}
\usepackage[makeroom]{cancel}
\usepackage{pifont}
\usepackage{circledsteps} 
\usepackage{multirow}
\usepackage{subcaption}
\usepackage{color,soul}

\usepackage{bm}

\DeclareSymbolFont{matha}{OML}{txmi}{m}{it}
\DeclareMathSymbol{\varv}{\mathord}{matha}{118}

\def\inner#1{\langle \, #1 \, \rangle}

\def\E{$\mathscr{E}\,$}
\def\O{$\mathscr{O}\,$}
\def\pE{$p_{\mathscr{E}}\,$}
\def\pO{$p_{\mathscr{O}}\,$}
\def\DL{\boldsymbol{\Delta L}}

\def\X#1{\boldsymbol{X}_{#1}}

\def\bU{\boldsymbol{{U}}}

\newcommand{\U}[2]{\bU\left({#1}, {#2}\right)}
\newcommand{\UOO}[2]{\tensor{U}{^{\mathscr{O}}^{\mathscr{O}}_{\Vec{#1}}_{\Vec{#2}}}}
\newcommand{\UOE}[2]{\tensor{U}{^{\mathscr{O}}^{\mathscr{E}}_{\Vec{#1}}_{\Vec{#2}}}}
\newcommand{\UEO}[2]{\tensor{U}{^{\mathscr{E}}^{\mathscr{O}}_{\Vec{#1}}_{\Vec{#2}}}}
\newcommand{\UEE}[2]{\tensor{U}{^{\mathscr{E}}^{\mathscr{E}}_{\Vec{#1}}_{\Vec{#2}}}}

\newcolumntype{P}[1]{>{\centering\arraybackslash}p{#1}}
\newcolumntype{M}[1]{>{\centering\arraybackslash}m{#1}}

\begin{document}

\preprint{APS/123-QED}

\title{Algebraic symmetries of the observables on the sky: Variable emitters and observers}

\author{Mikołaj Korzyński}
\email{korzynski@cft.edu.pl}
\author{Nezihe Uzun}%
 \email{nuzun@cft.edu.pl}
\affiliation{%
 Center for Theoretical Physics, Polish Academy of Sciences, Al. Lotników 32/46, 02-668 Warsaw, Poland
}%

\date{\today}

\begin{abstract}
In this paper we prove a number of exact relations between optical observables, such as trigonometric parallax, position drift and the proper motion of a luminous source in addition to the variations of redshift and the viewing angle. These relations are valid in general relativity for any spacetime and they are of potential interest for astrometry and precise cosmology.
They generalize the well--known Etherington's reciprocity relation between the angular diameter distance and the luminosity distance. Similar to the Etherington's relation, they hold independently of the spacetime metric, the positions and  the motions of a light source or an observer. We show that those relations follow from the symplectic property of the bi--local geodesic operator, i.e., the geometric object that describes the light propagation between two distant regions of a spacetime. 
The set of relations we present is complete in the sense that no other relations between those observables should hold in general. In the meantime, we develop the mathematical machinery of the bi--local approach to light propagation in general relativity and its corresponding Hamiltonian formalism. 
\end{abstract}

\maketitle


\section{\label{sec:Introduction}Introduction}
The concept of parallax has been used throughout the history of astronomy as an important tool to identify distances. The change of an observable at the linear order as the observer undergoes spatial translations incorporates important information about the underlying light propagation problem. The variations of the apparent positions under spatial translations of the source, on the other hand, are related to the image magnification and deformation, which play important roles in the gravitational lensing theory. Moreover, for the last few decades, continuous observations have been found instrumental in identifying features of our universe. The change of an observable at the linear order with respect to the observer's proper time, registered  over a finite time interval, is usually referred to as the \textit{drift effect}. Among such observables, the drift data of the redshifts, positions and distances are known to provide significant information about the underlying cosmological geometry and kinematics. In this work, we investigate the variations of observables in the widest scope. Essentially, we study the variations of observables under the variations of both the source and the observer, both in space and in time. 

Variational observables may appear unrelated at a first glance as their measurements involve rather different techniques. Our aim here is to show that from the theoretical perspective, once those observables are studied under a rigorous mathematical formalism, they form a set of closely related quantities. This follows from the fact that they are all related to the derivatives of the null
tangent vectors at the two endpoints of a null geodesic that connects the emitter and the observer. The first order dynamics of light propagation can then be studied with the help of the geodesic deviation equation (GDE) at the first order. This equation is a linear ordinary differential equation defined on a symplectic phase space. As a result of the corresponding algebraic symmetries and the reciprocity relations, one can construct exact relations between seemingly unrelated observables.

This argument is important not only for mathematical, but also for physical understanding of light propagation in general. For example, the redshift drift was studied intensively as a tool to test different cosmological scenarios. In the case of Friedmann--Lema{\^i}tre--Robertson--Walker
(FLRW) geometry, it provides direct information about the history of the Hubble parameter \cite{Sandage:1962,McVittie:1962,Loeb:1998,Quercellini:2010}. It is used to analyse the dark energy content within the standard model \cite{Lobo:2020hcz,Geng:2015ara,Guo:2015gpa}.  It has also been analysed for different geometries such as Lema{\^i}tre-Tolman-Bondi (LTB) \cite{Yoo:2010hi,Koksbang:2022upf}, Szekeres \cite{Mishra:2012vi,Mishra:2014vga}, Stephani \cite{Balcerzak:2012bv}, Bianchi I \cite{Fleury:2014rea} and generic spherically symmetric geometries \cite{Uzan:2008qp}. In addition, it was suggested as a means to differentiate alternative cosmological models, like the Timescape \cite{Wiltshire:2009db}. Note that all of the aforementioned studies focus on derivation of the redshift drift on specific geometries. More inclusive and holistic approaches were developed in later years in order to distinguish the effects of averaging and backreaction in a generic setting \cite{Koksbang:2015ctu,Koksbang:2019glb}. In addition, the expansion history of the universe was investigated independent of the underlying cosmological model  in which a general inhomogeneous and anistropic model was expressed via the the Taylor expansion of the kinematic variables of the observer's congruence \cite{Heinesen:2021nrc,Heinesen:2021qnl}. This method is referred to as cosmography in various contexts \cite{Visser:2004bf,Martins:2016bbi,Li:2019qic,Capozziello:2021xjw,Rocha:2022gog,Adamek:2024hme}. Note that the standard formalism of Sachs, which was originally developed for momentary observations \cite{Sachs:1961zz}, was extended with a covariant geometric formalism to include the drift and parallax effects in \cite{Korzynski:2017nas}. 

The optical  drift effects are also of importance in gravitational wave physics. Namely, the signal measured by interferometric detectors may in fact be expressed in terms of the redshift drift \cite{PhysRevD.90.062002, doi:10.1142/S021827181741022X}. Moreover, it is  possible to use the astrometric measurements of the position drifts (proper motions) to constrain the ultra--low frequency gravitational waves 
\cite{10.1093/mnras/stad2141, PhysRevD.34.1759, braginsky1990, Fakir:1993bj, Pyne:1995iy, Kaiser_1997, PhysRevD.48.2389, Damour_1998, Jaffe_2004, Schutz_2009, Book_Flanagan_2011, Qin_2019}.

The position drift, being essentially related to the redshift drift, is less studied in the community. Its relation to the late time anisotropies was investigated by a few groups \cite{Quercellini:2008ty,Quercellini:2010,Fontanini:2009qq,Krasinski:2012ty}. Its covariant formulation was developed in \cite{Korzynski:2017nas} within the same framework for the redshift drift. The distance drift, on the other hand, is the least studied one, due to the difficulty in the determination of angular diameter distances via analytical calculations.

Position drift and parallax effects \footnote{Note that the terminology varies among authors, many call parallax what is here referred to as the position drift.} were studied mostly due to their relevance for the anisotropies in the universe. Previously, cosmic parallax measured by an off--center observer in an LTB geometry was calculated to put constraints on cosmic anisotropy \cite{Quercellini:2008ty,Quercellini:2009ni}. Axisymmetric, homogeneous, Bianchi I metric was analysed to discuss whether the cosmic parallax can be used to differentiate between different anisotropic models \cite{Fontanini:2009qq}. An ellipsoidal Bianchi I model was also studied to set some constraints on anisotropies \cite{Campanelli:2011wz}. Whether the cosmic parallax can distinguish between the LTB void models and the dark energy theories in reproducing the apparent accelerated expansion of the universe was investigated in \cite{Quartin:2009xr}. Cosmic parallax was studied to identify the peculiar and proper acceleration in clustered systems to determine the gravitational potential \cite{Quercellini:2010zr}. Parallax data of quasars was analysed to distinguish between isotropic and anisotropic models \cite{Amendola:2013bga}. In addition, a consistency test for the FLRW model was suggested by using the comparison of parallax and angular diameter distances \cite{Rasanen:2013swa}. 
In \cite{Grasso:2018mei} it was proven that the relative difference between the parallax distance and the angular diameter distance can be used to measure the matter density along the line of sight irrespectively of the spacetime metric, while in \cite{Korzynski:2019oal} the signal was calculated in the standard homogeneous cosmology. The relation between the parallax distance and angular diameter distance (or luminosity distance) can also be used to test the null energy condition in any spacetime  \cite{PhysRevD.105.084017}. 

Some of the variational observables seem very difficult or impossible to measure with the current instruments in an astronomical setting. This is especially true for the
variations of the viewing direction, i.e., the spatial direction from which an anisotropic source is observed at a given moment. Measurements of viewing direction variations would require resolving a well--understood, anisotropic source of light. On the other hand,  the secular variations of the viewing angle  due to the proper motion were considered in the context of binary pulsar timing \cite{Kopeikin_1996, Splaver_2005, Arzoumanian:1996qq, PhysRevX.11.041050}. We nevertheless include those observables as the theoretical formalism appears incomplete without them. 

In the current work, we focus on optical drift effects and parallax--like effects as well. Specifically, we investigate the underlying algebraic symmetries of variational observables on the sky. The reason for such an inquiry is to relate certain seemingly unrelated types of observations within a rigorous mathematical framework. We believe that with such a framework, the properties of the background geometry can be better understood. The underlying formalism that is used in this work was developed in a series of papers \cite{Korzynski:2017nas,Grasso:2018mei,Korzynski:2019oal,Korzynski:2021aqk}. This framework is distinct in the sense that it combines multiple concepts in a novel way. Namely, there are four main elements which distinguish it from the standard usage of the geometric optics literature within an astrophysical or cosmological setting. Those are 
\begin{itemize}
    \item [(i)] circumvention of the concept of congruence of null geodesics,
    \item [(ii)] application of the geodesic deviation equation in order to study the drift and parallax effects in a covariant manner,
    \item [(iii)] implementation of bi--local tensors and vectors,
    \item [(iv)] introduction of a double tangent space and its corresponding symplectic formulation to study light propagation.
\end{itemize}

The paper is constructed as follows. In Section~\ref{sec:Preliminaries}, we reintroduce the aforementioned construction necessary to build our framework. For this, we explain why the concept of a congruence is too limiting for the purpose of studying all possible variational observations. Then, we give a summary of the symplectic phase space dynamics of light propagation in relation to the bi--locality and the geodesic deviation equation. We show that the canonical initial value problem of light rays can be transformed into a boundary value problem between the emitter and the observer. This means that the symplectic matrices of the linear initial value problem can be traded for symmetric matrices which are the solutions of the corresponding boundary value problem. Although switching between the boundary data and the initial data formulation is merely a mathematical transformation, it allows us to demonstrate the relationships between different variational observables more easily. 

In Section~\ref{sec:Scalar observables and bi--local variations}, we introduce a symmetric operator associated with the symmetric solution matrices of the boundary value problem. One can construct different observables resulting from the end point variations of the null geodesics with this operator. We then argue that those end point variations are not completely arbitrary. In other words, the variations at the observer and the variations at the emitter are not always independent from each other if we assume the endpoints to be connected by a null geodesic at all times. Obviously, this assumption is crucial when we consider a measurement of finite duration: we need to make sure that the observer keeps on receiving signals from the source during the course of the  measurement. Hence, we introduce the concept of \textit{permissible bi--local variations}.

We start Section~\ref{sec:The observables on the sky} by introducing two types of observables: (i) momentary observables such as the redshift, the position vector and the viewing direction vector which can be measured locally by an observer, (ii) variational observables such as the drift effects and the effects of transverse variations on the aforementioned momentary observables. Those include magnifications and parallax--like variations of the redshift, the position vector and the viewing direction vector. We show that one can relate those observables to the variables describing the kinematics of the observer and the source, in addition to the solution matrix of the boundary value problem corresponding to the geodesic deviation equation. The solution matrix can be seen as a functional of the curvature along the line of sight. This allows us to relate the variational observables to the spacetime geometry in a covariant way. Speaking in more informal terms, we express a broad set of  observables via appropriate integrals of Riemann curvature tensor along the line of sight as well as momentary 4--velocities and 4--accelerations.

In Section~\ref{sec:Physical outcomes}, we outline the relationships between those seemingly unrelated variational observables. We show that the underlying reason for those relations to occur are the algebraic symmetries of the boundary value problem corresponding to the geodesic deviation equation. Equivalently, those relationships can be viewed as the result of symplectic symmetries of the corresponding initial value problem. In particular, we show that Etherington's well--known reciprocity relation \cite{Etherington:1933} between the angular diameter distance and the luminosity distance is only one of the many consequences of the underlying symmetries. In this work, we provide the full set of relationships between the drift and the parallax effects, generalizing the Etherington's relation.

Finally, in Section~\ref{sec:Summary and conclusions} we provide a summary and demonstrate those physical relations in a more compact form. We conclude with certain remarks which might be useful for future light propagation studies.

In this work, we adopt geometrized units, i.e., $c$ and $G$ are set to one. The metric signature of the spacetime is chosen to be $(-,+,+,+)$.

\section{\label{sec:Preliminaries}Preliminaries}
The mathematical formalism of this paper was largely laid out in \cite{Grasso:2018mei, Korzynski:2021aqk}. We include a broad discussion of the formalism below, although without most derivations, referring the readers to those publications for details.
\subsection{\label{Abandoning the ``congruence'' concept} Abandoning the congruence concept}
In the standard usage of geometric optics limit for astronomical or cosmological observations, one investigates the dynamics of a null congruence in order to understand certain properties of the background spacetime. This method provides a geometric means to link the observed distances, brightnesses and image distortions on the sky to our theoretical calculations. In this paper, we would like to argue that such a practice is relevant for \textit{momentary} observations. Meaning, one needs to use a family of null geodesics around a \textit{fixed} central geodesic with an anchored tangent vector to study the dynamics of the observations.

In order to be precise, let us consider an emitter \E, also referred to as the source, and an observer \O performing \textit{instantaneous measurements} at two distinct spacetime points \pE and \pO respectively. In  this situation it is customary to consider thin light pencils shot from the emission and observation points in each other's direction. Both pencils form parts of congruences with vertices at \pO and \pE respectively.   The pencils are assumed to share the same central null geodesic $\eta\left(\lambda \right)$ which is parameterized by the affine parameter $\lambda$. Its tangent vector will be denoted by $\Vec{l}$, where the over--arrow is used to represent 4--dimensional vectors in their abstract form. This vector satisfies 
\begin{eqnarray}\label{eq:definition_l}
\qquad \nabla _{\vec{l}}\,\vec{l}=0,\qquad   \inner{\vec{l}, \vec{l}\,}=0,\qquad l_\mu=\nabla _\mu S,
\end{eqnarray}
where $\nabla$ represents the covariant derivative operator, $S$ is a function known as the \textit{ phase function} and $\Vec{l}$ is a gradient field in the geometric optics limit. For brevity we denote the inner product related to the spacetime metric $g$ with the angle brackets, $\inner{\cdot,\cdot}$.

\begin{figure}[!tbp]
  \begin{subfigure}[b]{0.40\textwidth}
    \includegraphics[width=\textwidth]{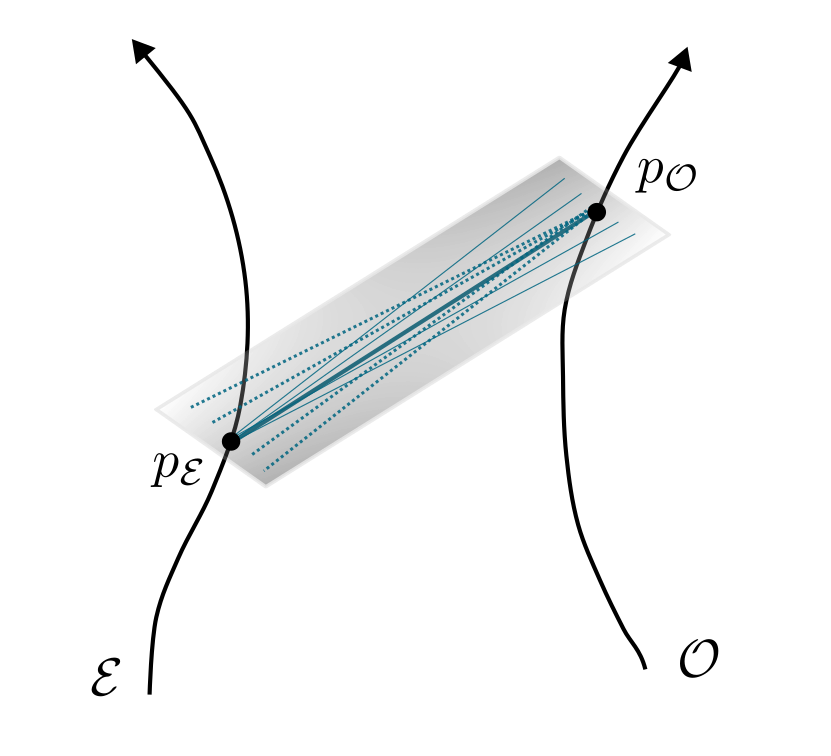}
    \caption{Congruences of null geodesics defined for momentary observations: from the observer to the emitter, $\mathbb{C}_{\mathscr{O}\mathscr{E}}$, and from the emitter to the observer, $\mathbb{C}_{\mathscr{E}\mathscr{O}}$.   In each congruence the null geodesics cross at a vertex. The deviation vector is orthogonal to the tangent vector $\Vec{l}$ at every point of the central null geodesic. Both congruences can only describe momentary observations, as they both lie in a null subspace represented by the shaded surface. One spatial dimension is suppressed, while the time dimension is retained.}
    \label{fig:congruence}
  \end{subfigure}
  \hfill
  \begin{subfigure}[b]{0.40\textwidth}
    \includegraphics[width=\textwidth]{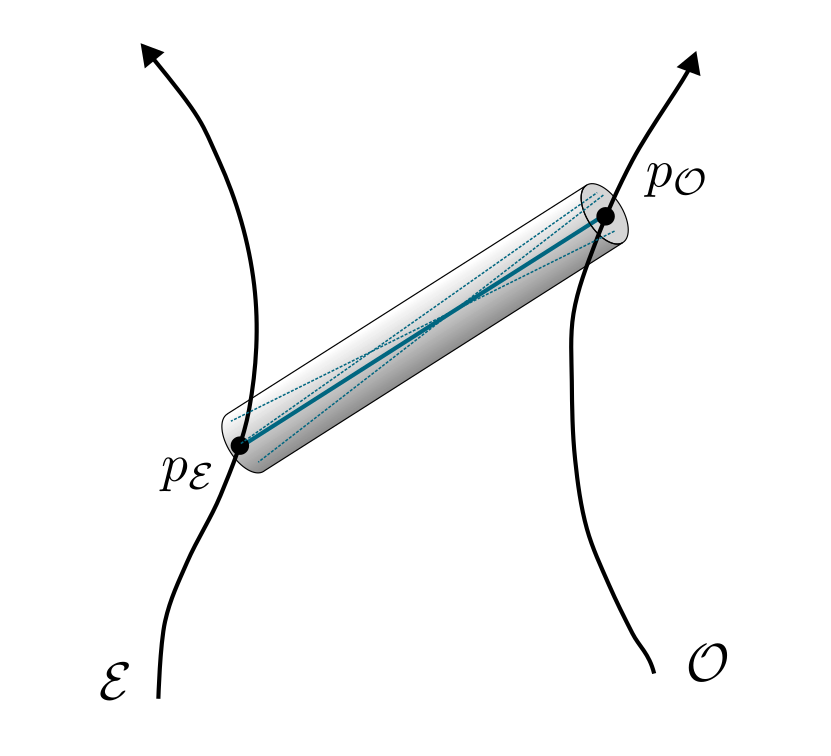}
    \caption{All possible deviations of the fiducial null geodesic, contained in a narrow tube connecting the neighbourhood of $p_{\mathscr{E}}$  with the 
    neighbourhood of $p_\mathscr{O}$. The spacetime geometry inside the tube is almost flat, with the exception of small curvature corrections influencing the behaviour of the perturbed geodesics. }
    \label{fig:tubes}
  \end{subfigure}
  \caption{Comparison of momentary observations to variable observations.}
\end{figure}

We now introduce a tetrad adapted to the direction of light propagation. We begin by picking a timelike, future--pointing and normalized 4--vector $u^\mu$, representing the frame of an observer, not necessarily coinciding with \O.
The null tangent vector, $\vec{l}$, can be decomposed into two portions aligned with a timelike 4--velocity vector, $\Vec{u}$, and a spacelike observation direction vector, $\Vec{r}$, as 
\begin{eqnarray}
    \Vec{l}=\omega\left(-\Vec{u}+\Vec{r}\right), \label{eq:om}
\end{eqnarray}
where $\omega$ is the frequency of light. Vector $\vec u$ may be related to a physical observer or to the source, but it may also be just a normalized timelike vector we pick for convenience. The position vector, $\Vec{r}$ may now serve as one of the spatial vectors of the tetrad. One can complete the tetrad by picking two transverse screen vectors $\vec e_A$ with $A, B \in \{2, 3\}$, satisfying
\begin{eqnarray}\label{eq:Sachs_basis}
 \inner{\Vec{l},\Vec{e}_A\,}=0,\,\,\,\, \inner{\Vec{e}_A,\Vec{e}_B\,}=\tensor{\delta}{_A_B},\,\,\,\, \inner{\Vec{e_A}, \Vec{u}} = 0.
\end{eqnarray}
It can be assumed that the tetrad $(\Vec{u},\Vec{r},\Vec{e_A})$ is parallel propagated along the null geodesic. The reader may check that $ \omega=\textrm{constant}$ holds
due to the relation (\ref{eq:om}) in this case. However, such an assumption is not necessary in general and it is possible to work with a non-parallel propagated tetrad as well.
One can then construct a null congruence, $\mathbb{C}_{\mathscr{E}\mathscr{O}}$, of light rays around $\eta$ from \E to \O,  and another one, $\mathbb{C}_{\mathscr{O}\mathscr{E}}$, from \O to \E (See Fig.~\ref{fig:congruence}.). 

For either of the cases, one needs to consider the geodesic deviation vector, $\delta \Vec{x}$, of the null bundle in order to study the dynamics of the corresponding momentary congruence. In general, the first order part of the dynamics is considered such that the deviation vector can be treated as a Jacobi field satisfying
\begin{eqnarray}\label{eq:first_ord_dev}
\nabla _{\vec{l}}\nabla _{\vec{l}}\,\delta{x}^\alpha=\tensor{R}{^\alpha_{\vec{l}}_{\,\vec{l}}_{\delta\vec{x}}}\,.
\end{eqnarray}
As the observation point is a vertex, the value of the deviation vector is zero at this point. It is easy to check that in this case we have
\begin{eqnarray}\label{eq:Orthog_l_deltax}
    \inner{\Vec{l},\delta \Vec{x}}=0.
\end{eqnarray}
along the null geodesic \cite{Grasso:2018mei, Korzynski:2017nas}. This property is preserved through out the propagation due to the first order deviation equation (\ref{eq:first_ord_dev}).
Then, one can decompose the deviation vector into portions parallel and orthogonal to the propagation vector as 
\begin{eqnarray}\label{eq:moment_dev_decomp}
    \delta \Vec{x}=\delta x^l \Vec{l}+\delta x^A \Vec{e}_A,
\end{eqnarray}
due to eq.~(\ref{eq:Orthog_l_deltax}). 
Moreover, the integrability conditions require the Lie dragging of the deviation vector and the null tangent vector along each other, i.e.,
\begin{eqnarray}\label{eq:Lie_dragging}
  \nabla _{\Vec{l}}\, \delta \Vec{x}= \nabla _{\delta \Vec{x}}  \Vec{l}.
\end{eqnarray}
Substituting the decomposition, (\ref{eq:moment_dev_decomp}), into the integrability condition, (\ref{eq:Lie_dragging}), and reading off the screen space components of the terms gives the well known Sachs' optical equations
\begin{eqnarray}\label{eq:Sachs_eqs}
\nabla _{\Vec{l}}\,\rho &=& \rho ^2+|\sigma|^2+\Phi _{00},\nonumber\\
\nabla _{\Vec{l}}\,\sigma &=& 2\rho \sigma+\Psi _{0},
\end{eqnarray}
where $\rho=-\theta$ represents the expansion scalar of the twist--free congruence and the shear, $\sigma=\sigma _R +i \sigma _I$, is a complex function with
\begin{eqnarray}\label{eq:defn_expansion_shear}
\theta&=&\frac{1}{2}\left(\inner{\nabla_{\Vec{e}_2}\Vec{l},\Vec{e}_2}+{\inner{\nabla_{\Vec{e}_3}\Vec{l},\Vec{e}_3}}\right),\\
\sigma_R&=&\frac{1}{2}\left(\inner{\nabla_{\Vec{e}_3}\Vec{l},\Vec{e}_3}-{\inner{\nabla_{\Vec{e}_2}\Vec{l},\Vec{e}_2}}\right),\\
\sigma_I&=&\frac{1}{2}\left(\inner{\nabla_{\Vec{e}_2}\Vec{l},\Vec{e}_3}+\inner{\nabla_{\Vec{e}_3}\Vec{l},\Vec{e}_2}\right).
\end{eqnarray}
The scalars corresponding to the Ricci tensor, $R_{\mu \nu}$, and the Weyl tensor, $C_{\mu \nu \alpha \beta}$, are respectively given as
\begin{eqnarray}\label{eq:Ricci_Weyl_scalars}
\Phi _{00}= \frac{1}{2}R_{\Vec{l}\,\Vec{l}},\qquad \Psi _{0}= C_{\Vec{l}\Vec{m}\Vec{l}\Vec{m}}
\end{eqnarray}
with $\Vec{m}=\left(\Vec{e}_2-i\Vec{e}_3\right)/\sqrt{2}$.
Thus, solving the Sachs' optical equations, (\ref{eq:Sachs_eqs}), is equivalent to solving the first order part of the geodesic deviation equation, (\ref{eq:first_ord_dev}), for a momentary congruence composed of \textit{abreast} rays \cite{Penrose:1984}. Those rays remain orthogonal to the connecting vector of the null congruence throughout the propagation. They span the same hypersurface when the congruence is non--twisting. This means eq.~(\ref{eq:Orthog_l_deltax}) holds and the geodesics of such a congruence belong to the same null hypersurface.

In the literature, it is a common practice to define congruences through the deviation of curves from a fixed, central curve under the perturbation of the position coordinates, $\Vec{x}\rightarrow \Vec{x}+\delta\Vec{x}$. For example, one way of deriving the geodesic deviation equation is to consider the geodesic deviation action which is obtained through the difference of the action of the central  geodesic with tangent vector $\Vec{l}\left(\Vec{x}\right)$ and the action of the outermost geodesic with tangent vector $\Vec{l}\left(\Vec{x}+\delta\Vec{x}\right)$. What is usually overlooked is the fact that in the case of prolonged observations we need to consider perturbations of a null geodesics corresponding to the variation of the emission time. These cannot be described by perturbations obeying equation (\ref{eq:Orthog_l_deltax}). Moreover, as we will see, determination of the observables related to the position variations requires the calculation of the perturbations of the tangent vector itself $\Vec{l}\rightarrow \Vec{l}+\delta\Vec{l}$. Those are considered under general variations of the geodesic endpoints while allowing the observer to receive signals from the emitter during the measurement process. 

We would like to argue here that the notion of a congruence is too limiting for our purposes. As each tangent vector is 4--dimensional at the two end points, the representation space of the family of all geodesics is 8--dimensional. If we include the condition for the geodesic to be null, then the dimension reduces to 7. On the other hand, a congruence is a family of geodesics in which one and only one geodesic passes through a single point of the spacetime. It is therefore \textit{at most} 3-dimensional. In addition, the transverse congruences 
$\mathbb{C}_{\mathscr{O}\mathscr{E}}$ and $\mathbb{C}_{\mathscr{E}\mathscr{O}}$ which are composed of abreast rays are only 2--dimensional. Therefore, a congruence cannot possibly describe \textit{all} admissible variations of the two endpoints. Thus, studying congruences would inevitably limit the considerations to a small sub--family of all possible variations of the fiducial null geodesic. We need to pass to a more general picture with as general variations as possible offered by the bi--local approach to light propagation.

In this paper, we distinguish \textit{momentary  observables} related to the properties of a single null geodesic connecting the source and the observer, and the \textit{variational observables}, defined via the variations of the instantaneous observables when we vary the geodesic endpoints. The examples of the former include the apparent position of the source's image on the observer's celestial sphere, the redshift and the momentary viewing direction.
The examples of the latter, on the other hand, include for example (i) the magnification matrix which relates the variation of the apparent position of the image on the observer's sky as we displace the source in a transverse direction; (ii) the  position drifts and the redshift drifts, i.e., the temporal variations of the apparent position and the redshift, with respect to the observer's proper time.

In this work, our main aim is to find algebraic relationships between the variational observables. Those relationships are in principle analogous to the distance reciprocity, also known as the Etherington's reciprocity relation, in general relativity. Note that, the angular diameter distance and the luminosity distance seem unrelated at first, because they are calculated for distinct congruences. One can relate the distances only when the first order geodesic deviation equation and the symmetries of the Riemann tensor are considered. Similarly, imagine a situation where we limit ourselves to null geodesics centered at the emitter's worldline or to the one centered at the observer's worldline. The former seems appropriate if we study observables related to the viewing direction, while the latter appears to be  relevant only if we are dealing with the apparent position on the observer's sky. Those two do not seem {\it a priori} related if we study them by considering distinct congruences.

This argument is relevant for the drift effects as well. For the  measurement of the drift effects, the  observer must register radiation from the emitter throughout the course of the observation. Therefore, we need to consider a family of null geodesics connecting \E and \O along their timelike worldlines, $\gamma _{\mathscr{E}}$ and $\gamma _{\mathscr{O}}$, respectively. Note that, in general, \E and \O might have coordinate accelerations, therefore, $\gamma _{\mathscr{E}}$ and $\gamma _{\mathscr{O}}$ are not necessarily geodesics. Since the drifts require comparing null geodesics corresponding to different emission and observation times, studying only the dynamics of the aforementioned congruences $\mathbb{C}_{\mathscr{O}\mathscr{E}}$ and $\mathbb{C}_{\mathscr{E}\mathscr{O}}$ which consist of only abreast null rays with a fixed tangent vector fall short for the given observational set up. Accordingly, the Sachs' optical equations do not provide all the necessary information. 

In that case, one needs to develop a method that allows us to vary both ends of the connecting null geodesic throughout the course of the measurement. In the mean time, we should consider variations in timelike directions as well. With those developed tools, we can study light propagation with variations that are as general as possible. This framework allows us to relate all possible variational observables to the spacetime geometry. Moreover, it turns out that with such a formalism we can show that there are many more relations between the variational observables analogous to the distance reciprocity in relativity.

In order to achieve this, we consider the close neighbourhoods, $N_{\mathscr{E}}$ and $N_{\mathscr{O}}$ of \E and \O respectively (See Fig.~\ref{fig:tubes}.). Those regions are assumed to be small enough in size with respect to the characteristic curvature scale of the underlying spacetime. We consider them to be locally flat. As \E and \O evolve, the connecting  null geodesic $\eta$ links the worldlines  $\gamma _{\mathscr{E}}$ and $\gamma _{\mathscr{O}}$. This is true for each instant, even if \E and \O are varied by $\delta \Vec{x}_{\mathscr{E}}$ and $\delta \Vec{x}_{\mathscr{O}}$ from their fiducial locations. When \E sends a signal to \O for a finite interval of time, for example, we assume that the light ray is initiated from a point within the region $N_{\mathscr{E}}$. Obviously,  the connecting geodesics in this case are not confined to a single pair of pencils related to a single observation and emission moment, but rather they belong to many different pencils. This means that we need to relax the condition given in eq.~(\ref{eq:Orthog_l_deltax}). Meaning, from now on, we are not considering only those geodesic deviation vectors that are in orthogonal correspondence with the null ray. 

\subsection{\label{Synge's world-function, bi--tensors and bi--local variations}Synge's world-function, bi--tensors and bi--local variations}
In order to construct a formalism that suits our physical set up, we still need to define a reference null geodesic. Assume we pick two reference points $x_\mathscr{E} \in N_\mathscr{E}$ and $x_\mathscr{O} 
\in N_\mathscr{O}$ connected by a null geodesic 
$\eta$ which we call the \textit{fiducial null geodesic}. It serves
as a reference for other geodesics with endpoints 
in $N_\mathscr{O}$ and in $N_\mathscr{E}$. We need to impose the condition that the geodesics we consider are null. This is simple to achieve using the Synge's world-function \cite{Synge:1960}. In general, the world function, $\sigma \left(x,x'\right)$, corresponds to the proper length between two points, $x$ and $x'$, which are in convex neighbour of each other and which are connected by a unique geodesic, i.e.,
\begin{eqnarray}
    \sigma\left(x,x'\right)=\frac{\Delta \lambda}{2}\int ^{\lambda_x}_{\lambda_{x'}}g_{\alpha \beta}\frac{dx^\alpha}{d\lambda}\frac{dx^\beta}{d\lambda}d\lambda.
\end{eqnarray}
Here $\lambda$ is the affine parameter that parameterizes the geodesic $\eta$ as before and $\Delta\lambda=\lambda_x-\lambda_{x'}$.
In general, the value of $\sigma$ depends on the causal character of the corresponding connecting geodesic; it is positive for spatial ones, negative for timelike ones and zero for the null case. In our work, we consider only null geodesics, thus we demand $\sigma=0$. With the  caustic--free assumption, this condition defines a local co--dimension--1 hypersurface in $M \times M$, i.e., product of the spacetime by itself. In \cite{Korzynski:2021aqk}  this surface was referred to as \textit{local surface of communication} and its geometry was investigated in detail \footnote{Note, however, that the 0 level set of $\sigma$ is not globally a regular hypersurface as it contains singular points along the diagonal and near caustics.}.

The world function is a bi--local object, so are its derivatives. Let us define its first order derivatives as
\begin{eqnarray}\label{eq:world_fun_tangent_vec}
    l_{\mu'}:=-\frac{1}{\Delta \lambda}\frac{\partial \sigma}{\partial x^{\mu'}}, \qquad l_{\mu}:=\frac{1}{\Delta \lambda}\frac{\partial \sigma}{\partial x^{\mu}}.
\end{eqnarray}
It is well--known that both vectors defined this way are tangent to the geodesic connecting $x$ and $x'$ \cite{Poisson:2011nh}. Note that we use primed coordinates, $x^{\mu'}$, for the \textit{base point} and unprimed coordinates, $x^{\mu}$, for the \textit{field point} as in \cite{Poisson:2011nh}. In this work, the former corresponds to the coordinates in the observer's neighbourhood and the latter are the ones of the emitter. We should emphasize that the variational observables we consider in this work depend on both the observation and on the emission points. Essentially, they are obtained through the end point variations of the null geodesics. Hence, $l_{\mu'}$ and $l_{\mu}$ should be varied covariantly while preserving their bi--local character.

In order to achieve this task, bi--local variation vectors were presented in \cite{Korzynski:2021aqk}. Those are 8--dimensional vectors, $\X{}$, which combine the two 4--dimensional end point variation vectors. Vectors  $\X{}$ span the direct sum of the tangent spaces defined at \pO and \pE, or, equivalently, the  space tangent to of the product $M \times M$ at $(p_\mathscr{O}, p_\mathscr{E})$. They represent the end point variations combined into a single object from the vector space $T_{p_\mathscr{O}} M\oplus T_{p_\mathscr{E}} M \cong T_{(p_{\mathscr{O}},p_\mathscr{E})} M\times M$. We write them as 
\begin{align}\label{eq:bilocal_var}
        \X{}= 
        \begin{bmatrix}
         {\delta x}_{\mathscr{O}}^{\mu'} \\ {\delta x}_{\mathscr{E}}^{\mu}
        \end{bmatrix}. 
        \end{align}
Here, ${\delta \vec{x}}_{\mathscr{O}}$ is the 4--dimensional variation vector at the end point of the null geodesic where the observer is located and ${\delta \vec{x}}_{\mathscr{E}}$ is the one located at the other end point of the geodesic, i.e., where the emitter is located. 

Recall that $\Vec{l}_{\mathscr{O}}$ and $\Vec{l}_{\mathscr{E}}$ are genuine bi--local objects, i.e., they depend on two points on $M$ and they transform as vectors under coordinate transformations at the respective end points. Therefore, their differentiation with respect to the end point variations is not trivial. For instance, the components $l_{\mu'}$ act as co--vector components for the tensorial operations performed on the observer's side and they act as 4--scalars for operations on the emitter's side. Similarly, the components $l_{\mu}$ act as co--vector components for tensorial operations performed on the emitter's side and they act as scalars for operations on the observer's side \cite{Poisson:2011nh}. Accordingly, we define a bi--local variation operator, $\boldsymbol{\nabla}_{\X{}}$, with respect to the 8--dimensional variation vector in eq.~(\ref{eq:bilocal_var}) acting on an arbitrary bi--local vector, $\Vec{v}$ as
\begin{eqnarray}\label{eq:def_bilocal_var_v}
\boldsymbol{\nabla}_{\X{}}\Vec{v}
:=
 \begin{bmatrix}
         \Delta{\Vec{v}}_{\mathscr{O}} \\ \Delta{\Vec{v}}_{\mathscr{E}}
        \end{bmatrix}
=    
\begin{bmatrix}
 \delta x_\mathscr{O}^{\mu'}\,\nabla_{\mu'}\Vec{v}_\mathscr{O} + \delta x_\mathscr{E}^{\mu}\,\partial_{\mu}\Vec{v}_\mathscr{O} \\
x_\mathscr{O}^{\mu'}\,\partial_{\mu'}\Vec{v}_\mathscr{E} + \delta x_\mathscr{E}^{\mu}\,\nabla_{\mu}\Vec{v}_\mathscr{E}
\end{bmatrix}.
\end{eqnarray}
Thus, from now on, when we refer to the variation of the tangent vectors of the null geodesics at the end points, we refer to the covariant bi--local variation,
\begin{eqnarray}\label{eq:def_bilocal_var_l}
\boldsymbol{\nabla}_{\X{}}\Vec{l}
:=
 \begin{bmatrix}
         \Delta{\Vec{l}}_{\mathscr{O}} \\ \Delta{\Vec{l}}_{\mathscr{E}}
        \end{bmatrix}. 
\end{eqnarray}

Since the local covariant derivative operator $\nabla_\mu$ can act on tensors and scalars, this construction extends easily to tensors and scalars as well. For a generic bi--local tensor, $A_{\alpha'\beta'\dots\kappa\lambda\dots}$, we have 
\begin{eqnarray}\label{eq:def_bilocal_var_tensor}
    {\bm \nabla_{\bm X}} A_{\alpha'\beta'\dots\kappa\lambda\dots} = { \delta x}_{\mathscr{O}}^{\rho'}\,A_{\alpha'\beta'\dots\kappa\lambda\dots;\rho'} + { \delta x}_{\mathscr{E}}^\sigma\,A_{\alpha'\beta'\dots\kappa\lambda\dots;\sigma}\nonumber\\
\end{eqnarray}

Now, let us return to our physical problem and consider every emitter in $N_{\mathscr{E}}$ which can send a signal to all possible points in $N_{\mathscr{O}}$. Local surface of communication is defined as the zero level surface of the worldfunction $\sigma$. In this respect, $\sigma $ can be expanded through the variations $\delta \Vec{x}_{\mathscr{E}}$ and $\delta \Vec{x}_{\mathscr{O}}$ at the two end points. Then, at the first order approximation one gets \cite{Korzynski:2021aqk},
\begin{eqnarray}\label{eq:synge_wf_expansion}
\sigma \left(\Vec{x}_{\mathscr{E}}+\delta \Vec{x}_{\mathscr{E}},\Vec{x}_{\mathscr{O}}+\delta \Vec{x}_{\mathscr{O}}\right)=\left(\partial _{\mu }\sigma\right) {\delta x}{^{\mu }_{\mathscr{E}}}+\left(\partial _{\mu '}\sigma\right) {\delta x}{^{\mu '}_{\mathscr{O}}}.\nonumber \\
\end{eqnarray}

As the first derivative of the world function defines the tangent vector of the curve, weighted by the affine distance as in eq.~(\ref{eq:world_fun_tangent_vec}), we have 
\begin{eqnarray}\label{eq:synge_tangent_vector}
\partial _{\mu }\sigma=\Delta\lambda\,l_{\mu },\qquad \partial _{\mu '}\sigma=-\Delta\lambda\,l_{\mu '}.
\end{eqnarray}
with $\Delta\lambda = \lambda_\mathscr{E} - \lambda_\mathscr{O}$ and $l_\mu$ and $l_{\mu'}$
being the lower--index tangent vector components of the fiducial null geodesic at $p_\mathscr{O}$ and $p_\mathscr{E}$ respectively, i.e. $l_{\mu} = l^\mathscr{E}_\mu$ and $l_{\mu'} = l^\mathscr{O}_{\mu'}$.

Then, the null character of the world function, together with equations (\ref{eq:synge_wf_expansion}) and (\ref{eq:synge_tangent_vector}) require
\begin{eqnarray}\label{eq:criteria_local_surf_com}
 \inner{\tensor{\Vec{l}}{_{\mathscr{O}}},\delta \tensor{\Vec{x}}{_{\mathscr{O}}}} - \inner{\tensor{\Vec{l}}{_{\mathscr{E}}},\delta \tensor{\Vec{x}}{_{\mathscr{E}}}}=0.
\end{eqnarray}
The equation above defines the necessary criterion for signals sent from $N_{\mathscr{E}}$ to reach $N_{\mathscr{O}}$ in the leading order approximation of the deviations $\delta \Vec{x}_\mathscr{O}$ and $\delta \Vec{x}_\mathscr{E}$. In \cite{Grasso:2018mei} this is referred to as the \textit{time lapse formula}. We refer to this criterion in the main body of the paper in order to mathematically construct the observables and study their symmetries.

\subsection{\label{sec:Symplectic symmetries and the bi--locality}Symplectic symmetries and bi--locality}

Let us now consider a covariant variation, $\Delta= \delta x^\mu \nabla _\mu$, of $\Vec{x}$ and $\Vec{l}$ rather than the standard one, $\delta=\delta x^\mu \partial _\mu$. Then, we have
\begin{eqnarray}
\Delta x^\mu&=&\delta x^\mu,\label{eq:cov_stand_x}\\
\Delta l^\mu&=&\delta l^\mu + \tensor{\Gamma}{^\mu _\alpha _\beta}l^\alpha \delta x^\beta, \label{eq:cov_stand_l}
\end{eqnarray}
Eq.~(\ref{eq:cov_stand_x}) follows from coordinates acting as scalars under the covariant derivative and $\tensor{\Gamma}{^\mu _\alpha _\beta}$ are considered to be the components of a symmetric connection in this work. 

Previously, it was shown that the perturbations of $\Vec{x}$ and $\Vec{l}$ at the emission and at the observation points can be linked to each other via a linear transform \cite{Grasso:2018mei}. Here, we denote it as $\phi$. This transformation was later studied under the lifted geodesic flows on a tangent space, $TM$, of the base manifold $M$ \cite{Serbenta:2023glm}. Then, an 8--dimensional vector
\begin{eqnarray}
    \mathbf{\mathcal{Y}}:=\begin{bmatrix}
\delta {x}^\mu \\
\Delta {l}^\nu
\end{bmatrix}
\end{eqnarray}
was treated as an object defined on the tangent space of the tangent manifold, i.e., $\mathbf{\mathcal{Y}}\in \,TTM$.
Since the $TTM$ space can be identified with a simple sum of two copies of the tangent space $T_pM$ at each point $(p, \Vec{v})$,  the transformation derived in \cite{Grasso:2018mei} was viewed as a bilocal linear map that takes the double copy of the tangent space of \O to the one of \E, i.e.,
\begin{eqnarray}\label{eq:map_phi}
\phi:T_\mathscr{O}M\oplus T_\mathscr{O}M \rightarrow T_\mathscr{E}M\oplus T_\mathscr{E}M.
\end{eqnarray}
In general, $\Vec{v}$ is a 4--dimensional tangent vector defined at an arbitrary point $p$.
More specifically,
\begin{eqnarray}
  \phi: \mathbf{\mathcal{Y} '}\rightarrow \mathbf{\mathcal{Y}},
\end{eqnarray}
where $\mathbf{\mathcal{Y}'}$ is the deviation vector at $(p_\mathscr{O}, \Vec{l}_\mathscr{O}) \in TM$, while $\mathbf{\mathcal{Y}}$ is the corresponding vector at $(p_\mathscr{E}, \Vec{l}_\mathscr{E}) \in TM$.
As the vector $\mathbf{\mathcal{Y}}$ is Lie--dragged under the geodesic spray, it can be shown that there exists a symplectic structure,
\begin{eqnarray}\label{eq:symp_struct}
\mathcal{w}=d\left(\tensor{g}{_\mu _\nu}v^{\nu}\right)\wedge dx^\mu,
\end{eqnarray}
that is closed under the flow of the geodesic spray \cite{Serbenta:2023glm}.
In this work,
\begin{eqnarray}\label{eq:canonical_coordinates}
dx^\mu := \delta x^\mu,\qquad
d{v}^\nu := \Delta {l}^\nu.
\end{eqnarray}

Now, we would like to transform the evolution problem into a setting that is aligned with the Hamiltonian formulation. Thus, instead of working on the two copies of the tangent space we would like to define a linear map, that takes an object which spans  the direct sum of the tangent space, $TM$, and the cotangent space, $T^*M$, at one point to another,
\begin{eqnarray}\label{eq:map_psi}
\psi:T_\mathscr{O}M\oplus T^*_\mathscr{O}M \rightarrow T_\mathscr{E}M\oplus T^*_\mathscr{E}M.
\end{eqnarray}
The transformation into the new picture amounts simply to lowering the index in $\Delta l^\mu$, i.e.,
\begin{eqnarray}
  \psi: \mathbf{\mathcal{Z} '}\rightarrow\mathbf{\mathcal{Z}},\qquad {\rm{with}}\qquad
  \mathbf{\mathcal{Z}}:=\begin{bmatrix}
\delta {x}^\mu \\
\Delta {l}_\nu
\end{bmatrix}.
\end{eqnarray}
Moreover, rather than considering the coordinate components of the vectors defined here, we take their local frame components. For this, we construct an \textit{auxiliary tetrad} $\Vec{e}_{\tilde{a}}$ with $\tilde{a},\tilde{b} \in \{0,1,2,3\}$ via its parallel propagation along the null geodesic, i.e.,
\begin{eqnarray}\label{eq:parallel_prop_tetrad}
    \nabla _{\Vec{l}}\,\Vec{e}_{\tilde{a}}=0.
\end{eqnarray}
This tetrad does not need to be adopted to either the emitter's or to the observer's frame. The idea is to use it as an auxilliary tool to be able to compare vectors at different spacetime points. Hence the name.
Then, we consider the evolution $\nabla _{\Vec{l}}{\mathbf{\mathcal{Z}}}$ of $\mathcal{Z}$ written on the auxiliary tetrad basis, i.e.,
\begin{eqnarray} \label{eq:X_evolution}
  \frac{d\mathbf{Z}}{d\lambda}=\mathbf{H}\,{\mathbf{Z}}, 
\end{eqnarray}
with
\begin{eqnarray}\label{eq:phase_space_vector_Hamiltonian}
\mathbf{Z}:=\begin{bmatrix}
\delta {x}^{\tilde{a}} \\
\Delta {l}_{\tilde{b}}
\end{bmatrix},
\qquad
{\mathbf{H}}=
\left[
\begin{array}{c|c}
     \mathbf{0} &  \, \tensor{g}{^{\tilde{a}}^{\tilde{b}}} \\
     \hline
     \tensor{\mathcal{R}}{_{\,\tilde{a}}_{\tilde{b}}} & \, \mathbf{0}
 \end{array}
\right],
\end{eqnarray}
where $\mathbf{0}$ is a $4\times 4$ zero matrix, $\tensor{g}{^{\tilde{a}}^{\tilde{b}}}=\tensor{g}{^\mu^\nu}\tensor{e}{_\mu^{\tilde{a}}}\tensor{e}{_\nu^{\tilde{b}}}$ is the inverse of the induced metric,  $\tensor{\mathcal{R}}{_{\,\tilde{a}}_{\tilde{b}}}:=\tensor{R}{_\mu_{\vec{l}}_{\,\vec{l}}_{\nu}}\tensor{e}{^\mu_{\tilde{a}}}\tensor{e}{^\nu_{\tilde{b}}}$ and $\tensor{R}{_\mu_{\alpha}_{\beta}_{\nu}}$ are the Riemann tensor components of the spacetime. Now that the components of $\mathbf{Z}$ are written with respect to the tetrad components, they act like scalars under the covariant derivative and the operator $\nabla _{\Vec{l}}$ becomes a standard derivative with respect to the affine parameter, $\lambda$.
The solution to this linear problem can be represented by an $8\times 8$ matrix which is the representation of the map $\psi$ introduced in eq.~(\ref{eq:map_psi}) when written with respect to a parallel propagated tetrad.
We denote the solution matrix as $\mathbf{W}$ which takes the initial value of the $\mathbf{Z}$ to its final value, i.e.,
\begin{eqnarray}\label{eq:perturb_linear_tr}
    \begin{bmatrix}
    {\delta x}_{\mathscr{E}}^{{\tilde{a}}} \\
    {\Delta l}^{\mathscr{E}}_{{\tilde{b}}}
    \end{bmatrix}
=\mathbf{W}\left(\lambda _\mathscr{E},\lambda _\mathscr{O}\right)
\begin{bmatrix}
{\delta x}_{\mathscr{O}}^{{\tilde{a}}'}\\
{\Delta l}^{\mathscr{O}}_{{\tilde{b}'}}
\end{bmatrix},
\end{eqnarray}
with the initial conditions
\begin{eqnarray}
\mathbf{W}\left(\lambda _{\mathscr{O}},\lambda _{\mathscr{O}}\right)=
\left[
\begin{array}{c|c}
     \tensor{\delta}{^{{\tilde{a}} '}_{{\tilde{b}} '}} & \mathbf{0}  \\
     \hline
     \mathbf{0} & \tensor{\delta}{^{{\tilde{a}} '}_{{\tilde{b}} ' }}
 \end{array}
\right],
\end{eqnarray}
where $\tensor{\delta}{^a_{b}}$ is the Kronecker delta. 

Substitution of the solution given in the form of eq.~(\ref{eq:perturb_linear_tr}) into eq.~(\ref{eq:X_evolution}) results in an evolution equation for $\mathbf{W}$, i.e., 
\begin{eqnarray}\label{eq:evol_W_tetrad}
\frac{d\mathbf{W}}{d\lambda}=\mathbf{H}\,\mathbf{W}.
\end{eqnarray}
What is important to realize here is that the sub--blocks of the  $\mathbf{H}$ matrix are all symmetric matrices. Namely, the inverse of the induced metric, $\tensor{g}{^a^b}$, is  symmetric and $\tensor{\mathcal{R}}{_a_b}$ is a symmetric matrix due to the symmetries of the torsionless Riemann tensor,
\begin{eqnarray}
\tensor{R}{_\mu_{\vec{l}}_{\,\vec{l}}_{\nu}}=-\tensor{R}{_\mu_{\vec{l}}_{\nu}_{\vec{l}}}=-\tensor{R}{_{\nu}_{\vec{l}}_\mu_{\vec{l}}}=\tensor{R}{_{\nu}_{\vec{l}}_{\,\vec{l}}_\mu}.
\end{eqnarray}
Therefore, $\mathbf{H}$ is a Hamiltonian matrix, i.e.,
\begin{eqnarray}\label{eq:Hamiltonian_matrix_defn}
\mathbf{J}\mathbf{H}=\left(\mathbf{J}\mathbf{H}\right)^\intercal,    
\end{eqnarray}
with
\begin{eqnarray}
\mathbf{J}=\left[
\begin{array}{c|c}
      \mathbf{0} &  -\mathbf{1} \\
     \hline
     \mathbf{1} & \mathbf{0}
 \end{array}
\right],
\end{eqnarray}
being the standard symplectic matrix. Here, $\mathbf{1}$ is a $4\times 4$ identity matrix, $^\intercal$ denotes the transpose operator and $\mathbf{J}$ satisfies
\begin{eqnarray}
\mathbf{J}^\intercal = \mathbf{J}^{-1},\qquad{\rm{and}}\qquad\det \mathbf{J}=1,
\end{eqnarray}
in which the determinant of a matrix is denoted by ``$\rm{det}$''.
As the solution matrix $\mathbf{W}$ can be obtained by taking the exponential map of $\mathbf{H}$, $\mathbf{W}$ is a symplectic matrix satisfying
\begin{eqnarray}\label{eq:W_symplectic_standard}
{\mathbf{W}}^{\intercal} {\mathbf{J}} {\mathbf{W}}=\mathbf{J}.
\end{eqnarray}
What we want to highlight here is that as the symplectic structure, $(\ref{eq:symp_struct})$, is preserved under the propagation of the geodesic spray, the evolution can be studied on some symplectic vector space in general. In that case, one can find another matrix representation, $\mathcal{W}$, to the linear evolution. This matrix satisfies  ${\mathcal{W}}^{\intercal} {\Omega} {\mathcal{W}}={\Omega}$ similar to eq.~(\ref{eq:W_symplectic_standard}) where $\Omega$ is now some arbitrary, constant skew--symmetric matrix. It is known that any skew--symmetric matrix $\Omega$ can be transformed to $\mathbf{J}$ via some symplectic transformation. The new basis is called a \textit{symplectic basis}. However, only when one uses a symplectic basis, or the standard symplectic bi--linear form, the matrix $\mathbf{H}$ preserves its Hamiltonian character and $\mathbf{W}$ is represented under the special linear group, with
\begin{eqnarray}\label{eq:W_det_1}
    \det \mathbf{W}=1.
\end{eqnarray}
With this, we guarantee that the components of $\mathbf{Z}$ are given with respect to a symplectic basis.

We write the $\mathbf{W}$ in the block form as
\begin{eqnarray}\label{eq:W_block_form}
\mathbf{W}
 =
 \left[
 \begin{array}{c|c}
     \tensor{\mathbf{W}}{_X_X^{\tilde{a}}_{{\tilde{b}} '}} &  \, \tensor{\mathbf{W}}{_X_L^{\tilde{a}}^{{\tilde{b}} '}} \\
     \hline
     \tensor{\mathbf{W}}{_L_X_{\tilde{a}}_{{\tilde{b}} '}} & \, \tensor{\mathbf{W}}{_L_L_{\tilde{a}}^{{\tilde{b}} '}}
 \end{array}
 \right]=
 \left[
 \begin{array}{c|c}
     \tensor{\mathbf{W}}{_X_X^{\tilde{a}}_{{\tilde{b}} }} &  \, \tensor{\mathbf{W}}{_X_L^{\tilde{a}}^{{\tilde{b}} }} \\
     \hline
     \tensor{\mathbf{W}}{_L_X_{\tilde{a}}_{{\tilde{b}} }} & \, \tensor{\mathbf{W}}{_L_L_{\tilde{a}}^{{\tilde{b}} }}
 \end{array}
 \right]\nonumber\\
\end{eqnarray}
The sub--blocks $ \tensor{\mathbf{W}}{_X_X}, \tensor{\mathbf{W}}{_X_L}, \tensor{\mathbf{W}}{_L_X}$ and $ \tensor{\mathbf{W}}{_L_L}$ are all $4\times 4$ matrices. We remind that the second equality in the above follows from the fact that the value of the tetrad components are equal at the observation and at the emission point due to eq.~(\ref{eq:parallel_prop_tetrad}). Therefore, from now on, we do not distinguish the components of the parallel propagated tetrad at different points. Another important point we would like to emphasise is that the sub--blocks of $\mathbf{W}$ are genuine rank--2 tensorial objects. However, each of them are composed of different covariant, contavariant and mixed forms. Therefore, the matrix $\mathbf{W}$ should be considered as a mere array, rather than a tensor, which keeps track of the solutions of the covariant evolution equations. Same thing applies to the Hamiltonian matrix $\mathbf{H}$. Each of the sub--blocks on the other hand defines a bi--tensor, i.e., a linear mapping acting from the appropriate tangent/cotangent space at $\mathscr{O}$ to the appropriate tangent/cotangent space at $\mathscr{E}$. 

The symplecticity conditions, (\ref{eq:W_symplectic_standard}) and (\ref{eq:W_det_1}), of $\mathbf{W}$, require its sub--blocks to satisfy certain symmetry conditions. Essentially, substituting the block form (\ref{eq:W_block_form}) into eq.~(\ref{eq:W_symplectic_standard}) gives
\begin{eqnarray}
\tensor{\mathbf{W}}{^\intercal_X_X}\tensor{\mathbf{W}}{_L_X}&=& \tensor{\mathbf{W}}{^\intercal_L_X}\tensor{\mathbf{W}}{_X_X},\label{eq:symp_cond_1}  \\
\tensor{\mathbf{W}}{^\intercal_X_L}\tensor{\mathbf{W}}{_L_L}&=& \tensor{\mathbf{W}}{^\intercal_L_L}\tensor{\mathbf{W}}{_X_L},\label{eq:symp_cond_2}\\
\tensor{\mathbf{W}}{^\intercal_X_X}\tensor{\mathbf{W}}{_L_L}&-& \tensor{\mathbf{W}}{^\intercal_L_X}\tensor{\mathbf{W}}{_X_L}=\mathbf{1}\label{eq:symp_cond_3}, 
\end{eqnarray}
or equivalently,
\begin{eqnarray}\label{eq:W_subblocks_symp_symm2}
\tensor{\mathbf{W}}{_X_X}\tensor{\mathbf{W}}{^\intercal_X_L}&=& \tensor{\mathbf{W}}{_X_L}\tensor{\mathbf{W}}{^\intercal_X_X},\label{eq:symp_cond_4} \\
\tensor{\mathbf{W}}{_L_X}\tensor{\mathbf{W}}{^\intercal_L_L}&=& \tensor{\mathbf{W}}{_L_L}\tensor{\mathbf{W}}{^\intercal_L_X},\label{eq:symp_cond_5}\\
\tensor{\mathbf{W}}{_X_X}\tensor{\mathbf{W}}{^\intercal_L_L}&-&\tensor{\mathbf{W}}{_X_L}\tensor{\mathbf{W}}{^\intercal_L_X}=\mathbf{1}. \label{eq:symp_cond_6}
\end{eqnarray}

Now, we would like to show that the symplectic symmetries that rule the dynamics of the perturbations are inherently related to the geodesic deviation equation of the light bundles. This can be understood through showing that at any point, the directional covariant derivative of $\delta \Vec{x}$ along $\Vec{l}$ is equal to the covariant variation of $\Vec{l}$. This holds due to the Lie dragging condition (\ref{eq:Lie_dragging}). Therefore, it is natural to have the following set of relationships between the perturbations and the deviation variables at the emission and at the observation points.
\begin{eqnarray}\label{eq:ICs}
\delta\Vec{x}\left(\lambda _{\mathscr{O}}\right)= \Vec{\tensor{\delta x}{_{\mathscr{O}}}}, \qquad \nabla _{\Vec{l}} \,\delta\Vec{x}\left(\lambda _{\mathscr{O}}\right)=\Delta \Vec{l}_{\mathscr{O}},\nonumber \\
\delta\Vec{x}\left(\lambda _{\mathscr{E}}\right)=  \Vec{\tensor{\delta x}{_{\mathscr{E}}}}, \qquad \nabla _{\Vec{l}} \,\delta\Vec{x}\left(\lambda _{\mathscr{E}}\right)=\Delta \Vec{l}_{\mathscr{E}}.
\end{eqnarray}
Note that the deviation equation (\ref{eq:first_ord_dev}) is a second order linear differential equation. Therefore, its solution can be represented by a linear transform. Here, we remind that due to the set of relationships, (\ref{eq:ICs}), and eq.~(\ref{eq:perturb_linear_tr}), the solution is given by the aforementioned symplectic matrix, $\mathbf{W}$, in some parallel propagated tetrad basis. Then, we have 
\begin{eqnarray}\label{eq:Deviation_W_matrix}
    \begin{bmatrix}
    {\delta x}_{\mathscr{E}}^{{\tilde{a}}}  \\
    \nabla _{\Vec{l}} \,{\delta x}^{\mathscr{E}}_{{\tilde{a}}}
    \end{bmatrix}
=\mathbf{W}\left(\lambda _\mathscr{E},\lambda _\mathscr{O}\right)
\begin{bmatrix}
    {\delta x}_{\mathscr{O}}^{{\tilde{b}}} \\
    \nabla _{\Vec{l}} \,{\delta x}^{\mathscr{O}}_{{\tilde{b}}}
    \end{bmatrix}.
\end{eqnarray}
The equation above was first realized in \cite{Fleury:2013sna} for the screen space projections of the deviation variables. Fleury \textit{et al.} found the solutions of the 2--dimensional geodesic deviation variables via a Wro{\'n}ski matrix method. Its applications for different geometries were shown to be useful in light propagation calculations for various cosmological scenarios \cite{Fleury:2014gha,Fleury:2014rea,Fleury:2015rwa}. Later, in \cite{Uzun:2018yes}, it was shown that Fleury \textit{et al.}'s Wro{\'n}ski matrix is indeed a symplectic matrix. This was achieved by considering an action principle for the dynamics of the geodesic deviation. A Hamiltonian formulation was obtained on a reduced phase space together with its potential physical implementations. Here, we would like to present its general form in an 8--dimensional phase space.

\subsection{\label{sec:Canonical transforms and initial to boundary value problem}Canonical transforms and initial to boundary value problem}
Both the geodesic equation and the geodesic deviation equation can be derived from a variational principle based on an action functional. Let us rewrite the null geodesic deviation action presented in \cite{Uzun:2018yes}
following from the bi--local formalism of \cite{Vines:2014}. The action follows as
\begin{eqnarray}\label{eq:deviation_action}
S=\int \left(\frac{1}{2}\inner{\nabla _{\Vec{l}}\, \delta\Vec{x},\nabla _{\Vec{l}} \,\delta\Vec{x}}+\frac{1}{2}\tensor{R}{_{\delta\Vec{x}}_{\vec{l}}_{\,\vec{l}}_{\delta\Vec{x}}}\right)d\lambda,
\end{eqnarray}
where
\begin{eqnarray}
\tensor{R}{_{\delta\Vec{x}}_{\vec{l}}_{\,\vec{l}}_{\delta\Vec{x}}}=   \tensor{R}{_{\alpha}_{\beta}_{\mu}_{\nu}} l^\beta l^\mu \delta x^\alpha \delta x^\nu.\nonumber
\end{eqnarray}
This is the full geodesic deviation action written up to the quadratic order in geodesic deviation variables, $\nabla _{\vec{l}}\, \delta\Vec{x}$ and $\delta \vec x$, in order to study the first order part of the dynamics of a collection of null geodesics. We now write its equivalent form by making use of the relation between the covariant variation of $\Vec{l}$ and the derivative of the geodesic deviation vector, i.e., the relations given in eq.~(\ref{eq:ICs}). Then, we have
\begin{eqnarray}\label{eq:perturbation_action}
S=\int \left(\frac{1}{2}\inner{\Delta \Vec{l},\Delta \Vec{l}}+\frac{1}{2}\tensor{R}{_{\delta\Vec{x}}_{\vec{l}}_{\,\vec{l}}_{\delta\Vec{x}}}\right)d\lambda.
\end{eqnarray}
Once we treat the integrand of the action given above as the Lagrangian and $\lambda$ as the evolution parameter, we can switch to a Hamiltonian formalism by making use of the parallel propagated tetrad components of $\Delta \Vec{l}$ and $\delta\Vec{x}$. 
Particularly, once we choose $\delta {x}^{\tilde{a}}$ to be the canonical coordinates and $\Delta {l}_{\tilde{b}}$ to be their associated canonical momenta, the Hamiltonian of the system take the form of
\begin{eqnarray}\label{eq:Hamiltonian_geod_dev}
\mathcal{H}=\frac{1}{2}\left(\tensor{g}{^{\tilde{a}}^{\tilde{b}}}\Delta {l}_{\tilde{a}}\Delta {l}_{\tilde{b}}-\tensor{\mathcal{R}}{_{\,\tilde{a}}_{\tilde{b}}}\delta {x}^{\tilde{a}}\delta {x}^{\tilde{b}}\right).
\end{eqnarray}
 Then the 8--dimensional vector, $\mathbf{Z}$, given in eq.~(\ref{eq:phase_space_vector_Hamiltonian}) can be treated as a phase space vector and the evolution equations (\ref{eq:X_evolution}) are indeed matrix Hamiltonian equations. 

The function in eq.~(\ref{eq:Hamiltonian_geod_dev}) can be viewed as a perturbation Hamiltonian to the standard Hamiltonian of a null geodesic \cite{Perlick:2000}. Accordingly, there exists an associated symplectic form which is preserved throughout the Hamiltonian flow,
\begin{eqnarray}\label{eq:symp_struc_geod_dev}
d\left({\Delta l}^{\mathscr{E}}_{{\tilde{b}}}\right)\wedge d\left({\delta x}_{\mathscr{E}}^{{\tilde{b}}}\right)=d\left({\Delta l}^{\mathscr{O}}_{{\tilde{b}}}\right)\wedge d\left({\delta x}_{\mathscr{O}}^{{\tilde{b}}}\right).
\end{eqnarray}
When compared to the symplectic structure in eq.~(\ref{eq:symp_struct}), it is clear that the one in eq.~(\ref{eq:symp_struc_geod_dev}) is representing the symplectic structure of the perturbation of the system, i.e., the deviation of geodesics.
Then due to the Poincar{\'e}'s lemma, there exists a generating function $S$ where
\begin{eqnarray}\label{eq:gen_fun_1form}
{\Delta l}^{\mathscr{E}}_{{\tilde{b}}} d\left({\delta x}_{\mathscr{E}}^{{\tilde{b}}}\right)={\Delta l}^{\mathscr{O}}_{{\tilde{b}}} d\left({\delta x}_{\mathscr{O}}^{{\tilde{b}}}\right)+d S\left({\delta x}_{\mathscr{E}}^{{\tilde{a}}},{\Delta l}^{\mathscr{E}}_{{\tilde{b}}};\lambda \right).\nonumber\\
\end{eqnarray}
The generating function in question is the action $S\left(\delta {x}^{\tilde{a}},\Delta {l}_{\tilde{b}}\right)$ in eq.~(\ref{eq:perturbation_action}) and the symplectic matrix $\mathbf{W}$ represents the associated linear canonical transformation. 
The matrix $\mathbf{W}$ provides a solution to the initial value problem of the position and tangent vector perturbations. Namely, given the initial values ${\delta x}_{\mathscr{O}}^{{\tilde{b}}}$ and ${\Delta l}^{\mathscr{O}}_{{\tilde{b}}}$, one can find the final values ${\delta x}_{\mathscr{E}}^{{\tilde{b}}}$ and ${\Delta l}^{\mathscr{E}}_{{\tilde{b}}}$ through $\mathbf{W}$. In this work, we assume those to be \textit{free canonical transformations} \cite{Arnold:1978} which are represented by \textit{free symplectic matrices} \cite{deGosson:2006}. Such symplectic matrices are assumed to have a non--zero determinant for the upper right block, i.e.,
\begin{eqnarray}\label{eq:cond_free_canonical}
{\det}\tensor{\mathbf{W}}{_X_L}=\left|\frac{\partial {\delta x}_{\mathscr{E}}^{{\tilde{a}}}}{\partial {\Delta l}^{\mathscr{O}}_{{\tilde{b}}}}\right|\neq 0.
\end{eqnarray}
The physical importance of this condition is presented in the next sub--section. For now, let us consider it as a mathematical assumption. 

Under the condition (\ref{eq:cond_free_canonical}), and fixed initial conditions at the observer, one can locally represent the the generating function $S\left(\delta {x}^{\tilde{a}},\Delta {l}_{\tilde{b}};\lambda\right)$ in terms of the boundary values of the canonical coordinates. 
\begin{eqnarray}
 S\left({\delta x}_{\mathscr{E}}^{{\tilde{a}}},{\Delta l}^{\mathscr{E}}_{{\tilde{b}}};\lambda \right)\rightarrow  S\left({\delta x}_{\mathscr{E}}^{{\tilde{a}}},{\delta x}_{\mathscr{O}}^{{\tilde{b}}};\lambda \right).
\end{eqnarray}
This is achieved by considering ${\Delta l}^{\mathscr{E}}_{{\tilde{b}}}={\Delta l}^{\mathscr{E}}_{{\tilde{b}}}\left({\delta x}_{\mathscr{O}}^{{\tilde{b}}}\right)$. Then, the generating function can be written as a quadratic form which follows as \cite{deGosson:2006}
\begin{eqnarray}\label{eq:gen_func_quadratic}
S&=&
\frac{1}{2}{\mathbf{\delta x}}{^\intercal_{\mathscr{O}}}\left(\tensor{\mathbf{W}}{^{-1}_X_L}\tensor{\mathbf{W}}{_X_X}\right){\mathbf{\delta x}}{_{\mathscr{O}}}
-
{\mathbf{\delta x}}{^\intercal_{\mathscr{O}}}\left(\tensor{\mathbf{W}}{^{-1}_X_L}\right){\mathbf{\delta x}}{_{\mathscr{E}}}\nonumber\\
&\qquad&\qquad \qquad +\frac{1}{2}{\mathbf{\delta x}}{^\intercal_{\mathscr{E}}}\left(\tensor{\mathbf{W}}{_L_L}\tensor{\mathbf{W}}{^{-1}_X_L}\right){\mathbf{\delta x}}{_{\mathscr{E}}}.
\end{eqnarray}
Next, it can be shown that the Hamilton--Jacobi equation
\begin{eqnarray}\label{eq:Hamilton_Jacobi}
\frac{dS}{d\lambda}=-\mathcal{H},
\end{eqnarray}
follows from eqs.~(\ref{eq:gen_fun_1form}) and (\ref{eq:gen_func_quadratic}). The generating function $S$ satisfies
\begin{eqnarray}
-{\Delta l}^{\mathscr{O}}_{{\tilde{b}}}=\frac{dS}{d\left({\delta x}_{\mathscr{O}}^{{\tilde{b}}}\right)},\,\, {\Delta l}^{\mathscr{E}}_{{\tilde{b}}}=\frac{dS}{d\left({\delta x}_{\mathscr{E}}^{{\tilde{b}}}\right)},
\end{eqnarray}
where ${\Delta l}^{\mathscr{O}}_{{\tilde{b}}}$ and ${\Delta l}^{\mathscr{E}}_{{\tilde{b}}}$ play the role of the canonical momenta at the boundary points. The reader might refer to the Appendix of \cite{Uzun:2018yes} for explicit calculations to prove eq.~(\ref{eq:Hamilton_Jacobi}) for the case of reduced Hamiltonian dynamics.

The quadratic generating function (\ref{eq:gen_func_quadratic}) can be put in a more compact form
\begin{eqnarray}
    S\left({\delta x}_{\mathscr{E}}^{{\tilde{a}}},{\delta x}_{\mathscr{O}}^{{\tilde{b}}};\lambda \right)=\frac{1}{2} \X{}^\intercal\mathbf{U}^\intercal \X{},\,\, {\rm{with}} \,\,
    \X{}=\begin{bmatrix}
    {\delta x}_{\mathscr{O}}^{{\tilde{b}}} \\
    {\delta x}_{\mathscr{E}}^{{\tilde{b}}}
    \end{bmatrix}.
\end{eqnarray}
We write $\mathbf{U}$ in the block form as 
\begin{eqnarray}\label{eq:U_matrix_subblocks}
    \mathbf{U}=\left[
\begin{array}{c|c}
      \tensor{\mathbf{U^{\mathscr{O}\mathscr{O}}}}{_{\tilde{a}}_{\tilde{b}}} &  \tensor{\mathbf{U^{\mathscr{O}\mathscr{E}}}}{_{\tilde{a}}_{\tilde{b}}} \\
     \hline
     \tensor{\mathbf{U^{\mathscr{E}\mathscr{O}}}}{_{\tilde{a}}_{\tilde{b}}} & \tensor{\mathbf{U^{\mathscr{E}\mathscr{E}}}}{_{\tilde{a}}_{\tilde{b}}}
 \end{array}
\right].
\end{eqnarray}
The components of the $8\times 8$ matrix $\mathbf{U}$ involves the sub--blocks of the matrix $\mathbf{W}$ through
\begin{eqnarray}\label{eq:U_subblocks_W_subblocks}
\mathbf{U^{\mathscr{O}\mathscr{O}}}&=&-\mathbf{W}^{-1}_{XL}
\mathbf{W}_{XX},\nonumber \\
\mathbf{U^{\mathscr{O}\mathscr{E}}}&=&\mathbf{W}^{-1}_{XL},\nonumber\\
\mathbf{U^{\mathscr{E}\mathscr{O}}}&=&
\mathbf{W}_{LL}\mathbf{W}^{-1}_{XL}\mathbf{W}_{XX}-\mathbf{W}_{LX},\nonumber\\
\mathbf{U^{\mathscr{E}\mathscr{E}}}&=&-\mathbf{W}_{LL}\mathbf{W}^{-1}_{XL}.
\end{eqnarray}
What we are ultimately doing here is to switch from an initial value problem to a boundary value problem. Specifically, we consider the first row of the symplectic matrix in eq.~(\ref{eq:perturb_linear_tr}) and re--write it for ${\Delta l}^{\mathscr{O}}_{{\tilde{b}}}$. Then, we substitute this term in the second row eq.~(\ref{eq:perturb_linear_tr}) (See Appendix C of \cite{Korzynski:2021aqk} for a similar calculation.). This is how we obtain the relations in eq.~(\ref{eq:U_subblocks_W_subblocks}).

Essentially, the matrix $\mathbf{U}$ transforms the boundary value of the canonical coordinates to the boundary value of the momenta, i.e.,
\begin{eqnarray}\label{eq:U_matrix_boundary}
    \begin{bmatrix}
    {\Delta l}^{\mathscr{O}}_{{\tilde{a}}} \\
    -{\Delta l}^{\mathscr{E}}_{{\tilde{a}}}
    \end{bmatrix}
=\mathbf{U}
\begin{bmatrix}
    {\delta x}_{\mathscr{O}}^{{\tilde{b}}} \\
    {\delta x}_{\mathscr{E}}^{{\tilde{b}}}
    \end{bmatrix}.
\end{eqnarray}

This linear transformation defines a map that sends the direct sum of the tangent space of the emitter and the observer to the direct sum of their cotangent spaces. Namely,
\begin{eqnarray}
\tilde{\psi}: T_\mathscr{O}M\oplus T_\mathscr{E}M \rightarrow T^*_\mathscr{O}M\oplus T^*_\mathscr{E}M.
\end{eqnarray}

Previously, we mentioned that $\mathbf{W}$ is a symplectic matrix and therefore its sub--blocks satisfy certain symplecticity conditions given in eqs.~(\ref{eq:symp_cond_1})-(\ref{eq:symp_cond_6}). Then, the sub--blocks of the $\mathbf{U}$ matrix are not independent from each other either. In fact, it is easy to show that 
\begin{eqnarray}
\mathbf{U^{\mathscr{O}\mathscr{O}}}&=&\mathbf{U^{\mathscr{O}\mathscr{O}}}^\intercal,  \qquad {\rm{(due\,\,to\,\,eq.~(\ref{eq:symp_cond_4}))}} \\
\mathbf{U^{\mathscr{O}\mathscr{E}}}&=&\mathbf{U^{\mathscr{E}\mathscr{O}}}^\intercal,  \qquad {\rm{(due\,\,to\,\,eqs.~(\ref{eq:symp_cond_4})\, {\rm{and}}\, (\ref{eq:symp_cond_6}))}} \\
\mathbf{U^{\mathscr{E}\mathscr{E}}}&=&\mathbf{U^{\mathscr{E}\mathscr{E}}}^\intercal.  \qquad {\rm{(due\,\,to\,\,eq.~(\ref{eq:symp_cond_2}))}} 
\end{eqnarray}
It follows that $\mathbf{U}$ is a symmetric matrix, i.e.,
\begin{eqnarray}
 \mathbf{U}^\intercal=\mathbf{U}.
\end{eqnarray}

Note that the relation between $\mathbf{W}$ and $\mathbf{U}$
is an invertible operation on matrices. It is easy to check that it maps symplectic matrices to symmetric matrices and vice versa. Accordingly, one can also represent the sub--clocks of $\mathbf{W}$ with respect to the sub--clocks of $\mathbf{U}$ through
\begin{eqnarray}
\mathbf{W}_{XX}&=&-\mathbf{U^{\mathscr{O}\mathscr{E}}}^{-1}\mathbf{U^{\mathscr{O}\mathscr{O}}},\nonumber \\
\mathbf{W}_{XL}&=&\mathbf{U^{\mathscr{O}\mathscr{E}}}^{-1},\nonumber \\
\mathbf{W}_{LX}&=&\mathbf{U^{\mathscr{E}\mathscr{E}}}\mathbf{U^{\mathscr{O}\mathscr{E}}}^{-1}\mathbf{U^{\mathscr{O}\mathscr{O}}}-\mathbf{U^{\mathscr{E}\mathscr{O}}},\nonumber \\
\mathbf{W}_{LL}&=&-\mathbf{U^{\mathscr{E}\mathscr{E}}}\mathbf{U^{\mathscr{O}\mathscr{E}}}^{-1},
\end{eqnarray}
i.e., same set of relations between the sub--blocks of two matrices as in eq.~(\ref{eq:U_subblocks_W_subblocks}). 

Let us just briefly mention that both $\mathbf{W}$ and $\mathbf{U}$ need to satisfy additional algebraic relations associated with the null tangent vector $\Vec{l}$ \cite{Grasso:2018mei, Korzynski:2021aqk}. They will not play an important role in this work, apart from the calculations in Appendix~\ref{sec:Proof that WXL is degenerate iff the Jacobi matrix D is degenerate}, so we will not refer to them here in detail.  

With these results, the generating function (\ref{eq:gen_func_quadratic}) of our Hamiltonian formulation can simply be re--written as
\begin{eqnarray}\label{eq:gen_func_U_block}
S&=&-\left(
\frac{1}{2}{\mathbf{\delta x}}{^\intercal_{\mathscr{O}}}\mathbf{U^{\mathscr{O}\mathscr{O}}}{\mathbf{\delta x}}{_{\mathscr{O}}}
+
{\mathbf{\delta x}}{^\intercal_{\mathscr{O}}}\mathbf{U^{\mathscr{O}\mathscr{E}}}{\mathbf{\delta x}}{_{\mathscr{E}}}
 +\frac{1}{2}{\mathbf{\delta x}}{^\intercal_{\mathscr{E}}}\mathbf{U^{\mathscr{E}\mathscr{E}}}{\mathbf{\delta x}}{_{\mathscr{E}}}\right),\nonumber\\
\end{eqnarray}
which can be viewed as the second order expansion of the world function, $\sigma$, of the null geodesic that connects the observer and the emitter.

\subsection{\label{sec:Symmetries and reciprocity relations}Symmetries and reciprocity relations}
Reciprocity relations in physics refer to the fact that certain measurement outcomes are unchanged under the change of input and output variables. An example for the case of relativistic observations is Etherington's distance reciprocity \cite{Etherington:1933}. This theorem states that the angular diameter distance is equal to the luminosity distance up to a redshift factor, i.e,
\begin{eqnarray}\label{eq:Etherington_recip}
D'_{L}=\left(1+z\right)\,D_{A}.
\end{eqnarray}
In addition, when the relationship between the relativistic, or corrected, luminosity distance, $D'_{L}$, and the bolometric or uncorrected luminosity distance, $D_{L}$, is taken into account one obtains the distance duality relation,
\begin{eqnarray}\label{eq:Bolometric_lum_dist}
D_{L}=\left(1+z\right)^2\,D_{A}.
\end{eqnarray}
In the literature, the proof of the distance reciprocity is shown for momentary observations where the observer is located at the vertex of a light bundle \cite{Ellis:1971pg}. One defines a $2\times 2$ Jacobi matrix, $\mathbf{D}$, that transforms the transverse components of the derivative of the geodesic deviation vector, $\delta\Vec{x}$, along $\Vec{l}$ at the initial point to the transverse components of the geodesic deviation vector at the output point \footnote{Note that in \cite{Korzynski:2017nas, Grasso:2018mei} we denoted the Jacobi matrix with $\cal D$ instead of $\mathbf{D}$. In the current paper, we reserve the symbol $\cal D$ to express the Fermi--Walker derivative operator as in Section~\ref{sec:Momentary observables}.}. In other words, it transforms the solid angle of the observer to the cross--sectional area calculated at the emitter. The determinant of the Jacobi matrix, thus gives the squared distances up to a frequency factor. The proof of reciprocity of distances follows from the fact that the value of the Jacobi matrix is equal to its minus transpose when the input and output variables are traversed. That is $\tensor{\mathbf{D}}{}\left(\lambda _\mathscr{E},\lambda _\mathscr{O}\right)=-\tensor{\mathbf{D}}{^\intercal}\left(\lambda _\mathscr{O},\lambda _\mathscr{E}\right)$. Then, their determinants are equal to each other. Hence, the distance reciprocity holds, with the redshift pre--factor appearing in (\ref{eq:Etherington_recip}) due to the differences in the solid angles at the two end points, caused by the stellar aberration effect. It is also possible to explain this difference as resulting from the differences in infinitesimal proper distances, $d\ell=\omega d\lambda$. In \cite{Uzun:2018yes}, it was shown that such a physical outcome follows from the symplectic symmetries of the geodesic deviation evolution defined on a reduced phase space. It was demonstrated that there exists a $4\times 4$ matrix which consists of the $2 \times 2$ Jacobi matrix $\mathbf{D}$ on its upper right block. It was argued that the distance reciprocity holds even for light propagation in multiple geometries due the symplectic symmetries. The underlying reason for such a symmetry essentially follows from the symmetries of the Riemann tensor \cite{Ellis:1971pg}.

Here, we consider the generalization of the symplectic matrix presented in \cite{Uzun:2018yes}. Compared to the one in \cite{Uzun:2018yes}, the symplectic matrix, $\mathbf{W}$, considered in the current work involves two additional degrees of freedom. As we discussed in the previous section, there exists a symmetric matrix, $\mathbf{U}$, associated with $\mathbf{W}$. Those two matrices carry the same set of information. This is also relevant for the reciprocity relations. In Section~\ref{sec:Physical outcomes}, we show that the symplectic symmetries of the underlying initial value problem, can be traded for the symmetries of the corresponding boundary value problem. Then, one can use the symmetry property of $\mathbf{U}$ in order to prove the reciprocity relations.
 
In the current work, we assume that there exists no light caustics between the emitter and the observer. For this, one requires $\det\mathbf{D}\neq 0$ such that the distances can be calculated \cite{Perlick:2004tq}. In the following sections we pick such a parallel propagated tetrad that the $2\times 2$ Jacobi matrix corresponds to the lower right block of our $4\times 4$ matrix $\tensor{\mathbf{W}}{_X_L}$. Namely,
\begin{eqnarray}
\tensor{\mathbf{W}}{_X_L}=
 \left[
 \begin{array}{c|c}
     \mathbf{A} & \, \mathbf{B} \\
     \hline
     \mathbf{C} & \, \mathbf{D}
 \end{array}
 \right]
\end{eqnarray}
For block matrices as above, if $\mathbf{D}$ is invertible, then the determinant of the matrix is given by 
\begin{eqnarray}
\det\tensor{\mathbf{W}}{_X_L}={\rm{det}}\left(\mathbf{A}-\mathbf{B}\mathbf{D}^{-1}\mathbf{C}\right) \det\mathbf{D}.
\end{eqnarray} 
For the matrix components calculated through our tetrad, it can be shown that ${\rm{det}}\left(\mathbf{A}-\mathbf{B}\mathbf{D}^{-1}\mathbf{C}\right)\neq 0$ at each point of the propagation. Thus, assuming $\det \mathbf{D}\neq 0$ guarantees that $\det\tensor{\mathbf{W}}{_X_L}\neq 0$ holds, see Appedix~\ref{sec:Proof that WXL is degenerate iff the Jacobi matrix D is degenerate} for the details of the proof. The physical interpretation of this result is the following: $\mathbf{W}_{XL}$ is invertible  if and only  if  the emitter is not situated on a caustic with respect to the observation point $p_\mathscr{O}$. In other words, $\mathbf{W}_{XL}$ can only become non--invertible on the spatial transverse subspace, or the screen space. Emergence of a singular direction in the screen space is by definition a sign of a caustic.

Recall that in Section~\ref{sec:Canonical transforms and initial to boundary value problem}, we argued that eq.~(\ref{eq:cond_free_canonical}) should hold in order to have an evolution problem represented by a free canonical transformation. Now it is clear that this condition is indeed equivalent to having a caustic--free assumption for the envelope of the family of light rays in question.

With this, we note that the aim of this paper is not only to outline the mathematical symmetries of variable observations, but also their physical outcomes. Switching from an initial value problem to a boundary value problem allows us to investigate the problem with matrices whose 
symmetries are easier to handle. In  addition, it allows us to find straightforward answers to questions regarding the physics of the problem. We are specifically interested in simultaneous variations that take place at the emitter's and at the observer's frame during the course of an observation that takes place within a finite interval of time. Switching to the boundary value problem allows us to asks questions such as ``Given that the emitter and the observer have non--zero position variations how would that affect the observer’s recordings of the redshifts?''. In return, the symmetry of  $\mathbf{U}$ allows us to find an answer to such questions which we  investigate in detail in Section~\ref{sec:Physical outcomes}.

\section{\label{sec:Scalar observables and bi--local variations}Scalar observables and bi--local variations}
Let us now identify the relevance of matrix $\mathbf{U}$ for generic observables. For this, we denote an operator $\bU\left(\cdot, \cdot \right)$ associated with the multiplication operator of the matrix $\mathbf{U}$. Its properties are as follows
\begin{itemize}
    \item [(i)] $\bU\left(\X{}, \cdot \right)$ creates an 8--dimensional covector,
    \begin{eqnarray}
    \left[\mathbf{U}\cdot\X{}\right]^\intercal:=\DL ^\intercal_{\X{}}=\left[{\Delta l}^{\mathscr{O}}_{{\tilde{a}}}\,\,\,\,\,-{\Delta l}^{\mathscr{E}}_{{\tilde{a}}}\right],
    \end{eqnarray}
    through eq.~(\ref{eq:U_matrix_boundary}). Here, $\X{}$ refers to a bi--local variation vector. The covector $\DL ^\intercal_{\X{}}$ represents how the tangent vectors at \pO and \pE change given the possible variations of positions at the corresponding end points by $\X{}$. The bi--local variation vector might represent the variations at \O only, at \E only or both. We write it in general as
        \begin{align}\label{eq:generic_bilocal_var}
        \X{}= 
        \begin{bmatrix}
         {\delta x}_{\mathscr{O}}^{{\tilde{a}}} \\ {\delta x}_{\mathscr{E}}^{{\tilde{b}}}
        \end{bmatrix}. 
        \end{align}
    
    Note that $\DL_{\X{}}$ also contains the information about how the deviation vectors at two points change along the null direction through the \textit{Lie dragging} condition. Namely, through $\nabla_{\X{}}\Vec{l}=\nabla _{\Vec{l}}\X{}$. With this notation, we imply  
    $\left(\nabla_{\delta \Vec{x}} \Vec l = \nabla_{\Vec {l}} \,\delta \Vec {x}\right) \Big|_\mathscr{O}$ and $\left(\nabla_{\delta \Vec{x}} \Vec l = \nabla_{\Vec {l}} \,\delta \Vec {x}\right) \Big|_\mathscr{E}$, where $\delta \Vec{x}$ is a solution of the first order geodesic deviation eq.~(\ref{eq:first_ord_dev}).
    
    \item [(ii)] $\bU\left(\X{}, \boldsymbol{Y} \right)$ creates a scalar, $\X{}^\intercal\mathbf{U}^\intercal\boldsymbol{Y}=\DL ^\intercal_{\X{}}\boldsymbol{Y}$, which is essentially the projection of $\DL ^\intercal_{\X{}}$ on some arbitrary 8--dimensional vector $\boldsymbol{Y}$. This scalar has the potential of identifying the observables affected by the variations, $\X{}$, that can be recorded by \E, by \O or both depending on the vector $\boldsymbol{Y}$. For the time being, one might think of representing $\boldsymbol{Y}$ through
     \begin{align}
     \boldsymbol{Y}=
        \begin{bmatrix}
             {Y}^{\mathscr{O}}_{\tilde{a}}  \\  
             {Y}^{\mathscr{E}}_{\tilde{b}}
        \end{bmatrix},
        \end{align}
        where ${Y}_{\mathscr{O}}^{\tilde{a}}$ and ${Y}_{\mathscr{E}}^{\tilde{b}}$ are components of some arbitrary projection vectors represented in a parallel propagated frame. Note that for all important observables we study in this paper, $\boldsymbol{Y}$ satisfies 
        \begin{eqnarray}\label{eq:permissibility_criteria_compact}
\mathbf{L}^\intercal\boldsymbol{Y}=0,\qquad {\rm{with}} \qquad
\mathbf{L}^\intercal= \left[{l}^{\mathscr{O}}_{\mu'}\,\,\,\,\,-{l}^{\mathscr{E}}_{\mu}\right].
\end{eqnarray}
        In other words, $\boldsymbol{Y}$ spans the tangent space to the local surface of communication mentioned before and described in \cite{Korzynski:2021aqk}.
        
        \item [(iii)] $\bU$ is a symmetric operator, i.e.,
        \begin{eqnarray}
            \bU\left(\X{}, \boldsymbol{Y} \right)=\bU\left( \boldsymbol{Y}, \X{} \right). \label{eq:U is symmetric}
        \end{eqnarray} This follows from
        \begin{eqnarray}
        \bU\left(\X{}, \boldsymbol{Y} \right)&=&\X{}^\intercal\mathbf{U}^\intercal\boldsymbol{Y}=\left(\X{}^\intercal\mathbf{U}^\intercal\boldsymbol{Y}\right)^\intercal\nonumber \\&=&\boldsymbol{Y}^\intercal \mathbf{U}\X{}=\boldsymbol{Y}^\intercal \mathbf{U}^\intercal \X{}=\bU\left( \boldsymbol{Y}, \X{} \right),
        \end{eqnarray}
        which holds due to $\mathbf{U}$ being a symmetric matrix.
        This means the variations represented by $\X{}$ and projected on $\boldsymbol{Y}$  results in the same observables as variations represented by $\boldsymbol{Y}$ and projected on $\X{}$. Then, in the case that both the variation vectors and the projection vectors are the solutions of the first order geodesic deviation equation for a null geodesic, the corresponding observables can be recorded covariantly during  observations that allow both ends to be varied. 

        \item [(iv)] $\bU\left(\X{}, \X{} \right)$ gives the generating function of the underlying dynamics through eq.~(\ref{eq:gen_func_U_block}). Its derivative along the propagation direction gives minus the Hamiltonian function of the system.

        \item  [(v)] The symmetry property of $\bU\left(\X{}, \boldsymbol{Y} \right)$ is unaffected even though one uses the coordinate components of bi--local vectors as inputs. For instance, we write
        \begin{eqnarray}
        \bU\left(\X{}, \boldsymbol{Y} \right)   
        = &\UOO{{Y}_{\mathscr{O}}}{{\delta x}_{\mathscr{O}}}+\UOE{{Y}_{\mathscr{O}}}{{\delta x}_{\mathscr{E}}}\nonumber \\
        &+\UEO{{Y}_{\mathscr{E}}}{{\delta x}_{\mathscr{O}}}+\UEE{{Y}_{\mathscr{E}}}{{\delta x}_{\mathscr{E}}}.
        \end{eqnarray}
Each term above refers to contraction of the submatrices of $\mathbf{U}$ with the corresponding vectors. If we consider the first term, for example, we have
\begin{eqnarray}
\UOO{{\delta x}_{\mathscr{O}}}{{Y}_{\mathscr{O}}}=  {\delta x}_{\mathscr{O}}^{\mu '}U^{\mathscr{O}\mathscr{O}}_{{\mu'}{\nu'}}{Y}_{\mathscr{O}}^{\nu '}={\delta x}_{\mathscr{O}}^{\tilde{a}}\mathbf{U}^{\mathscr{O}\mathscr{O}}_{{\tilde{a}}{\tilde{b}}}{Y}_{\mathscr{O}}^{\tilde{b}}.
\end{eqnarray}
Thus, even though it is $\mathbf{U}^{\mathscr{O}\mathscr{O}}_{{\tilde{a}}{\tilde{b}}}$ that is known to be a symmetric matrix, the resultant scalar is unaffected from the usage of coordinate components. Therefore, from now on, we use coordinate components of the bi--local vectors in the demonstration of symmetries in the rest of the paper. 
\end{itemize}
\subsection{\label{sec:Permissible bi--local variations}Permissible bi--local variations}
In order to identify the physical outcomes of the algebraic symmetries of observations, one should first make sure that the emitter and the observer can send each other signals under generic variations at the both ends. Accordingly, in this section, we identify under which conditions \E and \O can keep on communicating. For example, if there is a variation at the end point \pO by an amount equal to the 4--velocity of the observer, what should be the variation at \pE so that the communication channel is intact? What if we have a strictly spacelike variation at \pO, what should the variation at the other end be for a signal to reach \O? 

As we noted in the previous sections, the variations at the both ends are not completely arbitrary. Rather, there exists a criteria given as in eq.~(\ref{eq:criteria_local_surf_com}) that 
\begin{eqnarray}\label{eq:permissibility_criteria}
    \inner{\tensor{\Vec{l}}{_{\mathscr{O}}},\delta \tensor{\Vec{x}}{_{\mathscr{O}}}} - \inner{\tensor{\Vec{l}}{_{\mathscr{E}}},\delta \tensor{\Vec{x}}{_{\mathscr{E}}}}=0,
\end{eqnarray}
should hold. This guarantees that \E and \O have a \textit{null} connecting geodesic throughout their evolution. We refer to any set of bi--local variation, $\X{}$,
that guarantees the criteria in eq.~(\ref{eq:permissibility_criteria}) a \textit{permissible} bi--local variation. We re--write this criteria in a more compact form, throughout the paper as $\mathbf{L}^\intercal\X{}=0$.

We now introduce semi--orthonormal tetrads; $\Vec{{e}}_{\mathscr{E}}:=\{\vec{{u}}^{\mathscr{E}},\Vec{{l}}^{\mathscr{E}},\Vec{{e}}^{_{\mathscr{E}}}_2,\Vec{{e}}^{_{\mathscr{E}}}_3,\}$ at point \pE and $\Vec{{e}}_{\mathscr{O}}:=\{\vec{{u}}^{\mathscr{O}},\Vec{{l}}^{\mathscr{O}},\Vec{{e}}^{_{\mathscr{O}}}_2,\Vec{{e}}^{_{\mathscr{O}}}_3,\}$ at point \pO. Those satisfy  
\begin{eqnarray}\label{eq:tetrad_properties}
\inner{\Vec{{u}},\Vec{l}\,}&>&0,\qquad 
\inner{\Vec{l},\Vec{l}\,}=0,\qquad 
\inner{\Vec{{u}},\Vec{{u}}\,}=-1,\nonumber\\
\inner{\Vec{{u}},\Vec{{e}}_A\,}&=&0,\,\,\,\,\,\,\,
\inner{\Vec{l},\Vec{{e}}_A\,}=0,\,\,\,\,\,\,
\inner{\Vec{{e}}_A,\Vec{{e}}_B\,}=\tensor{\delta}{_A_B},
\end{eqnarray}
both for the emitter's and the observer's frame.
Note that the only assumption we make here is that the tangent vector of the fiducial null geodesic is parallel transported throughout the propagation, i.e., the value of $\Vec{{l}}^{\mathscr{E}}$ is equal to the one of $\Vec{{l}}^{\mathscr{O}}$. However, we do not assume that the tetrad at \pO is parallel transported to \pE or vice versa. Rather, we consider the aforementioned tetrad $\Vec{\tilde{e}}$ as an auxiliary tetrad, which is parallel propagated along $\vec{l}$, i.e., $\nabla _{\Vec{l}}\,\Vec{\tilde{e}}=0$. 
We remind that this tetrad does not have to be adopted either to the emitter's or to the observer's frame. The idea is to use it as a tool to be able to compare vectors at different spacetime points. Such a tetrad was previously introduced in \cite{Grasso:2018mei} and studied in full detail. It was later used in \cite{Serbenta:2021tzv} in order to calculate the bi--local geodesic operators in static spherically symmetric spacetimes. Here, we refer to it as a means to understand the underlying symmetries of some scalar observables. We come back to this point in a while.

Let us now return to the permissible bi--local variations. Keep in mind that some arbitrary bi--local variation as in eq.~(\ref{eq:generic_bilocal_var}) can be expanded with respect to the tetrad bases $\Vec{{e}}_{\mathscr{O}}$ and $\Vec{{e}}_{\mathscr{E}}$ at two points. For example,
\begin{eqnarray}\label{eq:X_expanded_tetrad}
\mathbf{X}
    =
    \begin{bmatrix}
    {\left(\delta x\right)}^{u'}{u}_{\mathscr{O}}^{\mu '}+\left(\delta x\right)^{l'}{l}_{\mathscr{O}}^{\mu '}+\left(\delta x\right)^{A'}{e}{^{\mu '}_{\mathscr{O}A'}} \\
    {\left(\delta x\right)}^{u}{u}_{\mathscr{E}}^\mu+\left(\delta x\right)^{l}{l}_{\mathscr{E}}^\mu +\left(\delta x\right)^{A}{e}{^\mu_{\mathscr{E}A}}
    \end{bmatrix}.
\end{eqnarray}
Therefore, we first identify all possible permissible bi--local variations of the tetrad basis of the observer and the emitter. Those are
\begin{align}\label{eq:permissible_var_tetrad}
     \begin{split}
     \boldsymbol{T}&=
     \begin{bmatrix}
     {u}_{\mathscr{O}}^{\mu '}   \\
     \left(1+z\right)^{-1}{u}_{\mathscr{E}}^{\mu }
     \end{bmatrix},\,\,\,\,\,
      \boldsymbol{K_{\mathscr{O}}}=
     \begin{bmatrix}
     {l}_{\mathscr{O}}^{\mu '}   \\
     0 
     \end{bmatrix},\,\,\,\,\, 
      \boldsymbol{K_{\mathscr{E}}}=
     \begin{bmatrix}
     0   \\
     {l}_{\mathscr{E}}^\mu 
     \end{bmatrix},
     \end{split} \nonumber \\
     \begin{split}
     \boldsymbol{E}_{A'}^{\mathscr {O}}&=
     \begin{bmatrix}
     {e}{^{\mu '}_{\mathscr{O}A'}}   \\
    0
     \end{bmatrix},\,\,\,\,\,
     \boldsymbol{E}_{A}^{\mathscr {E}}=
     \begin{bmatrix}
     0   \\
    {e}{^\mu_{\mathscr{E}B}}
     \end{bmatrix}.
     \end{split} 
\end{align}
Those 7 bi--local vectors form a  basis of the permissible variations space.
Here, z represents the redshift which is obtained through the ratio of the emitted frequency of light to its observed value, i.e.,
\begin{equation}\label{eq:redshift_def}
    z = \frac{\langle \vec l_{\mathscr{E}}, \vec u_{\mathscr{E}}\rangle}{\langle \vec l_{\mathscr{O}}, \vec u_{\mathscr{O}}\rangle} - 1.
\end{equation}
It is straightforward to verify that all of the bi--local vectors in eq.~(\ref{eq:permissible_var_tetrad}) satisfy the permissibility criteria (\ref{eq:permissibility_criteria}). This is due to the semi--orthonormal tetrad satisfying the properties presented in eq.~(\ref{eq:tetrad_properties}) at \pE and at \pO. 

Now that all possible permissible bilocal tetrad variations are obtained, one can consider possible observables corresponding to them. We mentioned in sub--section~\ref{sec:Scalar observables and bi--local variations} that one can compose scalar observables through the operator $\bU\left(\cdot, \cdot \right)$. Let us remind that this operator inputs two arbitrary bi--local variations, $\X{}$ and $\boldsymbol{Y}$, that are given in the form in eq.~(\ref{eq:X_expanded_tetrad}). This means $\X{}$ and $\boldsymbol{Y}$ can be expanded with respect to the permissible bi--local tetrad variations listed in eq.~(\ref{eq:permissible_var_tetrad}). 

For instance, let us consider a case where $\X{}=m\boldsymbol{E}_{A'}^{\mathscr {O}}$ with ${m}$ being an arbitrary function and $\boldsymbol{Y}=\boldsymbol{T}$. Then, $\U{m\boldsymbol{E}_{A'}^{\mathscr {O}}}{\boldsymbol{T}}$ inputs $m\boldsymbol{E}_{A'}^{\mathscr {O}}$ with the variation at the observer being $\delta \Vec{x}^{\mathscr{O}}=m\Vec{e}^{\mathscr{O}}_{A'}$ and the variation at emitter being $\delta \Vec{x}^{\mathscr{E}}=\Vec{0}$. It inputs $\boldsymbol{T}$ with the projection at the observer being $\Vec{Y}_{\mathscr{O}}=\Vec{u}_{\mathscr{O}}$ and the projection at emitter being $\Vec{Y}_{\mathscr{E}}=\left(1+z\right)^{-1}\Vec{u}_{\mathscr{E}}$. The result follows as
\begin{eqnarray}
    \U{m\boldsymbol{E}_{A'}^{\mathscr {O}}}{\boldsymbol{T}}&=&m\tensor{U}{^{\mathscr{O}}^{\mathscr{O}}_{\Vec{u}_\mathscr{O}}_{\Vec{e}^{\mathscr{O}}_{{A'}}}}+\tensor{U}{^{\mathscr{O}}^{\mathscr{E}}_{\Vec{u}_\mathscr{O}}_{\Vec{0}}}\nonumber\\
    &+&m\left(1+z\right)^{-1}\tensor{U}{^{\mathscr{E}}^{\mathscr{O}}_{\Vec{u}_\mathscr{E}}_{\Vec{e}^{\mathscr{O}}_{{A'}}}}+\left(1+z\right)^{-1}
    \tensor{U}{^{\mathscr{E}}^{\mathscr{E}}_{\Vec{u}_\mathscr{E}}_{\Vec{0}}}.\nonumber \\
    &=&m\left(\tensor{U}{^{\mathscr{O}}^{\mathscr{O}}_{\Vec{u}_\mathscr{O}}_{\Vec{e}^{\mathscr{O}}_{{A'}}}}+\left(1+z\right)^{-1}\tensor{U}{^{\mathscr{E}}^{\mathscr{O}}_{\Vec{u}_\mathscr{E}}_{\Vec{e}^{\mathscr{O}}_{{A'}}}}\right).
    \end{eqnarray}
This gives us an observable associated with the variations at the end points \pO and \pE by an amount equal to $m\Vec{e}^{\,\mathscr{O}}_{{A}'}$ and $\Vec{0}$ respectively, which is then measured by  an observer with 4--velocity $\Vec{u}_\mathscr{O}$ and an emitter with 4--velocity $\Vec{u}_\mathscr{E}$.

In the next section, we construct physical observables via such combinations of $\bU\left(\X{}, \boldsymbol{Y} \right)$. We then investigate the underlying symmetries of given observables in Section~\ref{sec:Physical outcomes} by making use of the symmetries of the matrix $\mathbf{U}$.

\section{\label{sec:The observables on the sky}The observables on the sky}
\subsection{\label{sec:Momentary observables}Momentary observables}
In this section, we recall the  momentary optical observables, related to the measurement of light emitted by distant sources in a curved spacetime. Those observables are recorded via instantaneous measurements. We simply list them as
\begin{itemize}
    \item [(i)] Redshift, which is defined through the ratio of emitted value of the frequency of light to the observed value. Here, for the sake of simplifying the expressions, we consider a logarithmic redshift 
    \begin{eqnarray}\label{eq:log_redshift_defn}
    \ln \left(1+z\right)=\ln \left(\frac{\inner{\Vec{l}_{\mathscr{E}},\Vec{u}_{\mathscr{E}}}}{\inner{\Vec{l}_{\mathscr{O}},\Vec{u}_{\mathscr{O}}}}\right).
\end{eqnarray}

    \item [(ii)] Position vector, which is defined by the observer. It can be written with respect to his/her 4--velocity and the null propagation vector as
    \begin{eqnarray}\label{eq:position_vec}
\Vec{r}_{\mathscr{O}}=\Vec{u}_{\mathscr{O}}+\omega_{\mathscr{O}}^{-1}\Vec{l}_{\mathscr{O}}, 
     \end{eqnarray}
     with $\omega_\mathscr{O} = \inner{{\Vec{l}_{\mathscr{O}},\Vec{u}_{\mathscr{O}}}}$.
     This vector satisfies
     \begin{eqnarray}
      \inner{\Vec{r}_{\mathscr{O}},\Vec{r}_{\mathscr{O}}}=1, \qquad   \inner{\Vec{r}_{\mathscr{O}},\Vec{u}_{\mathscr{O}}}=0,
     \end{eqnarray}
     i.e., it is normalized and spatial as viewed in the \textit{observer's} frame and it points in the direction from which the observer sees the light incoming.
      \item [(iii)] Viewing direction vector, which is defined by the emitter. It can be written with respect to his/her 4--velocity and the null propagation vector as
    \begin{eqnarray}\label{eq:viewing_direction_vec}
\Vec{r}_{\mathscr{E}}=\Vec{u}_{\mathscr{E}}+\omega_{\mathscr{E}}^{-1}\Vec{l}_{\mathscr{E}},
     \end{eqnarray}
     with $\omega_\mathscr{E} = \inner{{\Vec{l}_{\mathscr{E}},\Vec{u}_{\mathscr{E}}}}$.
      Similar to the case of the observer's position vector, this vector satisfies
     \begin{eqnarray}
      \inner{\Vec{r}_{\mathscr{E}},\Vec{r}_{\mathscr{E}}}=1, \qquad   \inner{\Vec{r}_{\mathscr{E}},\Vec{u}_{\mathscr{E}}}=0.  
     \end{eqnarray}
A more operational definition of this quantity would be as follows: it defines the direction from which we are observing the source. If the emission is not perfectly isotropic, the results obtained by the observer will depend on $\Vec{r}_{\mathscr{E}}$. The viewing direction is a normalized spatial vector in the \textit{emitter's} frame.
\end{itemize}

\subsection{\label{sec:Variational observables}Variational observables}

Now, we consider those observables which can not be recorded via momentary observations. We classify those observables in two main branches: (i) Drift effects, which require timelike variation of the observer and the emitter. The recordings of the observables are taken as the observer propagates on his/her own timelike worldline. (ii) Effects of spatial variations, which are caused by the end point variation of the observer and the emitter in transverse directions. The recordings of the observables are taken as the observer and the emitter perform spatial variations. We should add that one is free to add a constant variation along $\Vec{l}$ to arbitrary transverse variations of the emitter and the observer. The corresponding effect is just a reparametrization of the null geodesic without its alteration and thus it is not physically measurable in any case. Also, note that displacing an observer or a single point light source along a spacelike vector is obviously nonphysical. However, in practice, these quantities can be measured indirectly. For example, one can obtain them by comparing the measurement recordings of two nearby observers, or by observing an extended source of known shape and size.

In this section, we see that the variational observables can be obtained through the aforementioned symmetric operator $\bU\left(\cdot, \cdot \right)$ associated with the matrix $\mathbf{U}$. Later, we analyse the relationship between those observables and investigate some  relations similar to reciprocity relations via the symmetry properties of $\bU\left(\cdot, \cdot \right)$.

\subsubsection{\label{sec:Drift effects}Drift effects}

\begin{figure}[!tbp] \includegraphics[width=0.52\textwidth]{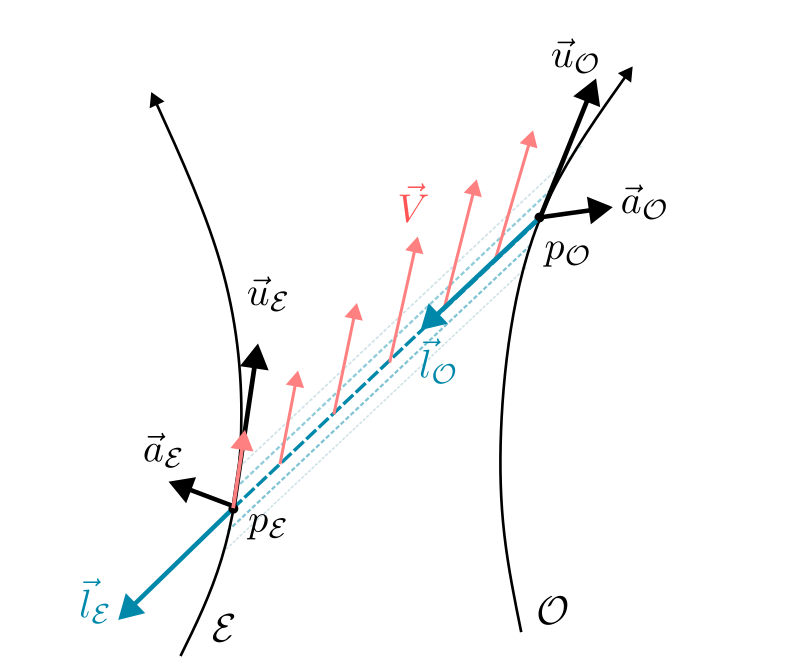}
    \caption{The observation time vector $\vec{V}$ records the variation of null geodesics as the emitter and the observer propagate on their own worldlines. It interpolates between the endpoint values $\Vec{u}_{\mathscr{O}}$ and $(1+z)^{-1}
    \,\Vec{u}_{\mathscr{E}}$ along a null geodesic connecting the worldlines of the observer $\mathscr{O}$ and the emitter $\mathscr{E}$. Vectors $\Vec{a}_{\mathscr{O}}$ and $\Vec{a}_{\mathscr{E}}$ denote momentary accelerations of the observer and the emitter respectively at the moments of observation at $p_{\mathscr{O}}$ and of emission at $p_{\mathscr{E}}$.}
\label{fig:observation_time_vector}
\end{figure}

Imagine a signal being sent by an emitter continuously in between the time interval $\tau = \tau _0$ and $\tau = \tau _0+\Delta \tau$ where $\tau$ is the emitter's proper time. Those signals will be continuously received by an observer in the time interval $\tau '= \tau ' _0$ and $\tau '= \tau ' _0+\Delta \tau '$ where $\tau '$ represents  the observer's proper time. In order to guarantee that each signal sent by the emitter is received by the observer during the given time interval, one needs to introduce a geometrical object, $\Vec{V}$, which records the variation of the null geodesic $\eta$ connecting the worldlines of the source and the observer as time passes (See Fig~\ref{fig:observation_time_vector}.).
This object was first introduced in \cite{Perlick:1990} and referred to as \textit{infinitesimal message}. Later, it was further developed in \cite{Korzynski:2017nas, Grasso:2018mei} where it was referred to as \textit{observation time vector}.

The vector $\Vec{V}$ has the following properties.
\begin{itemize}
    \item [(i)] It keeps track of the deviation of $\eta$ corresponding to the time interval of the observation. Therefore, it satisfies the geodesic deviation equation. In this work, we are concerned with the first order part of the dynamics. Then we have
    \begin{eqnarray}\label{eq:dev_eq_X}
\nabla _{\vec{l}}\nabla _{\vec{l}}V^\alpha=\tensor{R}{^\alpha_{\vec{l}}_{\,\vec{l}}_{\Vec{V}}}\,.
\end{eqnarray}
    \item [(ii)] The vector $\Vec{V}$ is necessarily  a solution to the deviation of only the null geodesics. Hence,
    \begin{eqnarray}
     \inner{\Vec{l},\nabla _{\Vec{l}}\Vec{V}}=0. \label{eq:ldotVlzero}
    \end{eqnarray}
    \item [(iii)] The boundary conditions of $\Vec{V}$ are given by 
    \begin{eqnarray}
        \Vec{V}\left(\lambda _{\mathscr{O}}\right)=\Vec{u}_{\mathscr{O}},\qquad \Vec{V}\left(\lambda _{\mathscr{E}}\right)=\left(1+z\right)^{-1}\Vec{u}_{\mathscr{E}},
    \end{eqnarray}
    where $\Vec{u}_{\mathscr{O}}$ is the 4--velocity of the observer, $\Vec{u}_{\mathscr{E}}$ is the one of the emitter and $z$ is the redshift as defined in eq.~(\ref{eq:redshift_def}). 
    Note that those boundary conditions follow from the fact that the null geodesic $\eta$ intersects the worldlines of the emitter and the observer at the end points, i.e., $\eta \left(\lambda _{\mathscr{O}}\right)=\gamma _{\mathscr{O}}\left(\tau '\right)$ and $\eta \left(\lambda _{\mathscr{E}}\right)=\gamma _{\mathscr{E}}\left(\tau \right)$. The need for the $(1+z)^{-1}$ prefactor in front of $\Vec{u}_{\mathscr{E}}$, on the other hand, follows from the null geodesic condition (\ref{eq:ldotVlzero}), see \cite{Korzynski:2017nas, Grasso:2018mei} for details. Physically, this pre-factor corresponds to the time dilation effect as the observer registers the emitter's proper time slowed down by the factor $\left(1+z\right)^{-1}$. Note that this fact was verified experimentally by making use of the Type--Ia supernovae data before \cite{Blondin:2008mz}.
\end{itemize}

It is then obvious that the vector $\Vec{V}$ evaluated at $p_\mathscr{O}$ and at $p_\mathscr{E}$ indeed defines a bi--local variation vector which causes the drift effects in general. Vector $\Vec{V}$ depends on both the 4--velocity of the source and of the observer. It tells us that the endpoint, \pO, of the null geodesic at the observer is varied by  $\delta \Vec{x}_{\mathscr{O}}=\Vec{u}_{\mathscr{O}}$ and the other endpoint is varied by $\delta \Vec{x}_{\mathscr{E}}=\left(1+z\right)^{-1}\Vec{u}_{\mathscr{E}}$ at \pE.
Those variations correspond to the changes that occur due to the kinematics of the observer and the emitter. Obviously,  such a bi--local variation of endpoints is defined through $\bm T$  of eq.~(\ref{eq:permissible_var_tetrad}). 
We may therefore think of $\vec V$ as the extension of $\bm T$ to the full connecting null geodesic. It follows that we can obtain the variations of $\vec l_\mathscr{O}$ and $\vec l_\mathscr{E}$ under $\vec V$ from $\bm U$. We use this fact in the derivation of the drifts below. 

\subsubsection*{\label{sec:Redshift drift}Redshift drift}

The redshift drift is the change in the redshift recording of an object on the sky during a continuous observation. Another way of putting this is that it is an observable corresponding to the relative total change in the light signal caused by $\Vec{V}$ at both ends and measured by an observer with 4--velocity $\Vec{u}_{\mathscr{O}}$ and an emitter with 4--velocity $\Vec{u}_{\mathscr{E}}$. Then, the endpoint variation is indeed equal to $\boldsymbol{T}$ from eq.~(\ref{eq:permissible_var_tetrad}). 

Let us now recall the covariant, bi--local variation operator, $\bm \nabla_{\X{}}$, defined in eqs.~(\ref{eq:def_bilocal_var_v}) and (\ref{eq:def_bilocal_var_tensor}) respectively acting on bi--local vectors and generic bi--local tensors.

We may apply this operator to the logarithmic redshift, $\ln \left(1+z\right)$, in order to see the effect of bi--local variation caused by $\boldsymbol{T}$. This variation is obviously equal to the derivative of the logarithmic redshift with respect to the observer's proper time $\tau'$. We begin by substituting $\boldsymbol{T}$ for the 8--dimensional generic variation vector $\X{}$ in eq.~(\ref{eq:def_bilocal_var_tensor}).  Then we obtain
\begin{align}
    &\frac{d}{d\tau'} \ln(1+z)  \equiv {\bm \nabla_{\bm T}} \ln \left(1+z\right)\nonumber \\
    &= \frac{1}{\inner{\Vec{l}_{\mathscr{E}},\Vec{u}_{\mathscr{E}}}}\Bigg( \Bigg.\inner{{\bm \nabla_{\bm T}} \Vec{l}_{\mathscr{E}},\Vec{u}_{\mathscr{E}}}+\inner{\Vec{l}_{\mathscr{E}},{\bm \nabla_{\bm T}} \Vec{u}_{\mathscr{E}}}\Bigg. \Bigg)\nonumber \\
    &-\frac{1}{\inner{\Vec{l}_{\mathscr{O}},\Vec{u}_{\mathscr{O}}}}\Bigg( \Bigg.\inner{{\bm \nabla_{\bm T}} \Vec{l}_{\mathscr{O}},\Vec{u}_{\mathscr{O}}}+\inner{\Vec{l}_{\mathscr{O}},{\bm \nabla_{\bm T}} \Vec{u}_{\mathscr{O}}}\Bigg. \Bigg).\nonumber
\end{align} 
Here,  we have 
\begin{eqnarray}
\begin{bmatrix}
         \boldsymbol{\nabla}_{\bm T}\Vec{l}_{\mathscr{O}} \\ \boldsymbol{\nabla}_{\bm T}\Vec{l}_{\mathscr{E}}
        \end{bmatrix} 
=
 \begin{bmatrix}
         \Delta{\Vec{l}}_{\mathscr{O}} \\ \Delta{\Vec{l}}_{\mathscr{E}}
        \end{bmatrix}, 
\end{eqnarray}
giving the the covariant variations of the null tangent vector at both ends, under the assumption that
\begin{eqnarray}
     \begin{bmatrix}
         \delta \Vec{x}_{\mathscr{O}} \\ 
         \delta \Vec{x}_{\mathscr{E}}
    \end{bmatrix} = {\bm T}.
\end{eqnarray}
 Then, once we use (i) eqs.  (\ref{eq:U_matrix_subblocks}) and (\ref{eq:U_matrix_boundary}) to identify the sub--blocks of the matrix $\mathbf{U}$ and (ii) the components of $\boldsymbol{T}= \left[\Vec{u}_{\mathscr{O}} \qquad (1+z)^{-1}\Vec{u}_{\mathscr{E}} \right]^\intercal$ at both ends, we obtain
\begin{eqnarray}\label{eq:redshift_drift_U}
  \frac{d}{d\tau'} \ln \left(1+z\right)=\omega _{\mathscr{O}}^{-1} \left(\frac{\inner{\Vec{l}_{\mathscr{E}},\Vec{a}_{\mathscr{E}}}}{(1+z)^{2}}-\inner{\Vec{l}_{\mathscr{O}},\Vec{a}_{\mathscr{O}}} - \U{\boldsymbol{T}}{\boldsymbol{T}}\right),\nonumber \\
\end{eqnarray}
where $\Vec{a}_{\mathscr{E}}={u}_{\mathscr{E}}^\mu\nabla_{\mu}\Vec{u}_{\mathscr{E}}$ and $\Vec{a}_{\mathscr{O}}={u}_{\mathscr{O}}^{\mu'}\nabla_{\mu'}\Vec{u}_{\mathscr{O}}$ are the coordinate accelerations of the emitter and the observer respectively. The redshift drift in eq.~(\ref{eq:redshift_drift_U}) involves the line of sight Doppler effect through the acceleration differences of the emitter and the observer, in addition to the effect of geometry which manifests itself in $\U{\boldsymbol{T}}{\boldsymbol{T}}$ term, i.e., through the integrated components of the Riemann tensor along the null geodesic.

We finally note that it is also possible to express eq.~(\ref{eq:redshift_drift_U}) with the help of the position $\Vec{r}_{\mathscr{O}}$ and the viewing direction $\Vec{r}_{\mathscr{E}}$ as
\begin{eqnarray}\label{eq:redshift_drift_U_r}
  \frac{d}{d\tau'} \ln \left(1+z\right)=\frac{\inner{\Vec{r}_{\mathscr{E}},\Vec{a}_{\mathscr{E}}}}{1+z}-\inner{\Vec{r}_{\mathscr{O}},\Vec{a}_{\mathscr{O}}} - \omega _{\mathscr{O}}^{-1} \,\U{\boldsymbol{T}}{\boldsymbol{T}},\nonumber \\
\end{eqnarray}
through eqs.~(\ref{eq:position_vec}) and (\ref{eq:viewing_direction_vec}).

\subsubsection*{\label{sec:Position drift}Position drift}
The position drift ${{\cal  D}}_{\tau'}\Vec{r}_{\mathscr{O}}$ 
is defined as the change of the transverse components of the position vector $\Vec{r}_{\mathscr{O}}$ given in eq.~(\ref{eq:position_vec}) with respect to the proper clocks of the observer and the emitter. The variation of the position is defined with respect to the Fermi--Walker transported spatial frame \cite{Korzynski:2017nas, Grasso:2018mei}. This variation can be expressed as the Fermi--Walker derivative $\cal D$ of $\Vec{r}_{\mathscr{O}}$
with respect to the observer's proper time $\tau'$:
\begin{align}
    &{\cal D}_{\tau'} \Vec{r}_{\mathscr{O}} \equiv \frac{\cal D}{d\tau'} \,\Vec{r}_\mathscr{O} \nonumber \\ & = \nabla_{\Vec{u}_{\mathscr{O}}}\,\Vec{r}_{\mathscr{O}} - \inner{\Vec{a}_{\mathscr{O}},\Vec{r}_{\mathscr{O}}} \,\Vec{u}_{\mathscr{O}} + \inner{\Vec{u}_{\mathscr{O}},\Vec{r}_{\mathscr{O}}} \,\Vec{a}_{\mathscr{O}}.
    \label{Fermi Walker}
\end{align}
The Fermi--Walker derivative is necessary in this context, because unlike the covariant derivative, it respects the decomposition into spatial vectors and $\Vec{u}_\mathscr{O}$ for accelerating observers as well \cite{Korzynski:2017nas, Grasso:2018mei}. Note that the last term in eq.~(\ref{Fermi Walker}) vanishes, so only the first two contribute.

It is a spatial and transverse vector, i.e.
\begin{eqnarray}
\inner{{\cal D}_{\tau'} \Vec{r}_{\mathscr{O}}, \Vec{r}_{\mathscr{O}}} = \inner{{\cal D}_{\tau'} \Vec{r}_{\mathscr{O}}, \Vec{u}_{\mathscr{O}}} = 0,
\end{eqnarray}
and therefore has only transverse components, related to the rate of change of the celestial coordinates of the source. In an adapted tetrad we obviously have
\begin{eqnarray}
\inner{{\cal D}_{\tau'} \Vec{r}_{\mathscr{O}}, \Vec{e}^{\,\mathscr{O}}_{A'}} =
\inner{\nabla_{\Vec{u}_{\mathscr{O}}}\Vec{r}_{\mathscr{O}}, \Vec{e}^{\,\mathscr{O}}_{A'}},
\end{eqnarray}
i.e., the second term in (\ref{Fermi Walker}) does not contribute and it is enough to calculate the transverse components of the covariant derivative of the position vector. Thus, the 
position drift is similar to the redshift drift, with an additional requirement that  only the observables projected on the observer's transverse screen are of interest. Then, it can be calculated using a similar variation under $\bm T$:
\begin{eqnarray}
\inner{\Vec{e}^{\,\mathscr{O}}_{A'},{\bm \nabla}_{\bm T}  \Vec{r}_{\mathscr{O}}}&=&
\inner{\Vec{e}^{\,\mathscr{O}}_{A'},{\bm \nabla}_{\bm T}\Vec{u}_{\mathscr{O}}}\nonumber
\\
&+&
\omega_{\mathscr{O}} ^{-1} \inner{\Vec{e}^{\,\mathscr{O}}_{A'},{\bm \nabla}_{\bm T} \Vec{l}_{\mathscr{O}}}
-   \frac{{\bm \nabla}_{\bm T}\, \omega_{\mathscr{O}}}{{\omega_\mathscr{O}}^2}\,\inner{\Vec{e}^{\,\mathscr{O}}_{A'},\Vec{l}_{\mathscr{O}}}.\nonumber \\
\label{eq:TrO}
\end{eqnarray}
The last term vanishes because $\Vec{e}^{\,\mathscr{O}}_{A'}$ is a transverse vector. The first term, just like in the redshift drift case, is simply equal to the observer's 4-acceleration $\Vec{a}_{\mathscr{O}}$.
Substitution of the covariant variation of the null tangent vector at the observer, ${\bm \nabla}_{\bm T} \Vec{l}_{\mathscr{O}}=\Delta \Vec{l}_{\mathscr{O}}$, in terms of the sub--blocks of the matrix $\mathbf{U}$ and the components of $\boldsymbol{T}$ gives
\begin{eqnarray}\label{eq:position_drift_U}
\inner{\Vec{e}^{\,\mathscr{O}}_{A'},{\cal D}_{\tau'} \Vec{r}_{\mathscr{O}}}&=&
\inner{\Vec{e}^{\,\mathscr{O}}_{A'},\Vec{a}_{\mathscr{O}}}
+\omega_{\mathscr{O}} ^{-1} {\U{\boldsymbol{T}}{\boldsymbol{E}_{A'}^{\mathscr {O}}}}.\nonumber \\
\end{eqnarray}
The position drift in eq.~(\ref{eq:position_drift_U}) involves an acceleration term which represents the non--gravitational aberration drift, i.e., the apparent position change of every source due to the gradual change of the frame caused by any non--gravitational acceleration. The effects of spacetime geometry are represented though $\U{\boldsymbol{T}}{\boldsymbol{E}_{A'}^{\mathscr {O}}}$, obtained by integrating the relevant components of the Riemann tensor.

\subsubsection*{\label{sec:Viewing direction drift}Viewing direction drift}
Let us recall that the viewing direction vector, $\Vec{r}_{\mathscr{E}}$, defined on the emitter's side is essentially the analogue of the position vector, $\Vec{r}_{\mathscr{O}}$, defined on the observer's side. Thus, we follow a similar method to derive the drift effect of the viewing direction vector. Namely, by using the definition of the viewing direction vector as in eq.~(\ref{eq:viewing_direction_vec}) we identify how it changes with respect to the timelike variations at the both ends. The result is again projected on a local screen belonging to the emitter, providing this way the two non--vanishing, transverse components of the Fermi--Walker derivative of $\Vec{r}_{\,\mathscr{E}}$ along the source's worldline. The construction is therefore in full analogy with the position drift. We may then derive the counterpart of eq.~(\ref{eq:TrO}) as
\begin{eqnarray}
\inner{\Vec{e}^{\,\mathscr{E}}_{A},{\bm \nabla}_{\bm T}  \Vec{r}_{\mathscr{E}}}&=&
\inner{\Vec{e}^{\,\mathscr{E}}_{A},{\bm \nabla}_{\bm T}\Vec{u}_{\mathscr{E}}}\nonumber
\\
&+&
\omega_{\mathscr{E}} ^{-1} \inner{\Vec{e}^{\,\mathscr{E}}_{A},{\bm \nabla}_{\bm T} \Vec{l}_{\mathscr{E}}}
-   \frac{{\bm \nabla}_{\bm T}\, \omega_{\mathscr{E}}}{{\omega_\mathscr{E}}^2}\,\inner{\Vec{e}^{\,\mathscr{E}}_{A},\Vec{l}_{\mathscr{E}}}.\nonumber \\
\end{eqnarray}
The last term vanishes as before. For the first term in the above, we  have ${\bm \nabla}_{\bm T} \Vec{u}_{\mathscr{E}} = \left(1+z\right)^{-1}\,\nabla_{\Vec{u}_\mathscr{E}} \Vec{u}_\mathscr{E} = (1+z)^{-1} \Vec{a}_\mathscr{E}$. We can also substitute
${\bm \nabla}_{\bm T} \Vec{l}_\mathscr{E} = \Delta \Vec{l}_\mathscr{E}$ and express it via the sub--blocks of $\mathbf{U}$, to obtain the analogue of eq.~(\ref{eq:position_drift_U}):
\begin{eqnarray}\label{eq:viewing_direction_drift_U}
\inner{\Vec{e}^{\,\mathscr{E}}_A,{\cal D}_{\tau'} \Vec{r}_{\mathscr{E}}}
&=&\frac{\inner{\Vec{e}^{\,\mathscr{E}}_A,\Vec{a}_{\mathscr{E}}}}{1+z}
-\omega_{\mathscr{E}}^{-1} {\U{\boldsymbol{T}}{\boldsymbol{E}_{A}^{\mathscr {E}}}}.\nonumber \\
\end{eqnarray}

As for the position drift, the operational definition of the viewing direction drift involves the Fermi--Walker transport of $\Vec{r}_{\mathscr{E}}$ along the source's timelike worldline. However, this does not affect the results as we are considering only the transverse components. We should also mention that the derivatives are taken again with respect to the \textit{observer's} proper time, as this seems more reasonable from the observational point of view. If one wants to obtain the drift of $\Vec{r}_{\mathscr{E}}$ with respect to the emitter's proper time, then one can adjust the result though the relationship $d\tau=\left(1+z\right)d\tau'$.

\subsubsection{\label{sec:Effects of transverse variations}Effects of transverse variations}
Until now we considered timelike bi--local variations defined through the 4--velocities of the observer and the emitter. Those resulted in drift effects. In this section, we define variational observables which are sourced by spacelike variations transverse to the null propagation direction.

The derivation process of those effects of the transverse variations is similar to the ones of drift effects. However, the variation bi--vector is different this time. Namely, instead of varying both ends with the bi--local vector $\boldsymbol{T}$, we consider variations of the form
\begin{eqnarray}\label{eq:variation_parallax_position}
    \X{}= \delta x ^{A '}_{\mathscr O}\,\boldsymbol{E}_{A'}^{\mathscr {O}} +
    \delta x ^{A }_{\mathscr E}\,\boldsymbol{E}_{A}^{\mathscr {E}},
\end{eqnarray}
where $\delta x^{A '}_\mathscr{O}$ and $\delta x ^{A }_{\mathscr E}$ represent the arbitrary values of the transverse variations at \pO and at \pE respectively. The bi--local vectors $\boldsymbol{E}_{A'}^{\mathscr {O}}$ and $\boldsymbol{E}_{A}^{\mathscr {E}}$ are the permissible bi--local tetrad variations given in eq.~(\ref{eq:permissible_var_tetrad}). 

\begin{figure}[!tbp]
  \begin{subfigure}[b]{0.45\textwidth}
    \includegraphics[width=\textwidth]{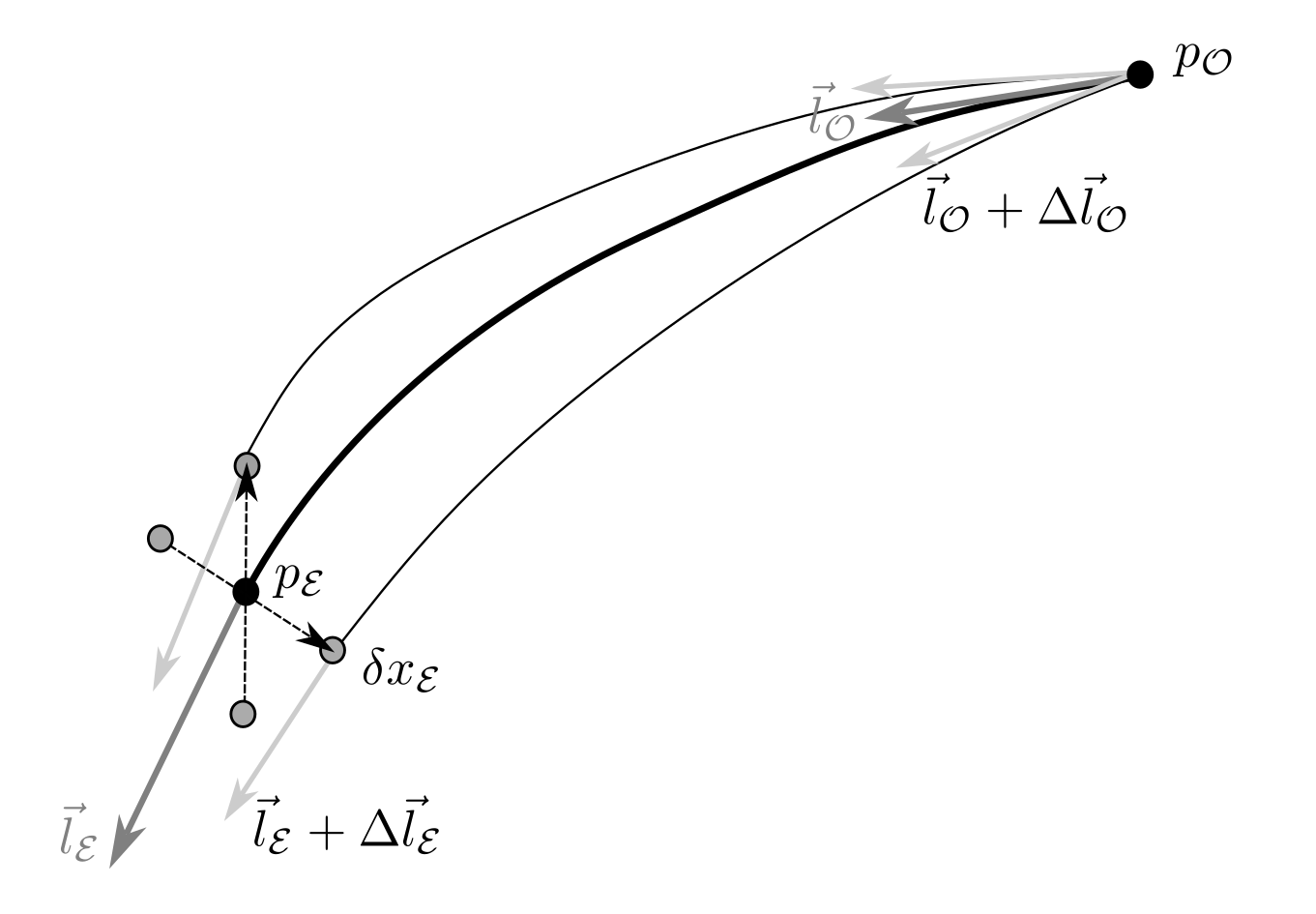}
    \caption{Transverse variations of the emission point by $\delta x_{\mathscr{E}}^A$. Those position variations lead to the variation of the null tangents at both endpoints. This results in the variation of the redshift, the apparent position and the viewing direction. In all cases the signal is received at the same position and at the same time, corresponding to the spacetime event $p_{\mathscr{O}}$. The time dimension is suppressed, while both transverse directions are present.}
    \label{fig:varying E}
  \end{subfigure}
  \hfill
  \begin{subfigure}[b]{0.45\textwidth}
\includegraphics[width=\textwidth]{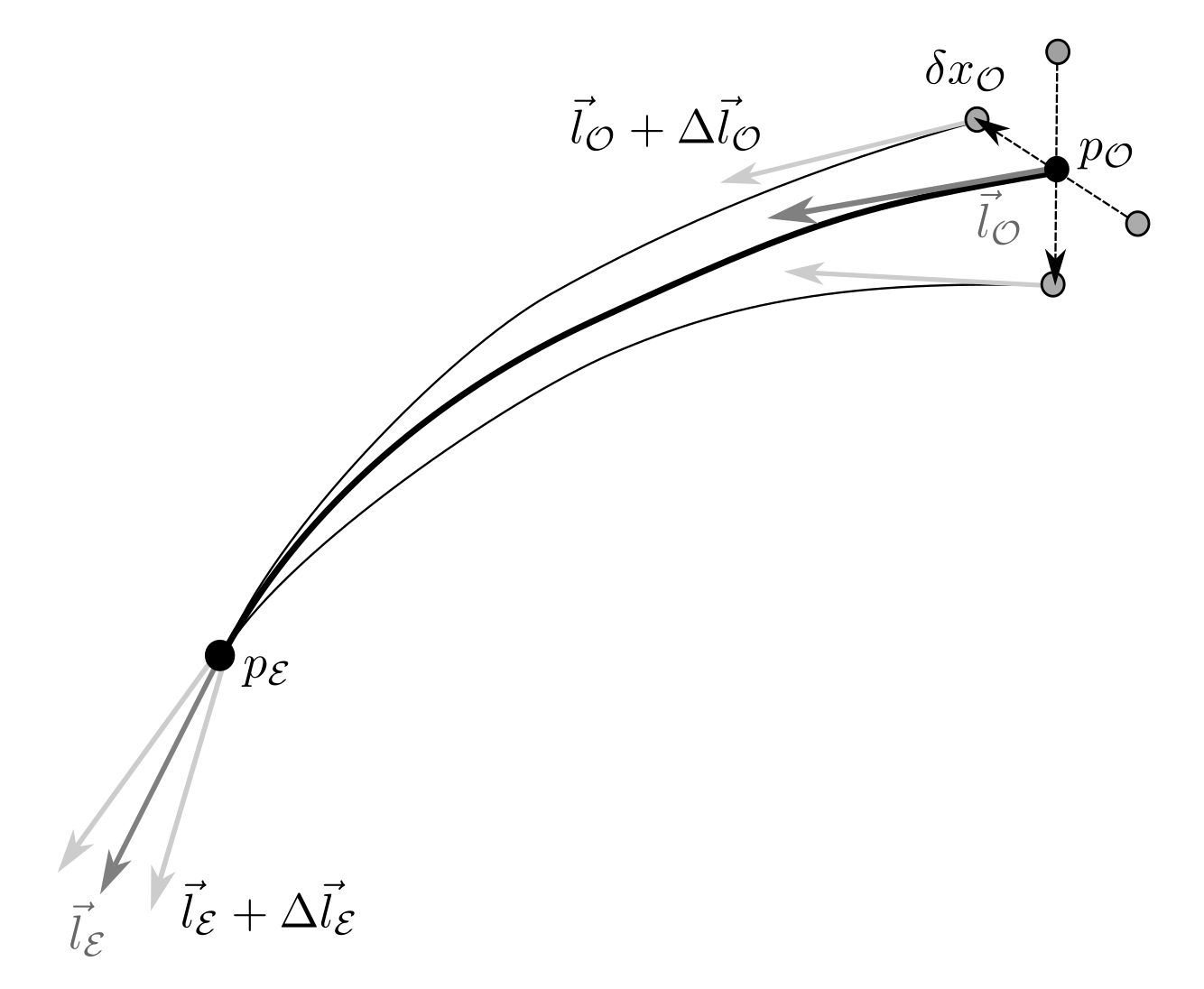}
    \caption{Transverse variations of the observation point $\delta x_{\mathscr{ O}}^{A'}$ that lead to the variations of the null tangents. All observations correspond to the same emission point at the same emission moment, represented by the event $p_{\mathscr{ E}}$. Time dimension is again suppressed, while both transverse directions are present.}
    \label{fig:varying O}
  \end{subfigure}
  \caption{Comparison of the transverse variations of the emitter and the observer.}
\end{figure}

In the calculations, we consider the position displacements of the emitter and the observer in tandem as in eq.~(\ref{eq:variation_parallax_position}). In the discussion of the physical interpretation of the results, on the other hand, we distinguish those displacements for a given scenario. The two physical setups we consider are presented in Fig.~(\ref{fig:varying E}) and (\ref{fig:varying O}), respectively.

\subsubsection*{\label{sec:Change in redshift under transverse variations}Redshift under transverse variations}
In this part, we investigate the variation of the redshift with respect to spatial variations of the emitter and the observer on the transverse directions. Here, we consider the changes in the logarithmic redshift again. We have a similar variational approach as in the drift case,
\begin{align}
    &{\bm \nabla}_{\bm X} \ln \left(1+z\right)\nonumber \\
    &= \frac{1}{\inner{\Vec{l}_{\mathscr{E}},\Vec{u}_{\mathscr{E}}}}\Bigg( \Bigg.\inner{{\bm \nabla_{\X{}}} \Vec{l}_{\mathscr{E}},\Vec{u}_{\mathscr{E}}}+\inner{\Vec{l}_{\mathscr{E}},{\bm \nabla_{\X{}}} \Vec{u}_{\mathscr{E}}}\Bigg. \Bigg)\nonumber \\
    &-\frac{1}{\inner{\Vec{l}_{\mathscr{O}},\Vec{u}_{\mathscr{O}}}}\Bigg( \Bigg.\inner{{\bm \nabla_{\X{}}} \Vec{l}_{\mathscr{O}},\Vec{u}_{\mathscr{O}}}+\inner{\Vec{l}_{\mathscr{O}},{\bm \nabla_{\X{}}} \Vec{u}_{\mathscr{O}}}\Bigg. \Bigg).\nonumber
\end{align}
with
\begin{eqnarray}
\begin{bmatrix}
         \boldsymbol{\nabla}_{\X{}}\Vec{l}_{\mathscr{O}} \\ \boldsymbol{\nabla}_{\X{}}\Vec{l}_{\mathscr{E}}
        \end{bmatrix} 
=
 \begin{bmatrix}
         \Delta{\Vec{l}}_{\mathscr{O}} \\ \Delta{\Vec{l}}_{\mathscr{E}}
        \end{bmatrix}, 
\end{eqnarray}
giving the the covariant variations of the null tangent vector at both ends. In order to investigate the effect of the transverse variations on the redshift, however, we take the bi--local variation vector to be in the form in eq.~(\ref{eq:variation_parallax_position}) this time, i.e.,
\begin{eqnarray}
     \begin{bmatrix}
         \delta \Vec{x}_{\mathscr{O}} \\ 
         \delta \Vec{x}_{\mathscr{E}}
    \end{bmatrix} = \delta x ^{A '}_{\mathscr O}
     \begin{bmatrix}
     {\Vec{e}}{^{\,\mathscr{O}}_{A'}}   \\
    0
     \end{bmatrix}+
     \delta x ^{B }_{\mathscr E}
     \begin{bmatrix}
     0   \\
    {\Vec{e}}{^{\,\mathscr{E}}_{B}},
     \end{bmatrix}.
\end{eqnarray}

This allows us to consider only the spatial, transverse variations. Then, we obtain
\begin{align}\label{eq:redshift_transverse_var_U}
&{\bm \nabla}_{\bm X} \ln \left(1+z\right) \nonumber \\
&=\omega_{\mathscr{E}}^{-1}\delta x^B_{\mathscr{E}}\inner{\Vec{l}_{\mathscr{E}},\nabla_B\Vec{u}_{\mathscr{E}}}-\omega_{\mathscr{O}}^{-1}\delta x^{A'}_{\mathscr{O}}\inner{\Vec{l}_{\mathscr{O}},\nabla_{A'}\Vec{u}_{\mathscr{O}}}\nonumber \\
&-\omega_{\mathscr{O}}^{-1}\delta x^B_{\mathscr{E}}{\U{\boldsymbol{E}_{B}^{\mathscr {E}}}{\boldsymbol{T}}}-\omega_{\mathscr{O}}^{-1}\delta x^{A'}_{\mathscr{O}}{\U{\boldsymbol{E}_{A'}^{\mathscr {O}}}{\boldsymbol{T}}}.
\end{align}
For the sake of simplicity we assume that the displaced source and observer are comoving with the original ones, i.e. ${\bm \nabla}_{\bm X} \Vec{u}_\mathscr{O} = 0 = {\bm \nabla}_{\bm X} \Vec{u}_\mathscr{E}$. Operationally, this means that we compare the results of the displacements after the astrometric reduction of the observed data to a common observer's frame. An example would be the barycentric frame of the Solar System. Moreover, if we restrict ourselves to situations where only the observer ($\delta x ^{\mathscr{E}}_B=0$) or only the emitter ($\delta x ^{\mathscr{O}}_{A'}=0$) has transverse variations, then we respectively obtain
\begin{itemize}
    \item [(i)] Parallax of the redshift: 
\begin{eqnarray}\label{eq:parallax_redshift_U}
&&{\bm \nabla}_{\bm X}\ln \left(1+z\right)\equiv
\delta x^{A'}_\mathscr{O}\,\partial_{A'}\ln (1+z) = \nonumber \\ && \qquad -\delta x^{A'}_{\mathscr{O}}\left(\omega_{\mathscr{O}}^{-1}\,{\U{\boldsymbol{E}_{A'}^{\mathscr {O}}}{\boldsymbol{T}}}\right).
\end{eqnarray}
Here we explicitly substitute $\bm X$ and use the fact that $\ln(1+z)$ is a bi--scalar. Meaning, the derivatives are taken with respect to both primed and unprimed coordinates.
This result can be viewed as the difference in the simultaneous redshift recordings of a point source by two observers: the fiducial observer and the one who is relatively displaced, but comoving with the first one. In analogy with the trigonometric parallax effect we call this effect the \textit{parallax of the redshift}.

    \item [(ii)] Reversed parallax of redshift:
    \begin{eqnarray}\label{eq:rev_parallax_redshift_U}
&&{\bm \nabla}_{\bm X} \ln \left(1+z\right)\equiv\delta x^{B}_\mathscr{E}\,\,\partial_{B}\ln (1+z) = \nonumber \\&&\qquad -\delta x^B_{\mathscr{E}}\left(\omega_{\mathscr{O}}^{-1}{\U{\boldsymbol{E}_{B}^{\mathscr {E}}}{\boldsymbol{T}}}\right)
.
\end{eqnarray}
This quantity can be interpreted as follows: consider an extended source whose parts are all comoving. The difference of the redshift recording across the image of this extended source is then given by eq.~(\ref{eq:rev_parallax_redshift_U}). This result is obviously dual (reciprocal) to the parallax of the redshift, i.e., they are related by swapping the roles of $\mathscr{O}$ and $\mathscr{E}$. Hence we propose the term \textit{reversed parallax of the redshift} for it.
\end{itemize}
The parallax of redshift and the reversed parallax of redshift are not standard names in astrometry, but in the context of all possible variational observables they seem appropriate.

\subsubsection*{\label{sec:Transverse variations of positions}Transverse variations of positions}
Now we consider the transverse variations of the position vector, $\Vec{r}_\mathscr{O}$. With transverse variations we refer to spatial changes with respect to the changes in the observer's and the emitter's worldlines. 
Then, considering the definition of the position vector as in eq.~(\ref{eq:position_vec}) and taking the variation with respect to the bi--local vector given in eq.~(\ref{eq:variation_parallax_position}) results in

\begin{eqnarray}\label{eq:Trig_parallax_positions_U}
&&\inner{\Vec{e}^{\,\mathscr{O}}_{A'},{\bm \nabla}_{\bm X}   \Vec{r}_\mathscr{O}}=\inner{\Vec{e}^{\,\mathscr{O}}_{A'},{\bm \nabla}_{\bm X}  \left(\Vec{u}_{\mathscr{O}}+\omega_\mathscr{O}^{-1}\,\Vec{l}_\mathscr{O}\right) }  \nonumber \\
&&\qquad =\omega_{\mathscr{O}} ^{-1} \delta x ^{B '}_{\mathscr{O}}\U{\boldsymbol{E}_{B'}^{\mathscr {O}}}{\boldsymbol{E}_{A'}^{\mathscr {O}}}\nonumber \\
&&\qquad+\omega_{\mathscr{O}} ^{-1}\delta x ^B_{\mathscr{E}}\U{\boldsymbol{E}_{B}^{\mathscr {E}}}{\boldsymbol{E}_{A'}^{\mathscr {O}}}.
\end{eqnarray}
We again assume that the observer's 4--velocity does not change, i.e., the fiducial and the displaced observers are both co--moving.

\subsubsection*{\label{sec:Transverse variations of viewing direction}
Transverse variations of the viewing direction}
Similar to the transverse variations of the position vector on the observer's side, one can define the transverse variations of the viewing direction vector, $\Vec{r}_{\mathscr{E}}$, on the emitter's side given in eq.~(\ref{eq:viewing_direction_vec}). The bi--local variation vector is again represented by the one presented in eq.~(\ref{eq:variation_parallax_position}). That means we are considering the spatial, transverse variations of both the emitter and the observer. Assuming that the emitter's 4--velocity $\Vec{u}_{\mathscr{E}}$ is covariantly constant, we have

\begin{eqnarray}\label{eq:Trig_parallax_viewing_dir_U}
&&\inner{\Vec{e}^{\,\mathscr{E}}_{A},{\bm \nabla}_{\bm X}   \Vec{r}_\mathscr{E}}=\inner{\Vec{e}^{\,\mathscr{E}}_{A},{\bm \nabla}_{\bm X}  \left(\Vec{u}_{\mathscr{E}}+\omega_\mathscr{E}^{-1}\,\Vec{l}_\mathscr{E}\right) }  \nonumber \\
&&\qquad =-\omega_{\mathscr{E}} ^{-1} \delta x ^{B '}_{\mathscr{O}}\U{\boldsymbol{E}_{B'}^{\mathscr {O}}}{\boldsymbol{E}_{A}^{\mathscr {E}}}\nonumber \\
&&\qquad-\omega_{\mathscr{E}} ^{-1}\delta x ^B_{\mathscr{E}}\U{\boldsymbol{E}_{B}^{\mathscr {E}}}{\boldsymbol{E}_{A}^{\mathscr {E}}}.
\end{eqnarray}

\subsubsection*{\label{sec:Parallax matrix, magnifications and the angular diameter distance}Parallax matrix, magnifications and angular diameter distances}
We now investigate how the apparent position on the observer's sky changes with respect to his/her (i.e., the observer's) own transverse, spatial variations. This corresponds to the well--known trigonometric parallax. On a curved background, the position on the sky is defined via the covariant changes of the position vector $\Vec{r}_\mathscr{O}$, i.e., ${\bm \nabla}_{\bm X} \Vec{r}_{\mathscr{O}}$. Now, we may also introduce orthogonal angular coordinates $\theta_\mathscr{O}^{A'}$ on the observer's celestial sphere, centered at the unperturbed position $\Vec r_\mathscr{O}$, projected along transverse vectors $\Vec{e}^{\,\mathscr{O}}_{A'}$ and measured in radians. Then, the variations of these coordinates $\delta\theta^\mathscr{O}_{A'}$ match exactly the components of ${\bm \nabla}_{\bm X} r^\mathscr{O}_{A'}$.

 We now relate the covariant variation of the position vector  to the variation of the null tangent $\Vec{l}_\mathscr{O}$. First, note that ${\bm \nabla}_{\bm X} \Vec{r}_\mathscr{O}$ is  equal to the derivative of the transverse component of the geodesic deviation vector with respect to the proper distance, i.e.,
 
\begin{eqnarray}
    {\bm \nabla}_{\bm X} r ^{A'}_{\mathscr{O}}=\frac{d(\delta x ^{A'}_{\mathscr{O}})}{d\ell}=\frac{d(\delta x ^{A'}_{\mathscr{O}})}{\omega _{\mathscr{O}} d\lambda}=\omega _{\mathscr{O}}^{-1}\nabla _{\vec{l}_{\mathscr{O}}}\delta x ^{A'}_{\mathscr{O}}.
\end{eqnarray}

Here, $d\ell$ is the infinitesimal proper length and its relation to the affine parameter, $\lambda$, and the observer's proper time, $\tau'$, follows as
\begin{eqnarray}
 d\ell= \omega_\mathscr{O}\, d\lambda=d\tau',
\end{eqnarray}
in geometrized units with the speed of light $c=1$.
On the other hand, as we discussed in Section~\ref{sec:Symplectic symmetries and the bi--locality}, we have
\begin{eqnarray}\label{eq:switch_geoddev_covvar}
\Delta l^{\mu}=\nabla _{\delta \vec{x}}l^{\mu}=\nabla _{\vec{l}}\,\delta x^{\mu}
\end{eqnarray}
in general due to the Lie dragging condition. This means the position variation can be defined through the covariant variation of the null tangent vector $\Delta \Vec{l}$ at \pO, with $\Delta \Vec{l}_\mathscr{O} \equiv {\bm \nabla}_{\bm X} \Vec{l}_{\mathscr{O}}$ in this case. Then, we write
\begin{eqnarray}\label{eq:angles_bilocal-parallax}
\delta \theta^\mathscr{O}_{A'} \equiv {\bm \nabla}_{\bm X} r^{\mathscr{O}}_{A'}=\omega _{\mathscr{O}}^{-1}\inner{\Vec{e}^{\,\mathscr{O}}_{\,A'},{\bm \nabla}_{\bm X}  \Vec{l}_{\mathscr{O}}}.
\end{eqnarray}
Now, we are interested in only the observer's transverse variations. Then, the bi--local variation vector is given by
\begin{eqnarray}\label{eq:variation_parallax_angle}
    \X{}=\delta x ^{A '}_{\mathscr{O}} \boldsymbol{E}_{A'}^{\mathscr {O}}.
\end{eqnarray}
Substitution of the sub--matrices of the matrix $\mathbf{U}$ and the components of $\X{}$ as in the above gives
\begin{eqnarray}
 \delta \theta^\mathscr{O}_{A'} \equiv {\bm \nabla}_{\bm X} r^{\mathscr{O}}_{A'}=-\tensor{\Pi} {_{A'}_{B'}} \delta x ^{B '}_{\mathscr{O}}.
\end{eqnarray}
with $\tensor{\Pi} {_{A'}_{B'}}$ being the parallax matrix, i.e.,
\begin{eqnarray}\label{eq:parallax_matrix_obs}
\tensor{\Pi} {_{A'}_{B'}}=-\frac{\partial \theta ^{\mathscr{O}}_{A'}}{\partial \delta x ^{B '}_{\mathscr{O}}}=-\omega _\mathscr{O}^{-1}\U{\boldsymbol{E}_{A'}^{\mathscr {O}}}{\boldsymbol{E}_{B'}^{\mathscr {O}}}.
\end{eqnarray}
The minus sign in the definition is chosen in order to be consistent with the definition of the trigonometric parallax in flat spacetime, i.e., the direction of the apparent position variation is \textit{opposite} to the direction of the observer's displacement.

One might also ask how the angles recorded at the observer changes with respect to the transverse variations at the emitter's side. This is obviously related to the image distortions and the magnifications. In the standard gravitational lensing theory, the magnification and the distortion effects are defined with respect to their expected values in flat background. We can relate our results to the one of the standard lensing theory by considering the apparent position variations at linear order. The definition of magnification is similar to the parallaxes, however, this time we are interested in the variations of type
\begin{eqnarray}\label{eq:variation_magnification_angle}
    \X{}=\delta x_{\mathscr{E}} ^{A }\, \boldsymbol{E}_{A}^{\mathscr {E}},
\end{eqnarray}
i.e., the one which involves transverse variations of the point source only,
rather than the variation in eq.~(\ref{eq:variation_parallax_angle}).
We can again make use of eq.~(\ref{eq:angles_bilocal-parallax}) for the variation of the apparent position, but with $\X{}$  given by eq.~(\ref{eq:variation_magnification_angle}).
The result can be written in the form
\begin{eqnarray}
 \delta \theta^\mathscr{O}_{A'} \equiv {\bm \nabla}_{\bm X} r^{\mathscr{O}}_{A'} =\tensor{M} {_{A'}_{B}}\delta x ^{B }_{\mathscr{E}},
\end{eqnarray}
where 
\begin{eqnarray}\label{eq:Magnification_matrix_observer}
\tensor{M} {_{A'}_{B}}=\frac{\partial \theta ^{\mathscr{O}}_{A'}}{\partial \delta x ^{B }_{\mathscr{E}}}=\omega _\mathscr{O}^{-1}\U{\boldsymbol{E}_{B}^{\mathscr {E}}}{\boldsymbol{E}_{A'}^{\mathscr {O}}},    
\end{eqnarray}
is the \textit{magnification matrix}. Note that in the standard lensing theory this matrix is usually rescaled by a transverse length scale.

The inverse of the magnification matrix tells us how the deviation vector at the emitter changes with respect to the angles recorded by the observer. In the literature, this matrix is known as the Jacobi matrix. In Section~\ref{sec:Symmetries and reciprocity relations} we referred to the Jacobi matrix defined from the observer to the emitter as $\tensor{\mathbf{D}}{}\left(\lambda _\mathscr{E},\lambda _\mathscr{O}\right)$. As we discussed before, its determinant is a measure of the ratio of the cross sectional area of an extended object at the emission point to the solid angles recorded by the observer. In this respect, it gives the square of the angular diameter distance $D_{A}$ up to a frequency factor, i.e.,

\begin{eqnarray}\label{eq:ang_diam_distance_U}
D_{A}^2=\left|{\rm{det}}\left({M} _{A'B}^{-1}\right)\right|=\omega _\mathscr{O}^{2}\,\left|{\rm{det}}\left[\U{\boldsymbol{E}_{B}^{\mathscr {E}}}{\boldsymbol{E}_{A'}^{\mathscr {O}}}\right]\right|^{-1}.\nonumber \\
\end{eqnarray}

Let us now compare the transverse variation of the position vector given in eq.~(\ref{eq:Trig_parallax_positions_U}) with the definitions of the parallax and the magnification matrices which are respectively given in eqs.~(\ref{eq:parallax_matrix_obs}) and (\ref{eq:Magnification_matrix_observer}). 
For this, we rewrite eq.~(\ref{eq:Trig_parallax_positions_U}) as
\begin{eqnarray}
    \inner{\Vec{e}^{\mathscr{O}}_{A'},{\bm \nabla}_{\bm X}  \Vec{r}_\mathscr{O}}&=&
\omega_{\mathscr{O}} ^{-1} \delta x ^{B '}_{\mathscr{O}}\U{\boldsymbol{E}_{B'}^{\mathscr {O}}}{\boldsymbol{E}_{A'}^{\mathscr {O}}}\nonumber \\
&\qquad&+\omega_{\mathscr{O}} ^{-1}\delta x ^B_{\mathscr{E}}\U{\boldsymbol{E}_{B}^{\mathscr {E}}}{\boldsymbol{E}_{A'}^{\mathscr {O}}}\nonumber \\
&\qquad&+\delta x ^{B '}_{\mathscr{O}}\inner{\Vec{e}^{\mathscr{O}}_{A'},\nabla _{B'} \Vec{u}_{\mathscr{O}}}.
\end{eqnarray}
Then, we immediately realize that the transverse variation of the position vector is given through the difference of the effect of the magnification matrix, $\tensor{M} {_{A'}_{B}}$, and the effect of the parallax matrix, $\tensor{\Pi} {_{A'}_{B'}}$, in addition to the transverse variations of the 4--velocity of the observer, i.e.,
\begin{eqnarray}\label{eq:Rel_trans_var_pos_magn_par}
\inner{\Vec{e}^{\mathscr{O}}_{A'},{\bm \nabla}_{\bm X}  \Vec{r}_\mathscr{O}}&=&\tensor{M} {_{A'}_{B}}\delta x ^B_{\mathscr{E}}-\tensor{\Pi} {_{A'}_{B'}}\delta x ^{B '}_{\mathscr{O}}\nonumber \\
&\qquad&+\delta x ^{B '}_{\mathscr{O}}\inner{\Vec{e}^{\mathscr{O}}_{A'},\nabla _{B'} \Vec{u}_{\mathscr{O}}}.
\end{eqnarray}
In the case where $\inner{\Vec{e}^{\mathscr{O}}_{A'},\nabla _{B'} \Vec{u}_{\mathscr{O}}}=0$, i.e., the transverse variations of observer's 4--velocity has no transverse components, the transverse variations of the position vector is precisely obtained through the effect of the parallax matrix and the magnification matrix. This is equivalent to assuming that we are comparing position measurements performed by displaced but strictly co--moving observers.

\subsubsection*{\label{sec:Reversed parallax matrix, reversed magnifications and luminosity distances}Reversed parallax matrix, reversed magnifications and luminosity distances}
In the previous subsection, we investigated how the vector defining the position of the image on the observer's sky changes with respect to the transverse variations of the observer or the emitter. Following a similar argument, one can investigate how the viewing direction  of the emitter would change with respect to the transverse variations of the emitter or the observer. The viewing direction  $\Vec{r}_\mathscr{E}$ and its variation are defined in full analogy with the previous subsection, i.e., we have
\begin{eqnarray}\label{eq:emmitter_var_angle}
    {\bm \nabla}_{\bm X} r_\mathscr{E}^A =\omega _{\mathscr{E}}^{-1}\nabla _{\vec{l}_{\mathscr{E}}}\delta x ^{A}_{\mathscr{E}}.
\end{eqnarray}
We use the relationship between the derivative of the deviation vector  $\nabla_{\Vec{l}_\mathscr{E}} \delta x_\mathscr{E}^A$ and the covariant variation of the null tangent vector ${\bm \nabla}_{\bm X} \Vec{l}_\mathscr{E} \equiv \Delta \Vec{l}_\mathscr{E}$, i.e., eq.~(\ref{eq:switch_geoddev_covvar}), again. We then write the variation of emitter's viewing direction as
\begin{eqnarray}\label{eq:angles_bilocal_magn_obs}
{\bm \nabla}_{\bm X} r^\mathscr{E}_A \equiv \delta \theta ^{\mathscr{E}} _{A}&=&\omega_{\mathscr{E}}^{-1}\inner{\Vec{e}^{\,\mathscr{E}}_A,{\bm \nabla}_{\bm X} \Vec{l}_{\mathscr{E}}}.
\end{eqnarray}
Here, we define the angular coordinates $\theta_{\mathscr{E}}^{A}$ related to the vectors $\Vec{e}^\mathscr{\,E}_A$ on the emitter's celestial sphere, repeating the construction on the observer's celestial sphere from the previous subsection.
Let us first investigate the trigonometric parallax effect on the emitter's side, i.e., the variation of the viewing direction when the observer is fixed and the position of the emitter varies in the transverse direction. Then, the bi--local variation vector, $\X{}$, should be given by eq.~(\ref{eq:variation_magnification_angle}) as we are interested in variations with respect to the emitter. Then, following the same procedure as in the previous section allows us to write the variation of the viewing direction in the form
\begin{eqnarray}
 \delta \theta ^{\mathscr{E}} _{A}=-\tensor{\tilde{\Pi}} {_{A}_{B}} \delta x ^{B }_{\mathscr{E}}.
\end{eqnarray}
with $\tensor{\tilde{\Pi}} {_{A}_{B}}$ being the parallax matrix of the emitter, 
\begin{eqnarray}\label{eq:rev_parallax_matrix_emitter}
\tensor{\tilde{\Pi}} {_{A}_{B}}=-\frac{\partial \theta ^{\mathscr{E}}_{A}}{\partial \delta x ^{B }_{\mathscr{E}}}=\omega _{\mathscr{E}}^{-1}\U{\boldsymbol{E}_{B}^{\mathscr {E}}}{\boldsymbol{E}_{A}^{\mathscr {E}}}.
\end{eqnarray}

We now demonstrate the analogue of the magnification matrix on the side of the emitter, i.e., how the viewing direction of the emitter changes with respect to the transverse variations of the observer. We refer to this effect as the \textit{reversed image magnification and deformation} effect. The variation of emitter's angles are defined as in eq.~(\ref{eq:emmitter_var_angle}), however, now we consider the variation vector to be restricted only to the observer's transverse plane. Thus, we substitute $\X{}$ as in form in eq.~(\ref{eq:variation_parallax_angle}) to eq.~(\ref{eq:emmitter_var_angle}). Then, variation of the angles can be given in the form
\begin{eqnarray}
 \delta \theta _{A}=\tensor{\tilde{M}} {_{A}_{B'}} \delta x ^{B'}.
\end{eqnarray}
with
\begin{eqnarray}\label{eq:rev_magnification_matrix_emitters}
\tensor{\tilde{M}} {_{A}_{B'}}=\frac{\partial \theta ^{\mathscr{E}}_{A}}{\partial \delta x ^{B '}}=-\omega _{\mathscr{E}}^{-1}\U{\boldsymbol{E}_{B'}^{\mathscr {O}}}{\boldsymbol{E}_{A}^{\mathscr {E}}}.
\end{eqnarray}
being the \textit{reversed magnification matrix}.

Note that the inverse of $\tensor{\tilde{M}} {_{A}_{B'}}$ is also a Jacobi matrix. However this time it is the Jacobi matrix, $\tensor{\mathbf{D}}{}\left(\lambda _\mathscr{O},\lambda _\mathscr{E}\right)$, defined from the emitter to the observer. Its determinant incorporates how the cross sectional area of a collection of light rays at the observer's location change with respect to the solid angles recorded by the emitter. This means, it is equivalent to the squared value of the luminosity distance, i.e.,
\begin{eqnarray}\label{eq:lum_diam_distance_U}
{D'}_{L}{^2}=\left|{\rm{det}}\left({\tilde{M}} _{AB'}^{-1}\right)\right|=\omega _\mathscr{E}^{2}\,\left|{\rm{det}}\left(\U{\boldsymbol{E}_{B'}^{\mathscr {O}}}{\boldsymbol{E}_{A}^{\mathscr {E}}}\right)\right|^{-1}.\nonumber \\
\end{eqnarray}

Let us now return to the transverse variations of the viewing direction vector, $\Vec{r}_\mathscr{E}$, which was previously presented in eq.~(\ref{eq:Trig_parallax_viewing_dir_U}). We rewrite it as
\begin{eqnarray}\label{eq:Trans_var_parallax_viewing_dir_U_special}
\inner{\Vec{e}^{\,\mathscr{E}}_{\,A},{\bm \nabla}_{\bm X}  \Vec{r}_\mathscr{E}}&=&
-\omega_{\mathscr{E}} ^{-1}\delta x ^B_{\mathscr{E}}\U{\boldsymbol{E}_{B}^{\mathscr {E}}}{\boldsymbol{E}_{A}^{\mathscr {E}}}\nonumber \\
&\qquad&-\omega_{\mathscr{E}} ^{-1} \delta x ^{B '}_{\mathscr{O}}\U{\boldsymbol{E}_{B'}^{\mathscr {O}}}{\boldsymbol{E}_{A}^{\mathscr {E}}}\nonumber \\
&\qquad&+\delta x ^{B }_{\mathscr{E}}\inner{\Vec{e}^{\,\mathscr{E}}_{\,A},\nabla _{B} \Vec{u}_{\mathscr{E}}}.
\end{eqnarray}
When we compare eqs.~(\ref{eq:Trans_var_parallax_viewing_dir_U_special}), (\ref{eq:rev_parallax_matrix_emitter}) and (\ref{eq:rev_magnification_matrix_emitters}) we observe that the transverse variation of the viewing direction vector is obtained through the differences of the effects of the reversed magnification matrix, $\tensor{\tilde{M}} {_{A}_{B'}}$, and the reversed parallax matrix, $\tensor{\tilde{\Pi}} {_{A}_{B}}$, in addition to the transverse variations of the 4--velocity of the emitter. In other words, we have
\begin{eqnarray}\label{eq:Rel_trans_var_view_dir_revmagn_revpar}
\inner{\Vec{e}^{\,\mathscr{E}}_{\,A},{\bm \nabla}_{\bm X}  \Vec{r}_\mathscr{E}}&=&\tensor{\tilde{M}} {_{A}_{B'}}\delta x ^{B '}_{\mathscr{O}}- \tensor{\tilde{\Pi}} {_{A}_{B}}\delta x ^B_{\mathscr{E}}\nonumber \\
&\qquad&+\delta x ^{B}_{\mathscr{E}}\inner{\Vec{e}^{\,\mathscr{E}}_{\,A},\nabla _{B} \Vec{u}_{\mathscr{E}}}.
\end{eqnarray}
For the case where transverse variations of $\Vec{u}_{\mathscr{E}}$ have no component in the transverse direction,
$\inner{\Vec{e}^{\,\mathscr{E}}_{\,A},\nabla _{B} \Vec{u}_{\mathscr{E}}}=0$, the viewing direction variation is given directly through the reversed magnification and reversed parallax matrices.

At this point, we exhausted all possible components of $\bm U$, considered as a quadratic form, restricted to the 7--dimensional subspace of the admissible vectors defined by (\ref{eq:criteria_local_surf_com}). It turns out that the optical properties of the spacetime are related \textit{only} to this restriction \cite{Korzynski:2021aqk}, while the components of $\bm U$ from outside that subspace are related to non--null geodesics, representing the propagation of massive fields or particles.

\section{\label{sec:Physical outcomes}Physical outcomes}
In the previous section, we derived variational observables by using the operator $\U{\cdot}{\cdot}$. This operator can be viewed as a tool to construct scalar observables which result from bi--local variations at the emitter and at the observer. It is essentially defined through a symmetric block matrix $\mathbf{U}$ which is the solution of the following boundary value problem: Given the position variation vectors at the both ends of a null geodesic, what is the resultant variation in its tangent vectors? In Sections~\ref{sec:Symplectic symmetries and the bi--locality} and \ref{sec:Canonical transforms and initial to boundary value problem} we showed that such a boundary value problem is equivalent to an initial value problem. In the equivalent form, the initial values of the position variation and the tangent vector variation defined at \pO are linearly transformed to give the final values at \pE. We showed that the corresponding linear transformation is indeed canonical and it is represented by a symplectic matrix $\mathbf{W}$. 

In this section, we present various relationships in between the variational observables which might seem unrelated at a first glance. We show that those relationships follow from the underlying symplecticity of the first order dynamics. Equivalently, they manifest themselves through the symmetric property of matrix $\mathbf{U}$ which is essentially the solution of the corresponding boundary value problem. In other words, operationally, the symplectic symmetries display themselves through the symmetric operator $\U{\cdot}{\cdot}$ which appears in all of the variational observables outlined in Section~\ref{sec:Variational observables}. Now, we investigate their relationships following from the symmetry of $\mathbf{U}$, or equivalently, from the symplectic property of $\mathbf{W}$. 
\subsection{\label{sec:Position drift--Parallax of redshift relationship} Position drift--parallax of the redshift relationship}

 Let us compare the results we obtained for position drift in eq.~(\ref{eq:position_drift_U}) and for parallax of the redshift in (\ref{eq:parallax_redshift_U}). Those are obtained from the operator $\U{\cdot}{\cdot}$. 
 By making use of the symmetric property (\ref{eq:U is symmetric}), it is straightforward to see that we have
 \begin{eqnarray} \label{eq:Rel_pos_drift_par_redshift}
&&\delta x_\mathscr{O}^{A'}\,\,\partial_{A'}\ln(1+z) = \nonumber\\
&&\qquad
\delta x_\mathscr{O}^{A'}\left(\inner{\Vec{e}_{A'}^\mathscr{O},\Vec{a}_{\mathscr{O}}}-\inner{\Vec{e}_{A'}^\mathscr{O},{\cal D}_{\tau'}\Vec{r}_{\mathscr{O}}}\right).
 \end{eqnarray}
 Obviously, we discovered here an exact relation between the position drift on the right hand side and the parallax of the redshift on the left hand side. The former has the contribution from the aberration drift due to the non--gravitational acceleration subtracted. From this, we observe that for a free--falling, geodesic observer the parallax of the redshift and the position drift components are identical with opposite signs.

\begin{figure}[!tbp] \includegraphics[width=0.52\textwidth]{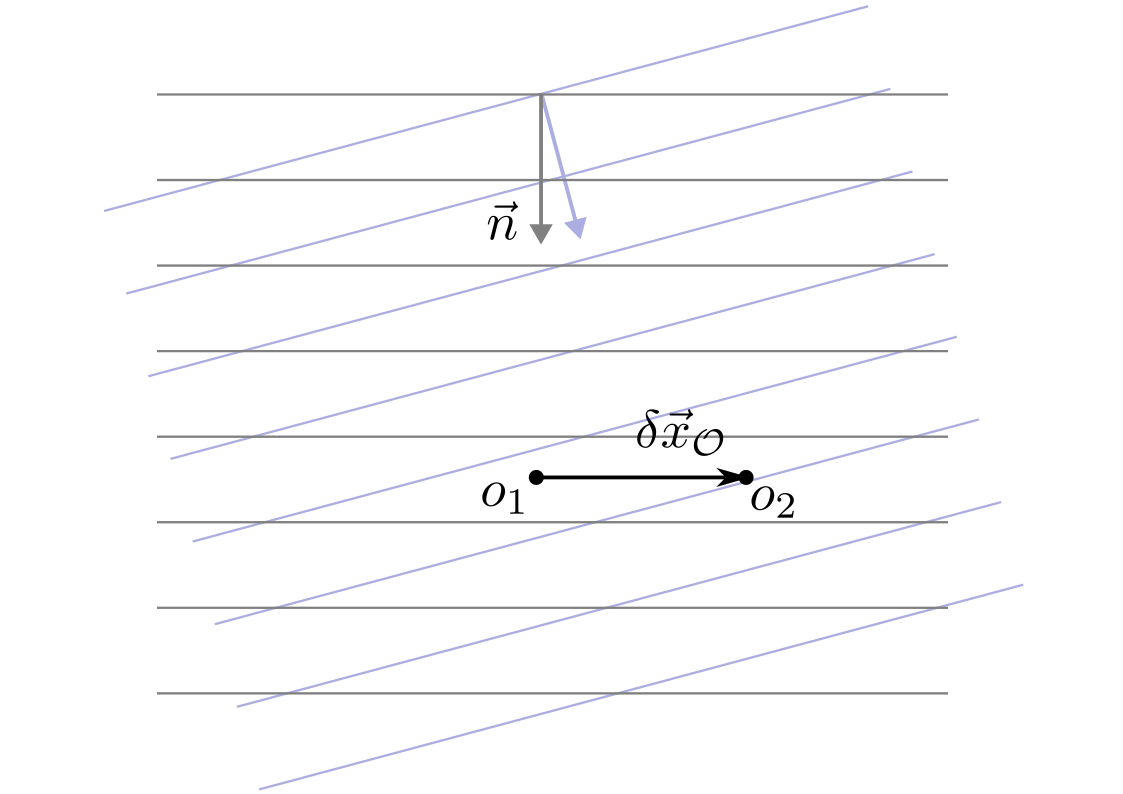}
    \caption{
    Sketch of a plane wave of constant frequency and slowly varying direction of propagation observed at two points $o_1$ and $o_2$ for two  fixed moments of time.}
\label{fig:variable-direction-wave}
\end{figure}

This equality is fairly easy to understand in the case of a distant source moving in a flat spacetime. 
Consider a plane wave of constant frequency $\omega$ with a slowly varying direction of propagation, $\vec n$, due to the source's motion (See Fig.~\ref{fig:variable-direction-wave}.). 
Then the wave function has the form of 
\begin{eqnarray}
    \Psi(t, \vec x) = \exp\left(i\phi\right)= \exp\left(i\omega[t - \vec n(t)\cdot\vec x]\right),
\end{eqnarray}
the dot denoting the 3--dimensional, spatial vector product.
The phase $\phi$ of the wave is measured  by stationary observers at two points, $o_1$ and $o_2$, which are respectively given by $\vec x = 0$ and $\vec x = \delta \vec x_{\mathscr{O}}$. The corresponding phases can be compared, for example, by the means of interferometry. The recorded phases are
\begin{eqnarray}
    \phi(t, 0) &=& \omega\,t \\
    \phi(t, \delta \vec x_{\mathscr{O}}) &=& \omega\,t - \omega\,\vec n(t)\cdot \delta \vec x_{\mathscr{O}}.\nonumber \\
\end{eqnarray}

By differentiating these expressions with respect to $t$ we obtain the frequencies of the wave measured  at the two points as $\omega(t, 0) = \omega$ and 
$\omega(t, \delta \vec x_{\mathscr{O}}) = \omega - \omega\,\dot {\vec n}(t)\cdot\delta \vec x_\mathscr{O}$.  Obviously they differ by a small amount $\Delta \omega = -\omega\,\dot {\vec n}(t) \cdot \vec \delta x_{\mathscr{O}}$, related to the rate of change of the direction of propagation. Since  the position on the observer's sky can be identified with $-\vec n$ and the observed frequencies are related to the redshifts as registered by both observers, it is straightforward to check that we have
\begin{eqnarray}
\Delta \ln(1+z) = \dot{\vec n}(t) \cdot \vec \delta x_{\mathscr{O}} = -\inner{\delta \vec x_{\mathscr{O}}, {\cal D}_t \,\vec r_{\mathscr{O}}}.
\end{eqnarray}
In other words, we  re--derived eq.~(\ref{eq:Rel_pos_drift_par_redshift}) for a non--accelerating observer in a flat spacetime this way.
In that case, the parallax of the redshift can be understood as the resultant effect of the slow drift of phases measured at two displaced points. The two phases drift away from each other  due to the slow drift of the direction of wave propagation. This is in line with the result of the standard theory of interferometric measurements of positions of distant light sources in astrometry.

\subsection{\label{sec:Viewing direction drift--Reversed parallax of the redshift relationship} Viewing direction drift -- reversed parallax of redshift relationship}
     A similar relationship can be demonstrated on the emitter's side as well. Namely, this time we consider  eqs.~(\ref{eq:viewing_direction_drift_U}) and (\ref{eq:rev_parallax_redshift_U}) for viewing direction drift and the reversed parallax of redshift. We combine them with the help of the symmetry property in eq.~(\ref{eq:U is symmetric}) and the relation $1 + z={\omega_\mathscr{E}}/{\omega_\mathscr{O}}$ in order to obtain:
     \begin{eqnarray}\label{eq:Rel_view_dir_drift_reversed_par_redshift}
&&\delta x_\mathscr{E}^{B}\,\,\partial_{B}\ln(1+z) = \nonumber\\
&&\qquad
\delta x_\mathscr{E}^{B}\left((1+z)\,\inner{\Vec{e}_{\,B}^{\,\mathscr{E}},{\cal D}_{\tau'}\Vec{r}_{\mathscr{E}}}-\inner{\Vec{e}_{\,B}^{\,\mathscr{E}},\Vec{a}_{\mathscr{E}}}\right).
 \end{eqnarray}
 This is a relation between the drift of the viewing angle as measured by the emitter and the variation of the redshift across the source's image when the aberration drift contribution is absent. It is  the exact reciprocal of relation (\ref{eq:Rel_pos_drift_par_redshift}). In fact, the reader may check that if we substitute the emitter's proper time $\tau$ for the observer's proper time $\tau'$ in the viewing angle drift, (\ref{eq:Rel_view_dir_drift_reversed_par_redshift}), then the resulting expression will be the exact dual of (\ref{eq:Rel_pos_drift_par_redshift}), with a global sign difference.

 Physically this relation means that for a perfectly static source, with no spin or non--gravitational acceleration whatsoever, any variation of the redshift accross the image of a source is precisely related to the variation of the viewing angle due to the motion of the source and the observer. 

Variation of this kind would be extremely difficult to measure in astronomy, since the gradient of redshift across the source's image is very small. Note that, just like the position drift--redshift parallax effect, this effect has fairly simple explanation in a flat spacetime: for a luminous body of finite size the direction of the line of sight varies slightly across the image, leading to a variation of the line--of--sight and transverse Doppler effects for a moving body. This in turn means that the redshift is slightly different in different parts of the image. 

\subsection{\label{sec:Angular shift--transverse coordinate shift relationship}Parallax matrix and reverse parallax matrix are both symmetric}
In Subsection~\ref{sec:Variational observables} we observed how the the transverse variations in the position and in the viewing direction vectors are related to their respective parallax and magnification matrices. In this subsection, we analyse physical outcomes of the symmetries of the parallax and the reversed parallax matrices themselves.   

For this, let us re--write the parallax and the reversed parallax matrices which we previously presented in eqs.~(\ref{eq:parallax_matrix_obs}) and (\ref{eq:rev_parallax_matrix_emitter}) respectively. This time, we present them explicitly by making use of the components of the $\mathbf{U}$ matrix. They respectively follow as
    \begin{eqnarray}
    \tensor{\Pi} {_{A'}_{B'}}&=&-\omega _\mathscr{O}^{-1}\UOO{e_{A'}}{e_{B'}}, \nonumber \\
    \tensor{\tilde{\Pi}} {_{A}_{B}}&=&\omega _\mathscr{E}^{-1}\UEE{e_A}{e_{B}}.
    \end{eqnarray}
We recall that $U^{\mathscr{O}\mathscr{O}}=U^{\mathscr{O}\mathscr{O}{\intercal}}$ and $U^{\mathscr{E}\mathscr{E}}=U^{\mathscr{E}\mathscr{E}{\intercal}}$ holds due the symmetry of $\mathbf{U}$. Then, we have
    \begin{eqnarray}
    \tensor{\Pi} {_{A'}_{B'}}&=& -\frac{\partial \theta ^{\mathscr{O}}_{A'}}{\partial \delta x ^{B '}_{\mathscr{O}}}= -\frac{\partial \theta ^{\mathscr{O}}_{B'}}{\partial \delta x ^{A '}_{\mathscr{O}}}=\tensor{\Pi} {_{B'}_{A'}},\label{eq:symm_parallax_mat_obs} \\
    \tensor{\tilde{\Pi}} {_{A}_{B}}&=& -\frac{\partial \theta ^{\mathscr{E}}_{A}}{\partial \delta x ^{B }_{\mathscr{E}}} = -\frac{\partial \theta ^{\mathscr{E}}_{B}}{\partial \delta x ^{A }_{\mathscr{E}}} =\tensor{\tilde{\Pi}} {_{B}_{A}}.\label{eq:symm_revparallax_mat_emit}
    \end{eqnarray}
We can interpret the equations above as the following. For this, we compare two situations. First, consider the variation of the observer's position $\delta \vec x_{\mathscr{O}}$ in one transverse direction, say $ \vec e^{\,\mathscr{O}}_{\,1'}$. This causes the geometric parallax effect in the form of the apparent displacement of the position of the image of a distant source in the direction of observer's displacement. In a curved spacetime  there might also be a  perpendicular component of the displacement of the source's image  along $\vec e^{\,\mathscr{O}}_{\,2'}$. Second, consider the variation of the observer's position by the same distance in the other transverse direction, i.e. $\vec e^{\,\mathscr{O}}_{\,2'}$. The apparent displacement of the source's image in direction $\vec e^{\,\mathscr{O}}_{\,2'}$ is unrelated to the previous case, but the perpendicular displacement in direction $\vec e^{\,\mathscr{O}}_{\,1'}$  must be exactly the same as the transverse displacement in the previous case. 
An analogous interpretation applies for the reversed situation as well, when we consider the celestial sphere of the emitter. This is implied by eq.~(\ref{eq:symm_revparallax_mat_emit}).

\subsection{\label{sec:Magnification matrix--Reversed magnification matrix relationship}Magnification matrix--Reversed magnification matrix relationship}   
Let us re--write the magnification matrix of the observer, $\tensor{M} {_{A'}_{B}}$, and the reversed magnification matrix of the emitter, $\tensor{\tilde{M}} {_{A}_{B'}}$, which were presented in eqs. (\ref{eq:Magnification_matrix_observer}) and (\ref{eq:rev_magnification_matrix_emitters}) respectively. This time, we explicitly write them by using the the sub--blocks of $\mathbf{U}$. Then, we have 
    \begin{eqnarray}
    \tensor{M} {_{A'}_{B}}&=&\omega _\mathscr{O}^{-1}\UOE{e_{A'}}{e_B}, \nonumber \\
    \tensor{\tilde{M}} {_{B}_{A'}}&=&-\omega _\mathscr{E}^{-1}\UEO{e_B}{e_{A'}}.
    \end{eqnarray}
    Recall that we have $U^{\mathscr{O}\mathscr{E}}=U^{\mathscr{E}\mathscr{O}{\intercal}}$. Substitution of this symmetry allows us to relate the magnification and reversed magnification matrix via
    \begin{eqnarray}\label{eq:recip_magnificaiton_matrices}
    \tensor{M} {_{A'}_{B}}=-\left(1+z\right)\tensor{\tilde{M}} {_{B}_{A'}}.
    \end{eqnarray}
    This result can be viewed as the reciprocity of the magnification matrices. It is equivalent to the well-known Etherington's reciprocity law for the Jacobi matrices \cite{Perlick:2004tq, Etherington:1933, Ellis:1971pg}, as we will see in the next section.
\subsection{\label{sec:Angular diameter distance--Luminosity distance relationship} Angular diameter distance--Luminosity distance relationship} 
   We mentioned in Section~\ref{sec:Parallax matrix, magnifications and the angular diameter distance} that the inverse of the magnification matrix gives the Jacobi matrix, $\mathbf{D}\left(\lambda_{\mathscr{E}},\lambda_{\mathscr{O}}\right)$, defined from the observer to the emitter. Similarly, the inverse of the reversed magnification matrix is equal to the Jacobi matrix, $\mathbf{D}\left(\lambda_{\mathscr{O}},\lambda_{\mathscr{E}}\right)$, defined from the emitter to the observer. Those Jacobi matrices were first introduced in Section~\ref{sec:Symmetries and reciprocity relations} in the context of cosmological distance calculations and their reciprocity. 
   
   Now, recall that the angular diameter distance is given in eq.~(\ref{eq:ang_diam_distance_U}) and the luminosity distance is given in eq.~(\ref{eq:lum_diam_distance_U}). They are obtained through the determinant of their respective Jacobi matrices. Then, if (i) we first take the inverse of the relation in eq.~(\ref{eq:recip_magnificaiton_matrices}) that tells us how the magnification matrix is related to the reversed magnification matrix, (ii) then take the determinant of both sides, we obtain
    \begin{eqnarray}
    {\rm{det}}\left[{M} _{A'B}^{-1}\right]=\left(1+z\right)^{-2}{\rm{det}}\left[\tensor{\tilde{M}} {_{B}_{A'}}^{-1}\right].    
    \end{eqnarray}
    Note that the left hand side of the above gives the square of the angular diameter distance $D_{A}$ and the right hand side involves the square of the corrected luminosity distance {$D'_{L}$}, weighted by the $\left(1+z\right)^{-2}$ term. This result is equivalent to
    \begin{eqnarray}\label{eq:recip_dist}
    D'_{L}=\left(1+z\right)\,D_{A},
    \end{eqnarray}
    and the distance reciprocity can be viewed as a simple consequence of the reciprocity of the magnification matrices represented in eq.~(\ref{eq:recip_magnificaiton_matrices}).
   Substituting the uncorrected, or bolometric luminosity distance $D_{L} = (1+z)\,D'_{L}$ yields the well--known distance duality relation,
    \begin{eqnarray}\label{eq:dist_rel_bol_ang}
         D_{L}=\left(1+z\right)^2\,D_{A}.
    \end{eqnarray}
\section{\label{sec:Summary and conclusions}Summary and conclusions}
In this work, we studied certain observables on the sky in the widest possible scope. In addition to the momentary observables such as the redshift, the position vector and the viewing direction vector; we considered variable observables such as the drift and parallax--like variations of those momentary observables. The investigation of such a large set of observables was conducted under a single, covariant, mathematical framework which was originally presented in a series of papers \cite{Korzynski:2017nas,Grasso:2018mei,Korzynski:2019oal,Korzynski:2021aqk}. Some of its applications were presented in \cite{Korzynski:2019oal, Serbenta:2021tzv}. 

In the current work, the aim was to unveil the underlying algebraic symmetries of variational observables. The reason for such an inquiry is not solely due to a mathematical concern but also due to our interest in the physical relations the symmetries provide. Namely, it is in virtue of the underlying symmetries that we were able to find the relationships between seemingly unrelated variational observables.

Let us now summarize the main layers of our construction.
\begin{itemize}
    \item [(i)] We allow both the emitter, $\mathscr{E}$, and the observer, $\mathscr{O}$, to have arbitrary variations both in space and in time throughout the course of the measurement process. 
    \item [(ii)] Accordingly, one has to abandon the \textit{congruence} concept to study the propagation of collection of light rays. The space of congruences are at most 3--dimensional, while the space of all possible families of geodesics is 8--dimensional due to the need of two 4--dimensional tangent vectors at the end points of a geodesic. The dimension of this space reduces to 7 once we assume that the geodesic is null. This follows from the fact that both the coordinate positions, $\Vec{x}$, and the tangent vectors of the null geodesics, $\Vec{l}$, undergo variations in a scenario where the emitter and the observer perform variations of their own. Hence, defining a congruence around a fixed null geodesic and studying its evolution falls short for the identification of our problem.
    \item [(iii)] We study the light propagation problem on a symplectic phase space. We achieve this via studying the initial value problem of the covariant variations of positions, $\delta\Vec{x}$, and covariant variations of the null tangent vectors, $\Delta\Vec{l}$. Note that we are interested in only the first order part of the dynamics in this work. In addition, we consider the tetrad components of the phase space variables. Accordingly, the solution to the corresponding Hamiltonian initial value problem is given by a symplectic matrix, $\mathbf{W}$. 
    \item [(iv)] Due to the Lie dragging condition, $\nabla _{\Vec{l}}\, \delta \Vec{x}= \nabla _{\delta \Vec{x}}  \Vec{l}$, covariant variations of the null tangent vectors satisfy $\Delta \Vec{l}=\nabla _{\Vec{l}}\, \delta\Vec{x}$. We use this fact to show that the matrix $\mathbf{W}$ is also the solution matrix of the first order part of the null geodesic deviation equation. As physically meaningful quantities are obtained through the solutions of the geodesic deviation equation in relativity, the observables are obtained through $\mathbf{W}$. Essentially, the aforementioned algebraic symmetries of variational observables are indeed symplectic symmetries.
    \item [(v)] We then switch to a formulation which allows us to study the initial value problem mentioned in (iii) via an equivalent boundary value problem. Thereby, the symplectic matrix $\mathbf{W}$ is interchangeable with the solution matrix, $\mathbf{U}$, of the corresponding boundary value problem. They are related to each other via an invertable transformation. It turns out that the symplectic symmetries of $\mathbf{W}$ dictate $\mathbf{U}$ to be a symmetric matrix  and vice versa.
    \item [(vi)] In order to define the variational observables we first consider bi--local variation vectors. Those are 8--dimensional vectors which contain the information about two 4--dimensional variation vectors, one for $\mathscr{E}$ and the other for $\mathscr{O}$, defined on distinct spacetime points. The endpoint variations of a null geodesic are not completely arbitrary though. Namely, in order for $\mathscr{E}$ and $\mathscr{O}$ to keep on receiving light--like signals while they perform variations, their position variation vectors have to satisfy $\inner{\tensor{\Vec{l}}{_{\mathscr{O}}},\delta \tensor{\Vec{x}}{_{\mathscr{O}}}} - \inner{\tensor{\Vec{l}}{_{\mathscr{E}}},\delta \tensor{\Vec{x}}{_{\mathscr{E}}}}=0$. We refer to any variation which satisfies this criteria as a \textit{permissible} variation.
    \item [(vii)] Next, we construct variable observables via the solution matrix $\mathbf{U}$ and the permissible bi--local variations. The matrix $\mathbf{U}$ gives the values of  $\Delta\Vec{l}$ at the emission and at the observation points which result from position variations, $\delta\Vec{x}$, at the two end points of the null geodesics. Note that all of the variational observables listed in this work follow from the projection of $\Delta\Vec{l}$ on some vector which is also the solution of the geodesic deviation equation. Therefore, they all involve the sub--blocks of $\mathbf{U}$ in their calculations. We list the variational observables that are constructed through the bi--local variations of the momentary observables in Table~\ref{table:variational_obs}. 
    \item [(viii)] Finally, by using the fact that $\mathbf{U}$ is a symmetric matrix, we construct relationships between the variable observables. The corresponding physical outcomes are explicitly given in Section~\ref{sec:Physical outcomes} and they are listed in Table~\ref{table:relationships_var_obs} along with the underlying symmetries. We should emphasise that the relationships between the variable observables are exact and valid in any spacetime for any source--observer pair as long as the geometric optics approximation works, i.e., the source is not passing through a caustic.
\end{itemize}

As we outlined in Section~\ref{sec:Introduction}, drift and parallax effects are instrumental for current cosmological studies. They provide important information about the inhomogeneity and anisotropy of our universe. Accordingly, drift and parallax effects are closely related to the latest challenges of the standard model such as the dark energy problem and the Hubble parameter tension. 

Those problems are usually attacked with methods that involve data sets obtained from different types of measurements which are analysed separately. However, once equivalent data sets are considered simultaneously, interesting insights are obtained. For example, cosmological distance estimations are conducted either by (i) standard candles to obtain luminosity distances, or by (ii) standard rulers to obtain angular diameter distances. The reason we treat those two types of measurements in an equal footing is not trivial. It is due to the Etherington's distance reciprocity that we treat them to be equal (up to a redshift factor). Previously, we showed that distance reciprocity follows from the symplectic symmetries of the light propagation on a reduced phase space \cite{Uzun:2018yes}. Thus, different types of distance measurements can indeed be used interchangably due the symplectic symmetries of the null geodesic deviation dynamics.

In the current work, we enlarged the light propagation phase space and we studied variable emitters and observers. Studying the underlying algebraic symmetries of the light propagation allowed us to find analogous relationships between drift and parallax effects similar to the distance reciprocity. Hence, just as distance reciprocity sets certain constrains on cosmological models, the additional relationships we provide might be as useful in cosmological data analysis in the future. We believe this is a reasonable goal to achieve with ever improving accuracy in  cosmological experiments. Our relations may also be of interest in the fields of  precise astrometry or binary pulsar timing.

\begin{table*}[h]
\begin{tabular}{ |P{4cm}|P{3cm}||P{3cm}|P{3cm}|P{4cm}|  }
 \hline
 &&\multicolumn{3}{|c|}{Observables} \\
 \hline
 Physical variation
 & 
 Permissible variation bi--vectors
 & 
 logarithmic redshift, $\ln(1+z)$  
 & 
 position vector, $\Vec{r}_{\mathscr{O}}$ 
 & 
 viewing direction vector, $\Vec{r}_{\mathscr{E}}$ \\
 \hline
Timelike vectors at the two end points  & 
$\boldsymbol{T}=
     \begin{bmatrix}
     {u}_{\mathscr{O}}^{\mu '}   \\
     \left(1+z\right)^{-1}{u}_{\mathscr{E}}^{\mu }
     \end{bmatrix}$ &
     redshift drift & 
     position drift &
     viewing direction drift   \\
    \hline
 Observer's position in transverse directions 
 & 
 $\boldsymbol{E}_{A'}^{\mathscr {O}}=
     \begin{bmatrix}
     {e}{^{\mu '}_{\mathscr{O}A'}}   \\
    0
     \end{bmatrix}$  
     
 & 
 parallax of the redshift   
 & 
 parallax matrix
 & 
 reversed magnification matrix \\
 \hline
 Emitter's position in transverse directions 
 & 
 $\boldsymbol{E}_{A}^{\mathscr {E}}=
     \begin{bmatrix}
     0   \\
    {e}{^\mu_{\mathscr{E}A}}
     \end{bmatrix}$  
 &
 reversed parallax of the redshift 
 & 
 magnification matrix 
 &  
 reversed parallax matrix\\
 \hline
\end{tabular}
\caption{Variational observables derived from bi--local variations of momentary observables. Those momentary observables are the logarithmic redshift, $\ln(1+z)$, the position vector, $\Vec{r}_{\mathscr{O}}$, and the viewing direction vector, $\Vec{r}_{\mathscr{E}}$. Variable observables are obtained through the bi--local variations of the momentary observables. For instance, the bi--local timelike variation, $\boldsymbol{T}$, of the position vector results in the position drift or the spatial transverse variation of the redshift with $\boldsymbol{E}_{A'}^{\mathscr {O}}$ gives the parallax of redshift. Swapping the last two rows and the last two columns corresponds to swapping the role of the observer and the source.}\label{table:variational_obs}
\end{table*}

\begin{table*}[h]
\begin{tabular}{ |P{7cm}||P{2cm}|P{3cm}|P{5cm}|}

 \hline
 Physical outcomes & Mathematical relationship & Underlying symmetries of $\mathbf{U}$ & Underlying symmetries of $\mathbf{W}$ \\
 \hline
Position drift can be obtained through the parallax of redshift. 
& 
Eq.~(\ref{eq:Rel_pos_drift_par_redshift}) 
& 
\begin{eqnarray}
\mathbf{U^{\mathscr{O}\mathscr{O}}}&=&\mathbf{U^{\mathscr{O}\mathscr{O}}}^\intercal,\nonumber  \\
\mathbf{U^{\mathscr{O}\mathscr{E}}}&=&\mathbf{U^{\mathscr{E}\mathscr{O}}}^\intercal.  \nonumber
\end{eqnarray} 
& 
\begin{eqnarray}
\tensor{\mathbf{W}}{_X_X}\tensor{\mathbf{W}}{^\intercal_X_L}&=& \tensor{\mathbf{W}}{_X_L}\tensor{\mathbf{W}}{^\intercal_X_X},\nonumber \\
\tensor{\mathbf{W}}{_X_X}\tensor{\mathbf{W}}{^\intercal_L_L}&-&\tensor{\mathbf{W}}{_X_L}\tensor{\mathbf{W}}{^\intercal_L_X}=\mathbf{1}.\nonumber
\end{eqnarray}  \\
\hline
Viewing direction drift can be obtained through the reversed parallax of redshift. 
& 
Eq.~(\ref{eq:Rel_view_dir_drift_reversed_par_redshift}) 
& 
\begin{eqnarray}
\mathbf{U^{\mathscr{O}\mathscr{E}}}&=&\mathbf{U^{\mathscr{E}\mathscr{O}}}^\intercal,  \nonumber \\
\mathbf{U^{\mathscr{E}\mathscr{E}}}&=&\mathbf{U^{\mathscr{E}\mathscr{E}}}^\intercal. \nonumber
\end{eqnarray}
& 
\begin{eqnarray}
\tensor{\mathbf{W}}{_X_X}\tensor{\mathbf{W}}{^\intercal_X_L}&=& \tensor{\mathbf{W}}{_X_L}\tensor{\mathbf{W}}{^\intercal_X_X},\nonumber \\
\tensor{\mathbf{W}}{_X_X}\tensor{\mathbf{W}}{^\intercal_L_L}&-&\tensor{\mathbf{W}}{_X_L}\tensor{\mathbf{W}}{^\intercal_L_X}=\mathbf{1},\nonumber\\
\tensor{\mathbf{W}}{^\intercal_X_L}\tensor{\mathbf{W}}{_L_L}&=& \tensor{\mathbf{W}}{^\intercal_L_L}\tensor{\mathbf{W}}{_X_L}.\nonumber
\end{eqnarray}  \\
\hline
The perpendicular components of the  variations of the image position (i.e. the parallax) for identical displacements of the observer in two orthogonal transverse directions are equal.
& 
Eq.~(\ref{eq:symm_parallax_mat_obs})
& 
\begin{eqnarray}
\mathbf{U^{\mathscr{O}\mathscr{O}}}&=&\mathbf{U^{\mathscr{O}\mathscr{O}}}^\intercal.  \nonumber 
\end{eqnarray}
& 

\begin{eqnarray}
    \tensor{\mathbf{W}}{_X_X}\tensor{\mathbf{W}}{^\intercal_X_L}&=& \tensor{\mathbf{W}}{_X_L}\tensor{\mathbf{W}}{^\intercal_X_X}.\nonumber
\end{eqnarray}

\\
\hline
The perpendicular components of the viewing angle variations for identical displacements of the emitter in two orthogonal transverse directions are equal.
& 
Eq.~(\ref{eq:symm_revparallax_mat_emit})
& 
\begin{eqnarray}
\mathbf{U^{\mathscr{E}\mathscr{E}}}&=&\mathbf{U^{\mathscr{E}\mathscr{E}}}^\intercal.  \nonumber 
\end{eqnarray}
& 

\begin{eqnarray}
    \tensor{\mathbf{W}}{^\intercal_X_L}\tensor{\mathbf{W}}{_L_L}&=& \tensor{\mathbf{W}}{^\intercal_L_L}\tensor{\mathbf{W}}{_X_L}.\nonumber
\end{eqnarray}

\\
\hline
There exists a reciprocity relation between the magnification matrix and the reversed magnification matrix.
& 
Eq.~(\ref{eq:recip_magnificaiton_matrices}) 
& 
\begin{eqnarray}
\mathbf{U^{\mathscr{O}\mathscr{E}}}&=&\mathbf{U^{\mathscr{E}\mathscr{O}}}^\intercal.  \nonumber 
\end{eqnarray}
&  
\begin{eqnarray}
\tensor{\mathbf{W}}{_X_X}\tensor{\mathbf{W}}{^\intercal_X_L}&=& \tensor{\mathbf{W}}{_X_L}\tensor{\mathbf{W}}{^\intercal_X_X},\nonumber\\
\tensor{\mathbf{W}}{_X_X}\tensor{\mathbf{W}}{^\intercal_L_L}&-&\tensor{\mathbf{W}}{_X_L}\tensor{\mathbf{W}}{^\intercal_L_X}=\mathbf{1}.\nonumber
\end{eqnarray} 
\\
\hline
There exists a reciprocity relation between angular diameter distances and luminosity distances. 
& 
Eq.~(\ref{eq:recip_dist}) 
& 
\begin{eqnarray}
\mathbf{U^{\mathscr{O}\mathscr{E}}}&=&\mathbf{U^{\mathscr{E}\mathscr{O}}}^\intercal.  \nonumber 
\end{eqnarray}
& 
\begin{eqnarray}
\tensor{\mathbf{W}}{_X_X}\tensor{\mathbf{W}}{^\intercal_X_L}&=& \tensor{\mathbf{W}}{_X_L}\tensor{\mathbf{W}}{^\intercal_X_X},\nonumber \\
\tensor{\mathbf{W}}{_X_X}\tensor{\mathbf{W}}{^\intercal_L_L}&-&\tensor{\mathbf{W}}{_X_L}\tensor{\mathbf{W}}{^\intercal_L_X}=\mathbf{1}.\nonumber
\end{eqnarray} 
\\
 \hline
\end{tabular}
\caption{Relationships between variational observables and the corresponding algebraic symmetries.}\label{table:relationships_var_obs}
\end{table*}
\section*{Acknowledgements}
This research is part of the project No. 2021/43/P/ST2/01802 co-funded by the
National Science Centre and the European Union Framework Programme for Research
and Innovation Horizon 2020 under the Marie Skłodowska-Curie grant agreement No.
945339. For the purpose of Open Access, the author (NU) has applied a CC-BY public
copyright licence to any Author Accepted Manuscript (AAM) version arising from this
submission.

\begin{appendix}

\section{Proof that $\mathbf{W}_{XL}$ is degenerate iff the Jacobi matrix $\mathbf{D}$ is degenerate} \label{sec:Proof that WXL is degenerate iff the Jacobi matrix D is degenerate}

For convenience we introduce  an auxiliary, parallel-transported and semi-orthonormal vector tetrad $\left(\vec f_{\tilde{a}}\right)$, defined just like in Section~\ref{sec:Permissible bi--local variations}. It is given by $\left(\vec u, \vec l, \vec e_A\right)$, with $\vec u$ being again a 4-velocity, $\vec{e}_A$ denoting the transverse vectors and $\vec l$ being the usual null tangent to the fiducial geodesic. The products of the basis vectors are given by eq.~\eqref{eq:tetrad_properties}.

Its dual co--vector tetrad $(h^{\tilde a})$ reads
$\left(l^\flat, \frac{u^\flat}{\omega} + \frac{l^\flat}{\omega^2}, e_A^\flat \right)$, where $\flat$ denote the index lowering operation using the spacetime metric $g$. 

In \cite{Grasso:2018mei, Korzynski:2021aqk}, a number of identities relating $\mathbf{W}$ and the null vector $\Vec{l}$ were proven. In particular, it was shown that
\begin{eqnarray}
{\mathbf{W}_{XL}}^{\tilde a \tilde b} \, l_{\mathscr{O}\,\tilde b} = \Delta \lambda\cdot \,{l_\mathscr{E}}^{\tilde a}, \qquad
l_{\mathscr{E}\,\tilde a}{\mathbf{W}_{XL}}^{\tilde a \tilde b} = \Delta\lambda\cdot {l_{\mathscr{O}}}^{\tilde b}.\nonumber
\end{eqnarray}
By expressing these identities in the language of $(\vec f_{\tilde{a}})$ and $(h^{\tilde a})$  tetrads, we obtain 
$\mathbf{W}_{XL}\,\,h^{\tilde 0} = \Delta\lambda \cdot f_{\tilde 1}$ and $h^{\tilde 0}\circ \mathbf{W}_{XL}(\cdot)  = \Delta\lambda \cdot f_{\tilde 1}(\cdot)$. In the matrix notation this is in turn equivalent to
\begin{eqnarray}
\mathbf{W}_{XL}\,
\begin{bmatrix}
 1 \\ 0 \\ 0 \\ 0
\end{bmatrix} &=& \Delta\lambda \,\begin{bmatrix}0 \\ 1 \\ 0 \\ 0\end{bmatrix}, \\
&\rm{and}&\nonumber\\
\begin{bmatrix} 1 & 0 & 0 & 0 \end{bmatrix} \mathbf{W}_{XL} &=& \Delta\lambda\,\begin{bmatrix} 0 & 1 & 0 & 0 \end{bmatrix}.
\end{eqnarray}
It follows that the matrix $\mathbf{W}_{XL}$ needs to have the form of
\begin{eqnarray}
\mathbf{W}_{XL}&=&
 \left[
 \begin{array}{cc|ccc}
     0 & \Delta\lambda & 0& 0 \\
    \Delta\lambda & * & * & *\\
    \hline
    0 & * & \multicolumn{2}{c}{\multirow{2}{*}{$\mathbf{D}_{AB'}$}} &\\
    0 & * &&&\\
 \end{array}
 \right]
\end{eqnarray}
with $\mathbf{D}_{AB'}$ being the Jacobi matrix elements and ``$*$'' denoting an arbitrary value of the matrix element on the given slot. 
We can now calculate the determinant of $\mathbf{W}_{XL}$ with the help of the Laplace expansion with respect to the first row and then the first column:
\begin{eqnarray}\label{eq:Det_WXL_Det_Jacobi}
\det {\mathbf{W}_{XL}} &=& -\Delta\lambda^2\,\det \mathbf{D}.
\end{eqnarray}
It follows that $\det {\mathbf{W}_{XL}} = 0$ iff $\det \mathbf{D} = 0$, i.e., the emitter sits on a caustic. 
\newpage

\end{appendix}

%

\bibliography{references}

\end{document}